\documentclass[aps,reprint,twocolumn,pra,groupedaddress,floatfix]{revtex4-2}
\usepackage{amsmath,amssymb,amsfonts}
\usepackage{mathrsfs}
\usepackage{stackengine}
\usepackage{soul}
\usepackage{graphicx}
\usepackage{color}
\begin{document}
	\title{Unique multistable states in periodic structures with saturable nonlinearity. I. Conventional case and unbroken $\mathcal{PT}$-symmetric regime}
	\author{S. Vignesh Raja}
	\author{A. Govindarajan}
	\email[Corresponding author: ]{govin.nld@gmail.com}
	\author{M. Lakshmanan}
	\affiliation{$^{*}$Department of Physics, Pondicherry University, Puducherry, 605014, India}
	
	\affiliation{$^{\dagger,\mathsection}$Department of Nonlinear Dynamics, School of Physics, Bharathidasan University, Tiruchirappalli - 620 024, India}
	\begin{abstract}
		In this work, we predict that periodic structures without gain and loss do not exhibit an S-shaped hysteresis curve in the presence of saturable nonlinearity (SNL). Instead, the input-output characteristics of the system admit ramp-like optical bistability (OB) and multistability (OM) curves that are unprecedented in the context of conventional periodic structures in the literature. An increase in the nonlinearity (NL) or the gain-loss parameter increases the switch-up and down intensities of different stable branches in a ramp-like OM curve.  Revival of the typical S-shaped hysteresis curve requires the device to work under the combined influence of frequency detuning and $\mathcal{PT}$-symmetry.  An increase in the detuning, NL and gain-loss parameters reduces the switching intensities of the S-shaped OB (OM) curves. During the process, mixed OM curves that feature a fusion between ramp-like and S-shaped OM curves emanate at low values of the detuning parameter in the input-output characteristics. The detuning parameter values for which ramp-like, S-shaped, and mixed OM appear varies with the NL coefficient. For a given range of input intensities, the number of stable states admitted by the system increases with the device length or NL.  When the laser light enters the device from the opposite end of the grating, nonreciprocal switching occurs at ultra-low intensities via an interplay between NL, detuning, and gain-loss parameters. 
	\end{abstract}
	\maketitle
	\section{Introduction}
	\label{Sec:1}
	In optical fibers, nonlinearities (NLs) arise through intensity-dependent variation in the refractive index (RI). The induced nonlinearity depends on the laser power and the nature of the material. For instance, an ordinary silica fiber features a low nonlinear coefficient value and induces third-order cubic nonlinearity alone, and higher-order nonlinearities are uncommon in them \cite{agrawal2001applications}. Experimental measurements show that the fifth and seventh-order nonlinear coefficients have a pivotal role in controlling the nonlinear phase shift induced by the high-intensity laser in chalcogenide fibers \cite{harbold2002highly,chen2006measurement}. The variations in the doping concentration along the fiber length create inhomogeneous nonlinear profiles \cite{nobrega1998optimum,nobrega2000multistable}.  
	
	Sulfide- \cite{hall1989nonlinear,olbright1986interferometric}, and heavy-metal-doped oxide \cite{kang1995femtosecond} glasses exhibit nonlinear saturation response. These materials possess a faster nonlinear and slower thermal response than silica fibers \cite{yao1985ultrafast}. In such cases, cubic nonlinearity may not accurately
	characterize the induced RI variations \cite{langbein1985nonlinear} at higher input intensities ($P_0$) \cite{da1995dynamics}. In reality, the nonlinear response of such materials cannot increase infinitely, and it saturates beyond a value of input intensity known as critical intensity. In other words, there is a maximum limit for the intensity-dependent RI change. Beyond this limit, the variations in the intensity-dependent RI cease \cite{herrmann1991propagation}. The intensity at which the nonlinear response saturates varies for different materials \cite{abou2011impact}. For instance, the effect of saturable nonlinearity (SNL) comes into play at moderately high intensities in $CdS_{x}Se_{1-x}$ semiconductor-doped glasses fiber \cite{gatz1991soliton}.

	Periodic perturbations in the RI of the fiber in the form of Bragg gratings lead to the reflection of a band of input optical signals. The range of wavelengths reflected by the fiber Bragg grating (FBG) is called the photonic bandgap (PBG) or stopband \cite{erdogan1997fiber}. An association between PBG and intensity-dependent RI promotes the study of several nonlinear effects that include all-optical switching.  The anticipation for alternative solutions to control light with light has accelerated substantial research growth in nonlinear FBGs. Investigations on the steady-state switching dynamics of FBGs mainly target a reduction in the switching intensities \cite{agrawal2001applications}. Switching in an FBG is described by the OB or OM phenomenon \cite{winful1979theory}. As the name suggests, the transmission characteristics of nonlinear FBGs present two or more output states for a given input power. For studying OB in FBGs, researchers employed numerous materials with a variety of nonlinearities  \cite{karimi2012all,lee2003nonlinear,yosia2007double}. 
	
	An alternative way to control the steering dynamics of the FBG is to tailor the characteristics of the  input signal. For instance, tuning the operating wavelength of continuous wave (CW) far away from the Bragg wavelength ($\lambda_b$) reduce the switching intensities \cite{winful1979theory}. In time domain approach, the manipulation of the rise and fall time of the Gaussian and rectangular pulses leads to a reduction in the switching intensities \cite{lee2003nonlinear,ping2005bistability,ping2005nonlinear}.    Tapering the coupling coefficient \cite{radic1995analysis} along the direction of propagation ($z$) is the other method to control the steering dynamics of FBGs.  Structural modifications in the FBG's physical structure can also decrease the switching intensities to a large extent. In particular, a phase-shift region in the middle of the grating leads to the steering at ultra-low powers \cite{radic1995theory}. To date, phase-shifted FBGs are the first choice of researchers for realizing low-power all-optical switch (AOS) in the context of conventional FBGs \cite{suryanto2009numerical,radic1994optical}. In chirped FBGs, spatially varying the grating period ($\Lambda$) along $z$ leads to low-intensity switching \cite{maywar1998effect}.
	
	The primary theme of the article is the study of nonreciprocal switching dynamics in a grating structure in the presence of saturable nonlinearity (SNL). For the steering dynamics to be nonreciprocal, the nonlinear response of the device must be direction-dependent.  In other words, the input-output characteristic curves pertaining to the different light launching directions (left and right) should be distinguishable.  Acquiring asymmetric switching response requires the incorporation of parity and time ($\mathcal{PT}$) symmetry notions \cite{el2007theory, regensburger2012parity, kottos2010optical,feng2017non,ruter2010observation} into the traditional FBG structures \cite{raja2019multifaceted,PhysRevA.100.053806,sudhakar2022low,komissarova2019pt}. Nonreciprocal mode interaction occurs when the periodic perturbation is a complex function of the form $n_0+n_{1R} cos( 2 \pi z/\Lambda) + i sin ( 2 \pi z/\Lambda)$. Physical realization of a parity and time symmetric-FBG (PTFBG) requires gain and loss regions to be placed next to each other in one period $\Lambda$ (unit cell) and periodically repeating the unit cell for the entire device length ($L$). The cosine and sine terms in the expansion of this complex exponential function signify the modulation in the real ($n_{1R}$) and imaginary ($n_{1I}$) parts of the complex RI profile [$n(z)$]. Such an arrangement ensures that the system obeys the $\mathcal{PT}$-symmetric condition $n(z)=n^*(-z)$ \cite{phang2015versatile,phang2014impact,phang2015versatile,govindarajan2018tailoring,govindarajan2019nonlinear}. 
	
	In the context of FBGs, Poladian conceptualized the complex periodic perturbation of the RI profile \cite{poladian1996resonance}, and Kulishov \emph{et al.} carried out the first systematic study on the linear response of the device \cite{kulishov2005nonreciprocal}. The groundwork laid by Kulishov \emph{et al.} paved the way for non-Hermitian Physicists to accomplish significant discoveries in linear PTFBGs \cite{raja2020phase,raja2020phase,raja2021n}.  Lin \emph{et al.} reiterated the same study done by Kulishov \emph{et al.} \cite{kulishov2005nonreciprocal} and pointed out that operating the nonlinear PTFBG at the unitary transmission point leads to the loss of bistable or multistable switching behavior \cite{lin2011unidirectional}. Several years ago, researchers thought that the nonlinear FBG switches function in the unbroken $\mathcal{PT}$-symmetric regime alone \cite{1555-6611-25-1-015102}. Due to the limited range of operation, the field of all-optical switching in nonlinear PTFBGs was not receiving significant research interest in the past. 
	
	Contemporary works on the steady-state switching dynamics of PTFBGs validate that the gain and loss parameter ($g$) impacts in reducing the switch-up and down intensities provided that its magnitude is closer to the value of the coupling parameter ($\kappa$) \cite{raja2019multifaceted, PhysRevA.100.053806,sudhakar2022low}. Subsequently, it is  found that the PTFBGs inscribed on a chalcogenide fiber (that supports higher-order nonlinearities) feature lower switching intensities than the silica based PTFBGs (that supports cubic nonlinearity alone) \cite{raja2019multifaceted}. Operating the PTFBGs at wavelengths other than the Bragg wavelength aids in reducing these switching intensities further. The choice to work at shorter or longer wavelengths must concur with the sign of nonlinearity. In Ref. \cite{raja2019multifaceted}, the existence of ramp-like OB (OM) curves confirm that the broken regime is not an instability domain. Including a chirping inhomogeneity in a PTFBG opens the door for low-power switching as long as the signs of nonlinearity, chirping ($C$), and detuning ($\delta$) parameters match. The interplay between chirping and broken $\mathcal{PT}$-symmetry generates optical bistable and multistable states with high spectral uniformities. Suitable manipulation of the detuning parameter alters the spectral span over which OB and OM occur \cite{PhysRevA.100.053806,sudhakar2022low}. The concept of launching light from the rear end serves as a new route for realizing low-power AOS. Both homogeneous and inhomogeneous PTFBGs exhibit low-intensity switching in the presence of right light incidence conditions \cite{komissarova2019pt,raja2019multifaceted,PhysRevA.100.053806}. Under the modulation of Kerr nonlinearity, the switching intensities get reduced, provided that the sign of chirping and the detuning parameter is negative  ($C$ and $\delta<0$) \cite{sudhakar2022low}. Allowing the nonlinearity to vary inhomogeneously along the propagation length in a PTFBG also helps in the realization of ultralow-power AOS. Such a customization also leads to peculiar OB curves with zero switching intensities in the broken $\mathcal{PT}$-symmetric regime \cite{sudhakar2022inhomogeneous}.

Researchers have investigated different types of nonlinear FBGs and PTFBGs in the past from a switching viewpoint without including SNL. The scientific contributions that deal with the impact of SNL on the dynamics exhibited by non-periodic structures are many. Nevertheless, there seem to exist no works dealing with switching dynamics exhibited by nonlinear periodic structures with SNL. Therefore, we present the mathematical function that describes SNL in periodic structures in Sec. \ref{Sec:2}. Furthermore, this section also deals with the derivation of the first-order differential equations or coupled mode equations of the present system in detail.  The conventional model was used in the literature to study soliton dynamics in periodic structures \cite{malomed2005coupled}. Since the seminal proposal by Merhasin \emph{et al.} \cite{merhasin2007solitons}, the literature does not find any further systematic research on FBGs with SNL from switching or any other application perspectives. For the fist time, we now investigate the switching dynamics shown by a conventional FBG with SNL. We also wish to know whether the SNL parameter alters the characteristics of the hysteresis curves and the switching intensities. Therefore, Sec. \ref{Sec:3} -- deals with the transmission characteristics of conventional FBG with SNL. We take a step further toward the study of steering dynamics exhibited by a PTFBG with SNL. The gain-loss parameter serves as an additional degree of freedom to manipulate the OB curves. Further, it allows us to study the switching in two different $\mathcal{PT}$-symmetric operating regimes, namely, the unbroken and broken $\mathcal{PT}$-symmetric regimes. Section \ref{Sec:4} deals with the input-output characteristics of the proposed system in the unbroken regime for the left light incidence condition.   Inclusion of $\mathcal{PT}$-symmetry also facilitates the study of nonreciprocal switching in FBGs under a reversal in the direction of light incidence (right) in the presence of SNL, which is dealt by Sec. \ref{Sec:5}. The OB curves, stemming from the input-output characteristics of a conventional FBG and an unbroken PTFBG influenced by the SNL term, exhibit peculiar behaviors. Notably, the characteristics of the OB curves differ significantly between existing systems studied in the literature and the proposed ones. This underscores the importance of conducting a comprehensive study of both regimes to understand their responses to variations in system parameters. However, this article, at present, refrains from discussing the outcomes related to the system's operation in the broken regime ($g > \kappa$), considering the unique and extensive dynamics of the system in the absence of gain and loss and in the unbroken regime, which will be addressed separately. Section  \ref{Sec:6} summarizes the important results of the present work.

	\section{Theoretical framework}
	\label{Sec:2}
	The RI distribution [$n(z)$]  of a PTFBG that includes the SNL effect reads as
	\begin{gather}
		\nonumber n(z)=n_0+n_{1R} cos(2 \pi z/\Lambda)+in_{1I} sin(2 \pi z/\Lambda)\\-{n_2}f(|E|^2).
		\label{Eq:1}
	\end{gather}
	Squaring Eq. (\ref{Eq:1}) and neglecting higher-order terms in $n_{1I}$, $n_{1R}$, and $n_2$ results in 
	\begin{gather}
		\nonumber n^2(z)=n_0^2+2 n_o n_{1R} cos(2 \pi z/\Lambda)+2i n_0 n_{1I} sin(2 \pi z/\Lambda)\\+2 n_0 {n_2}f(|E|^2).
		\label{Eq:2}
	\end{gather}
	The function $f(|E|^2)=\cfrac{1}{1+|E|^2}$ better describes the SNL in periodic structures and has a nonlinear term in the denominator \cite{malomed2005coupled,merhasin2007solitons,yulin2008discrete,melvin2006radiationless}. The proposed model is analogous to the discrete version of the Vinetskii–Kukhtarev equation that accounts for the SNL in 1-dimensional optical lattices and waveguide arrays \cite{hadvzievski2004power,vicencio2006discrete}. In Eq. (\ref{Eq:2}), $n_0$ and $n_2$ represent the constant RI and nonlinear coefficient of the FBG, respectively. The RI perturbations ($n_{1R}$, $n_{1I}$ and $n_2$) are small compared to the core RI ($n_0$). At a given operating wavelength ($\lambda$), the coupling parameter ($\kappa$) dictates the amount of coupling  between the counter-propagating fields, and its mathematical representation reads $\kappa=\cfrac{\pi n_{1R}}{\lambda}$. The balanced gain and loss levels ($g$) supplied to achieve $\mathcal{PT}$-symmetry depend on the modulation of $n_{1I}$, and the relation between these two parameters reads as  $g=\cfrac{\pi n_{1I}}{\lambda}$. The saturable nonlinearity parameter ($S$) is mathematically related to the nonlinear coefficient of the material ($n_2$) via the mathematical expression, $S =\cfrac{2 \pi n_2}{\lambda}$ \cite{raja2019multifaceted,PhysRevA.100.053806,sarma2014modulation}. The incident optical field ($E$) is the superposition of the forward ($E_f$) and backward ($E_b$) field distributions, and it reads as \cite{erdogan1997fiber}
	\begin{gather}
		E(z)=E_f \exp(ikz)+E_b \exp(-ikz),
		\label{Eq:3}
	\end{gather}
	where $k$ signifies the magnitude of the wave vector. Obtaining the governing equations that describe the propagation of fields in a PTFBG requires the substitution of squared RI given in Eq. (\ref{Eq:2}) and electric field distribution (Eq. (\ref{Eq:3})) in the time-independent Helmholtz equation given below \cite{raja2020phase}:
	\begin{gather}
		\cfrac{d^2E}{dz^2}+k^2\left(\cfrac{n^2(z)}{n_0^2}\right)E=0.
		\label{Eq:4}
	\end{gather} 
	While expanding Eq. (\ref{Eq:4}), derivative terms like $E_f^{''}$ and $E_b^{''}$ can be neglected using the slowly varying envelope approximation (SVEA) \cite{lin2011unidirectional,miri2012bragg}. Along these lines, the rapidly oscillating  exponential terms of the form $\exp[\pm i(2 \pi z/\Lambda+kz)]$ also get neglected \cite{raja2020phase}. The four-wave mixing (FWM) terms $E_f^* E_b$ and $E_b^* E_f$ are assumed to have no significant impact on the propagation \cite{komissarova2019pt,sudhakar2022low,malomed2005coupled}. Under these circumstances, retaining the self-phase modulation (SPM) and cross-phase modulation (XPM) terms is sufficient while expanding the nonlinearity. The ratio between the SPM and XPM terms is 1:1 (mathematically) \cite{malomed2005coupled,merhasin2007solitons}. Also, the equations are further rearranged for the forward and backward propagating fields with retention of the phase mismatch and effective feedback terms. With these assumptions, the resulting equation reads
	
	\begin{gather}
		\nonumber	iE_f^{'}\exp(ikz)-iE_b^{'}\exp(-ikz)\\\nonumber+(\kappa+g)E_b\exp[ i(2 \pi z/\Lambda-kz)]\\\nonumber+(\kappa-g)E_f\exp[-i(2 \pi z/\Lambda-kz)]\\-S\cfrac{E_f \exp(ikz)+E_b \exp(-ikz)}{1+|E_f|^2+|E_b|^2}=0.
		\label{Eq:5}
	\end{gather}
	
	The nonlinear first-order differential equations for the transmitted and reflected waves read as
	\begin{gather}
		\nonumber iE_f^{'}+(\kappa+g)E_b\exp[ i(2 \pi z/\Lambda-2 kz)]\\-\cfrac{S E_f}{(1+|E_f|^2+|E_b|^2)}=0,
		\label{Eq:6}
	\end{gather}
	\begin{gather}
		\nonumber iE_b^{'}+(\kappa-g)E_f\exp[ -i(2 \pi z/\Lambda-2 kz)]\\-\cfrac{S E_b}{(1+|E_f|^2+|E_b|^2)}=0.
		\label{Eq:7}
	\end{gather}
	
	From the fundamentals of FBGs, the detuning parameter ($\delta$) that indicates the deviation in the operating wavelength ($\lambda$) from the Bragg wavelength ($\lambda_b$) reads as $\delta=k-\pi/\Lambda=2\pi n_0\left(\cfrac{1}{\lambda}-\cfrac{1}{\lambda_b}\right)$. Numerically, it is possible to separate the detuning parameter from the coupling term by adopting a transformation $u,v=E_{f,b} \exp(\mp i \delta z)$ \cite{porsezian2005modulational}, and the resulting equations read
	\begin{gather}
		\frac{d u}{dz}=i\delta u+i \left(k + g\right)v-\cfrac{i S u}{(1+ |u|^{2}+|v|^{2})},
		\label{Eq:8}
	\end{gather}
	\begin{gather}
		-	\frac{d v}{dz}=i\delta v+i \left(k - g\right)u-\cfrac{i S v}{(1+ |u|^{2}+|v|^{2})}.
		\label{Eq:9}
	\end{gather}
	
	These equations are valid for left light incident conditions. Under a reversal in the direction of light incidence (right), the term $\kappa + g$ in Eq. (\ref{Eq:8}) changes to $\kappa - g$. Similarly, the term $\kappa - g$ in Eq. (\ref{Eq:9}) is replaced by $\kappa + g$.
	
	Before delving into the theoretical investigation of the system based on the above governing model, it is crucial to address why its consideration holds merit. To do so, we first highlight how this mathematical model distinguishes itself from those already discussed in the existing literature \cite{malomed2005coupled,merhasin2007solitons}. Previous studies in the literature focus on exploring the dynamics of solitons and their interesting collision properties in photorefractive crystal based one dimensional optical lattices and bulk longitudinal gratings. Although the fundamental study pertaining to the stable nature of solitons with phase-matched condition has been quite extensively investigated in these systems \cite{malomed2005coupled,merhasin2007solitons}, both the fundamental and application perspectives of continuous wave (CW) remain unexplored to date. In the present investigation, we analyze the switching properties of CW states in the FBGs with an additional term known as the detuning parameter indicating a substantial difference between the Bragg wavelength and the wavelength of the input light.

	The inclusion of $\mathcal{PT}$-symmetry terms sets the current governing model apart from those found in the literature, presenting a distinctive feature that diverges from established formulations and introducing a unique dimension to the theoretical framework, enabling the study of OB in diverse operating conditions, including both unbroken and broken regimes, under two different light incidence conditions (left and right). This multifaceted approach, made possible by the presence of $\mathcal{PT}$ symmetry terms, is notably impossible in existing models where the absence of such terms limits the feasibility of exploring such scenarios. Similarly, the incorporation of the detuning parameter into the system further distinguishes the present model from existing ones, enabling the study of nonlinear characteristics at nonsynchronous wavelengths. This addition not only enhances the practicality of the approach at phase mismatched conditions but also provides an additional degree of freedom to manipulate the characteristics of OB/OM curves.

	We use the well-known implicit Runge-Kutta fourth-order method to solve the system of coupled equations in (\ref{Eq:8}) and (\ref{Eq:9}) with the following boundary conditions
	\begin{gather}
		u(0)=u_0 \quad \text{and} \quad v(L)=0.
		\label{Eq:10}
	\end{gather}
	The input and output intensities read as $P_0 = |u_0|^2$ and $P_1(L)=|u(L)|^2$, respectively.
	\subsection{Overview of OB}
	Before we present the simulation results, we provide an overview of the OB and OM phenomena. As the input intensity ($P_0$) varies, the output intensity [$P_1(L)$] increases linearly. A sudden jump in the output intensity occurs at one particular value known as switch-up intensity ($P_{th}^{\uparrow}$) 
	indicating an onset of a second stable branch of the OB, and the mechanism is commonly known as switch-up action. A part of the input-output curve corresponding to input intensities lying between zero and switch-up represents ($0<P_0<P_{th}^{\uparrow}$)  the first stable branch of the OB curve. Once the system switches to the second branch, output remains in it for a given range of input intensities. In the case of OM, switching to the successive stable states happen at distinct switch-up intensity values.  Tuning the input intensity in the reverse direction completes the s-shaped hysteresis curve. During this process, the system does not return to the previous stable branch at the same switch-up intensity value.  But, it returns to its previous state at another critical intensity known as the switch-down intensity ($P_{th}^{\downarrow}$)
	during the switch-down mechanism. For any value of input intensities between the switch-up and down values ($P_{th}^{\downarrow}<P_0<P_{th}^{\uparrow}$), the output of the system is bistable. The difference between the critical switch-up and down intensities dictates the width of the hysteresis curve ($\Delta P_{th}$ = $P_{th}^{\uparrow} - P_{th}^{\downarrow} $).
	
	\subsection{Choice of device length and coupling coefficient}
	It is well-known that the feedback offered by the system is an essential ingredient besides the intensity-dependent RI for the OB/OM to occur in FBGs. Insufficient coupling strength inhibits the formation of OB. On the other hand, the value of $\kappa$ cannot be arbitrarily large in PTFBGs \cite{raja2019multifaceted, PhysRevA.100.053806,sudhakar2022low}. Optimizing the device length ($L$) is essential for the desirable OB/OM curves to appear.  In practice, the value of the coupling parameter ($\kappa$) ranges from 1 to 10 $cm^{-1}$ \cite{agrawal2001applications}. In the literature, we could find FBGs fabricated in a wide range of physical lengths ranging from 1 $mm$ to 20 $cm$ \cite{broderick2000nonlinear}. In our numerical experiments, we observe that desirable OB curves in the input-output characteristics of a FBG with SNL occur when the coupling coefficient is reduced to a value that is approximately ten times less than the values we used in our previous works \cite{raja2019multifaceted, PhysRevA.100.053806}. Therefore,  the coupling coefficient ($\kappa$) is assumed to have a value of 0.4 $cm^{-1}$ throughout this article (unless specified).  However, the product of these two parameters ($\kappa L$) will never go beyond the permitted numerical values (1 to 50) as the reduction in the coupling is compensated by the increment in the device length \cite{raja2019multifaceted}. With this note, we now look into the nonlinear transmission characteristics of the system under different operating conditions. The nonlinear PTFBG works in the unbroken $\mathcal{PT}$-symmetric regime under the condition $\kappa > g$ \cite{miri2012bragg,sarma2014modulation}. At the unitary transmission point, an inevitable phenomenon is the possibility of the breaking of the $\mathcal{PT}$-symmetry in the system, when $\kappa = g$. Above this condition, the system operates in the broken regime where $\kappa > g$. An alternative perspective on the boundary of $\mathcal{PT}$-symmetry breaking can be gained by investigating the dispersion curves supported by the same system (but with partial differential equations taking into account the time co-ordinate), which was recently done by Tamilselvan \emph{et al.} \cite{tamilselvan2023modulational} in the case of modulational instability analysis.   We investigate the proposed system for different values of NL parameters at two different lengths ($L = 20$ and 70 $cm$).
	
		\subsection{Some practical considerations}
	It is important to emphasize that although observing the OB and OM curves in FBGs is possible experimentally, it requires addressing several practical challenges, including the identification of a suitable fiber material that offers SNL at relatively low power.  From the available scientific literature, it is evident that semiconductor-doped glass having strong SNL can possess a third-order nonlinear coefficient ($n_2$) in a range varying from $10^{-15}$ to $10^{-13}$ $m^2/W$ \cite{ironside1988nonlinear,hall1989nonlinear}.  $CdS_{1-x}Se_{x}$ is an example of this type of glass material characterized by a fast nonlinear response time of $10^{-11}$ second and a significant nonlinearity value \cite{ironside1988nonlinear,jain1983degenerate,acioli1988measurement,roussignol1987new}. When dealing with physical units, it is important to take into account the effective area of the fiber ($A_{eff}$) in the mathematical formula for the calculation of saturable nonlinearity \cite{agrawal2000nonlinear}. With the value of $A_{eff}$ assumed to be 100 $\mu m^2$, the saturable nonlinearity ($S$) for the above material is found to be 0.5927 $W^{-1}/cm$ at the operating wavelength of 1060 $nm$ and $n_2 = 10^{-15} m^2/W$ \cite{coutaz1991saturation,ironside1988nonlinear,kang1995femtosecond} (note that a conventional silica fiber, on the other hand, exhibits a third-order nonlinearity value of $2.6 \times 10^{-20}$ $m^2/W$ \cite{agrawal2000nonlinear,yosia2007double,ping2005bistability,ping2005nonlinear}).
		
	Along these lines, the third-order susceptibility [$\chi^{(3)}$] of sulfide and heavy-metal doped oxide glass with a range of RI 2.19 -- 2.5 lies, respectively, within a range of 3.1 -- 5.6 $\times$ $10^{-13}$ esu at 1.06 $\mu$m \cite{kang1995femtosecond,borrelli1995resonant} and 1.2$\pm$0.4 -- 7.9$\pm$2.4 $\times$ $10^{-13}$ esu at 1.25 $\mu$m \cite{kang1995femtosecond}. For instance, the third-order nonlinear coefficient [$n_2 = 24 \pi/n_0$ $\chi^{(3)}$] of $PbO$(60)$TeO_2$(25)$SiO_2$(15) at 1.25 $\mu$m measures to be 3.322 $\times$ $10^{-12}$ $m^2/W$, provided that the RI is 2.27 and $\chi^{(3)} = 1 \times 10^{-13}$ \cite{kang1995femtosecond,borrelli1995resonant}.
		  Some other examples of this type of glass materials are $GeS_{2}(87.3)Ga_{2}S_{3}(13.7)$,  and $La_2S_3(35)Ga_2S_3(65)$ \cite{kang1995femtosecond}. 
	
In light of these facts, we present some practical values that can aid experimental physicists in creating low-power OB/OM curves influenced by SNL. These values are listed in Table \ref{tab7}. For comparative purposes, the values that were used for the formation of OB and OM curves are also presented from the already existing literature.
	
	\begin{table*}
		\caption{\centering{Comparison of various device parameters used in physical units }}
		\begin{center}
			\begin{tabular}{c c c c }
				\hline
				\hline
				{Symbol}&{Device } & {Physical} & {Values} \\
				{}&{parameter} & {values} & {used in the literature} \\
				\hline
				
				{$n_2$}&{third-order nonlinear}&{$10^{-15}$ -- $10^{-13}$ $m^2/W$ \cite{ironside1988nonlinear,kang1995femtosecond}}&{$2.6 \times 10^{-20}$ $m^2/W$ \cite{lee2003nonlinear,ping2005bistability,shum2007optical}}\\
				{}&{coefficient}&{}&{2.2 $\times 10^{-16}$ $m^2/W$ \cite{ping2005nonlinear}}\\
				{}&{}&{}&{2.7 $\times 10^{-13}$ $cm^2/W$ \cite{yosia2007double}}\\
				{}&{}&{}&{2 $\times 10^{-17}$ $m^2/W$ \cite{phang2013ultrafast}}\\
				{}&{}&{}&{}\\
				
				{$A_{eff}$}&{effective area of}&{100 $\mu m^2$ }&{1 -- 100 $\mu m^2$ \cite{agrawal2000nonlinear}}\\
				{}&{the fiber}&{}&{}\\
				{}&{}&{}&{}\\
				{}&{}&{}&{}\\

				{$S$}&{saturable nonlinearity}&{0.5 -- 6 $W^{-1}/cm$}&{1.0544 $\times$ $10^{-5}$ $W^{-1}/m$ \cite{lee2003nonlinear,ping2005bistability,shum2007optical}}\\
				{}&{}&{}&{}\\
				{}&{}&{}&{}\\

				{$L$}&{device length}& {20 $cm$} &{1 $cm$ \cite{ping2005bistability,ping2005nonlinear,yosia2007double,shum2007optical}, 3.5 $cm$ \cite{lee2003nonlinear}}\\
				{}&{}&{}&{}\\

				{$\kappa$}&{coupling coefficient}& {0.4 $cm^{-1}$} &{0.8 $cm^{-1}$ \cite{lee2003nonlinear}, 5 $cm^{-1}$\cite{ping2005bistability,shum2007optical}}\\
				{}&{}&{}&{ }\\
				{}&{}&{}&{} \\

				{$g$}&{gain-loss }& {0 (conventional)} &{}\\
				{}&{coefficient}&{0 -- 4 $cm$ $^{-1}$}&{0 - 1200 $cm$ $^{-1}$ \cite{phang2015versatile, phang2014impact}}\\
				{}&{}&{(unbroken regime)}&{} \\
				{}&{}&{}&{} \\
				
				{$n_0$}&{RI of the core }& {2.27} &{2.19 -- 2.5 \cite{kang1995femtosecond,borrelli1995resonant}}\\
				{}&{}&{}&{}\\
				
				{$\lambda_b$}&{Bragg wavelength} & {1060 nm} &{1000 nm \cite{phang2013ultrafast,phang2014impact}} \\
				{}&{}& {} & {}\\
				
				{$\lambda_b$$-$$\lambda$}&{variations in operating} & {$\pm$ 0.05 nm}&{-0.015 nm \cite{yosia2007double}, 0.125 nm \cite{shum2007optical,ping2005bistability}} \\
				{}&{ wavelength ($\lambda$) from $\lambda_b$}& {} &{}\\
				{}&{}& {}  &{}\\

				{$\delta$}&{detuning parameter}& {0 -- 2.5 $cm^{-1}$} &{0.005 $cm^{-1}$ \cite{lee2003nonlinear}, -0.9611 $cm^{-1}$ \cite{ping2005bistability},  }\\
				{}&{}&{}&{-1.8039 $cm^{-1}$ \cite{ping2005nonlinear}, 4.74 $cm^{-1}$ \cite{shum2007optical},  }\\
				{}&{}&{}&{4.75 $cm^{-1}$ \cite{lee2003nonlinear}}\\
				{}&{}&{}&{}\\

				{$P_0$ and $P_1(L)$ }&{input and output }&{0 -- 15 $MW/cm^2$,}&{0 -- 400 $MW/cm^2$ \cite{yosia2007double},}\\
				{}&{intensities}&{}&{0 -- 100 $GW/cm^2$ \cite{lee2003nonlinear}}\\
				{}&{}&{}&{0 -- 30 $GW/cm^2$ \cite{shum2007optical,ping2005bistability}}\\
				{}&{}&{}&{0 -- 4 $\times$ $10^{14}$ $W/m^2$ \cite{phang2014impact,phang2015versatile}}\\
				
				\hline\hline
			\end{tabular}
			\label{tab7}
		\end{center}
	\end{table*}
	
The table suggests that in the presence of SNL, the device length should be increased by at least ten times to observe OB and OM curves, while the coupling coefficient must be reduced in an appropriate way to maintain the feedback parameter ($\kappa L$) in the optimal range. For the conventional FBG without $\mathcal{PT}$-symmetry, the gain and loss parameters are zero, but in the unbroken regime of an FBG with $\mathcal{PT}$-symmetry, they can take values between 0--4 $cm^{-1}$. Note that the coupling and gain-loss coefficients at the unitary transmission point can be equal to 4 $cm^{-1}$.

The existing literature predominantly covers experimental studies on the nonlinear-optical effects of heavy-metal and sulfide-glass at near-infrared wavelengths between 1 and 1.250 $\mu$$m$. For instance, Borrelli  \emph{et al.} have carried out a systematic study on the nonlinear properties of these glasses at 1.06 $\mu$m \cite{borrelli1995resonant}. Kang's similar study \cite{kang1995femtosecond} on the nonlinear optical properties of these materials at 1.25 and 1.06 $\mu$$m$ has provided us an important information pertaining to the selection of the Bragg wavelength ($\lambda_b$) for the proposed system, which falls at a value of 1060 $nm$ wavelength. Given an assumed RI of 2.27 for the background material [$PbO(60)TeO_2(25)SiO_2(15)$], a difference of $0.05$ $nm$ between the operating wavelength of the system and $\lambda_b$ can induce a detuning value ($\delta$) of 6.3472 $cm^{-1}$, (for which the grating period is calculated to be, $\Lambda = 233.48$ $nm$).
	
Similarly, another essential experimental challenge is finding a suitable CW source that can deliver a high $kW$ power in the given wavelength range. YLR CW Ytterbium fiber lasers that are available (commercially) are highly suitable for this purpose because of their high stability, efficiency, beam quality, and long lifetime. They can provide an input power of up to 4 $kW$ to the system \cite{CWsource}. It is important to note that intensity is a measure of power confined per unit area of the core. For instance, if the effective area of the core is 100 $\mu$$m^2$ and the input power is in Watts, the input intensities ($P_0$) that are required to create OB and OM states can be in the order of $MW/cm^2$ \cite{shum2007optical, phang2014impact, phang2015versatile}.

Developing PTFBGs with complex RI profiles can be accomplished through external pumping in fibers doped with rare-earth dopants. Nonetheless, selecting a suitable dopant material that can generate the desired gain and loss regions is a notable challenge. To overcome this obstacle, adding $Er^{3+}$ and $Cr^{3+}$ dopants onto a suitable glass substrate to develop gain and loss regions, respectively, is a promising option.

	\section{OB/OM in conventional Bragg structures with SNL}
	\label{Sec:3}
	
	\begin{figure}
		\centering	\includegraphics[width=0.5\linewidth]{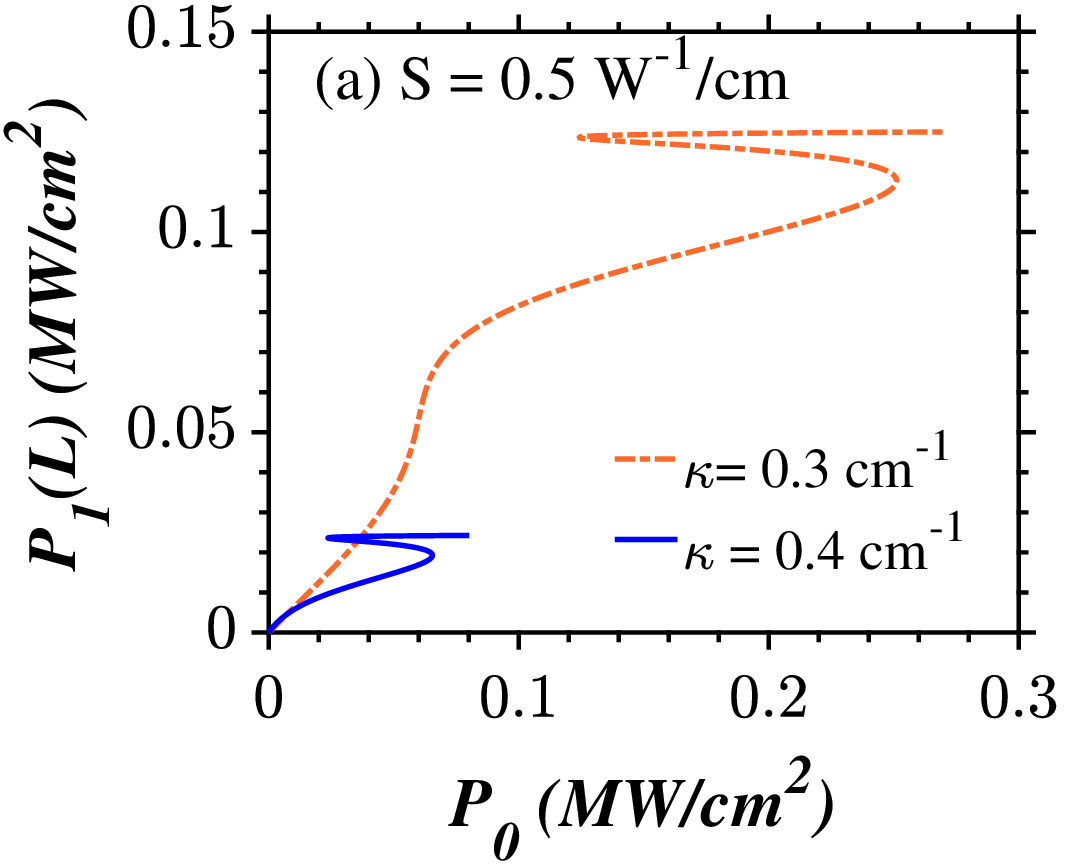}\includegraphics[width=0.5\linewidth]{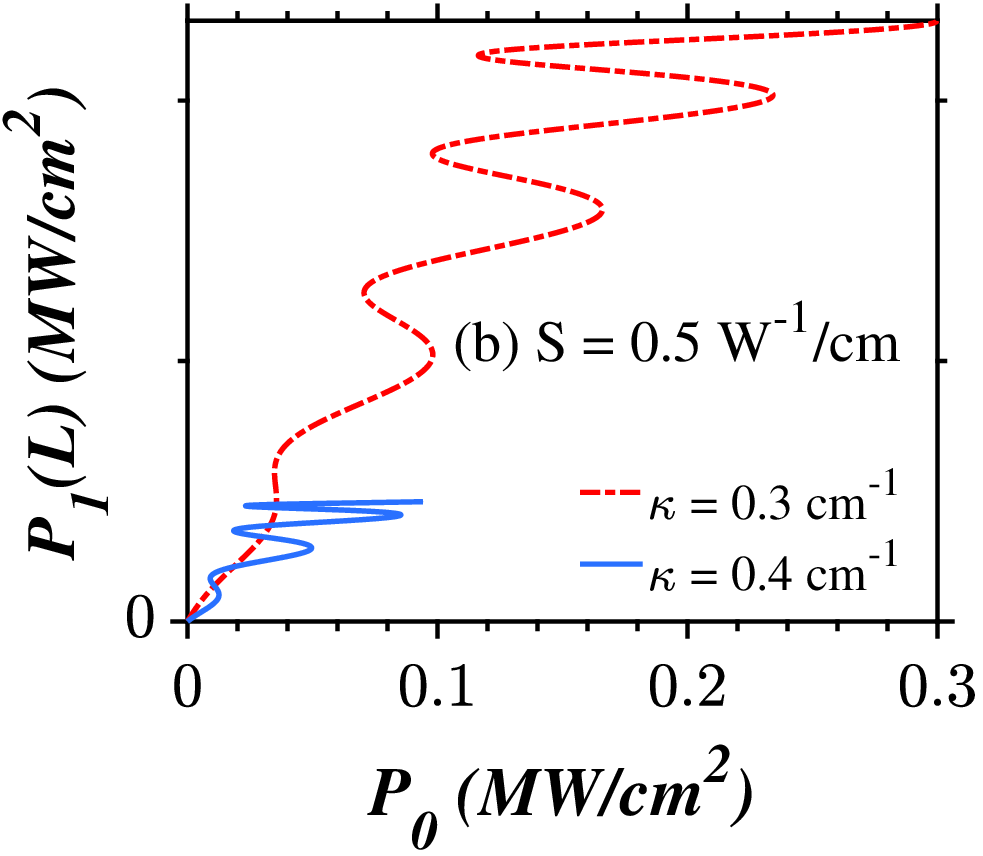}\\\includegraphics[width=0.52\linewidth]{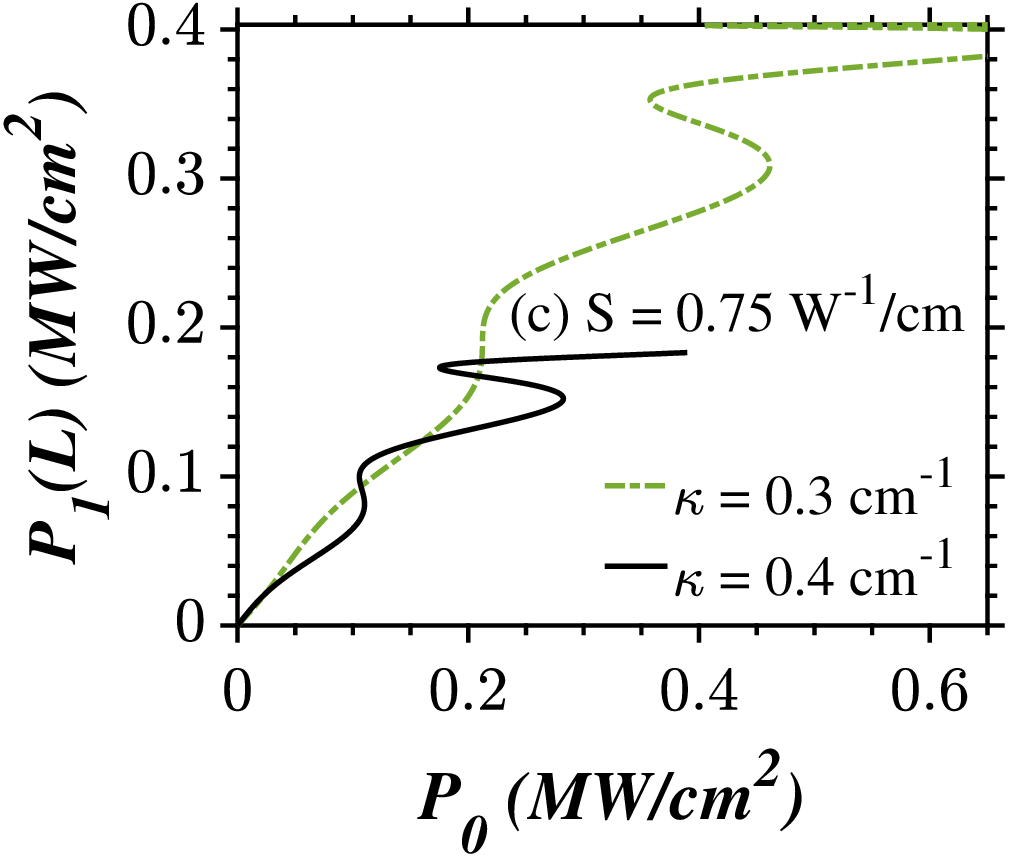}\includegraphics[width=0.52\linewidth]{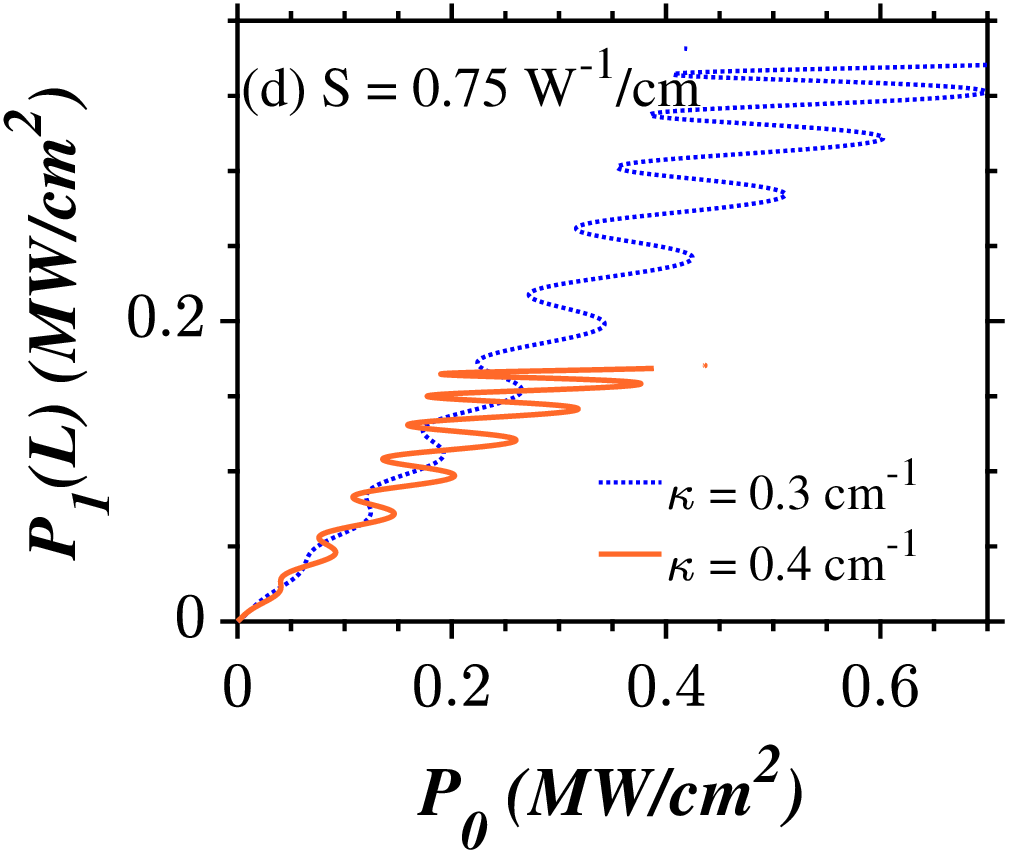}\\\includegraphics[width=0.5\linewidth]{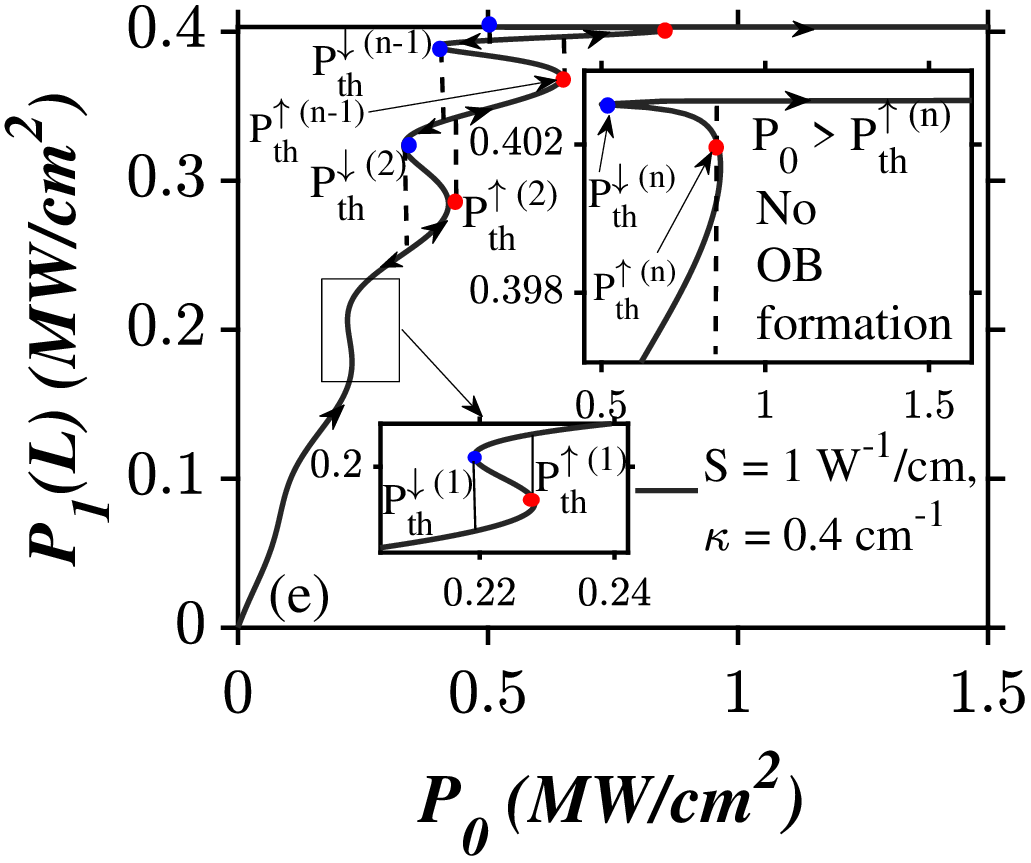}\includegraphics[width=0.5\linewidth]{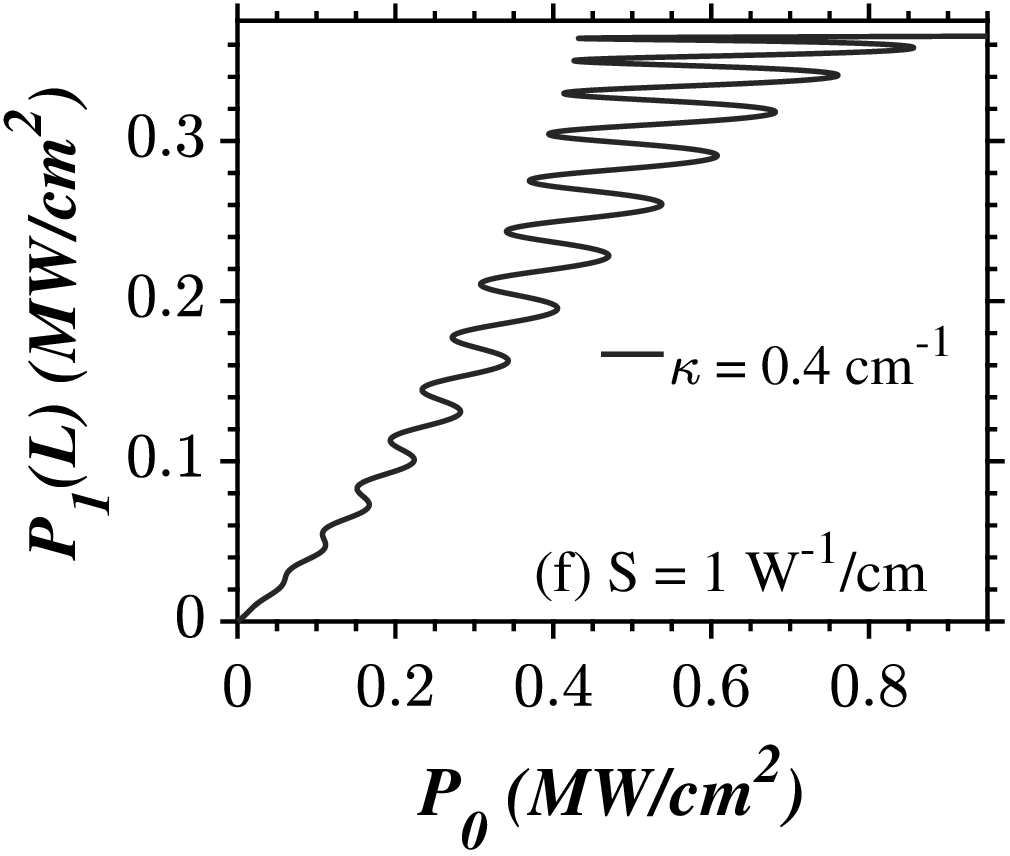}\\\includegraphics[width=0.5\linewidth]{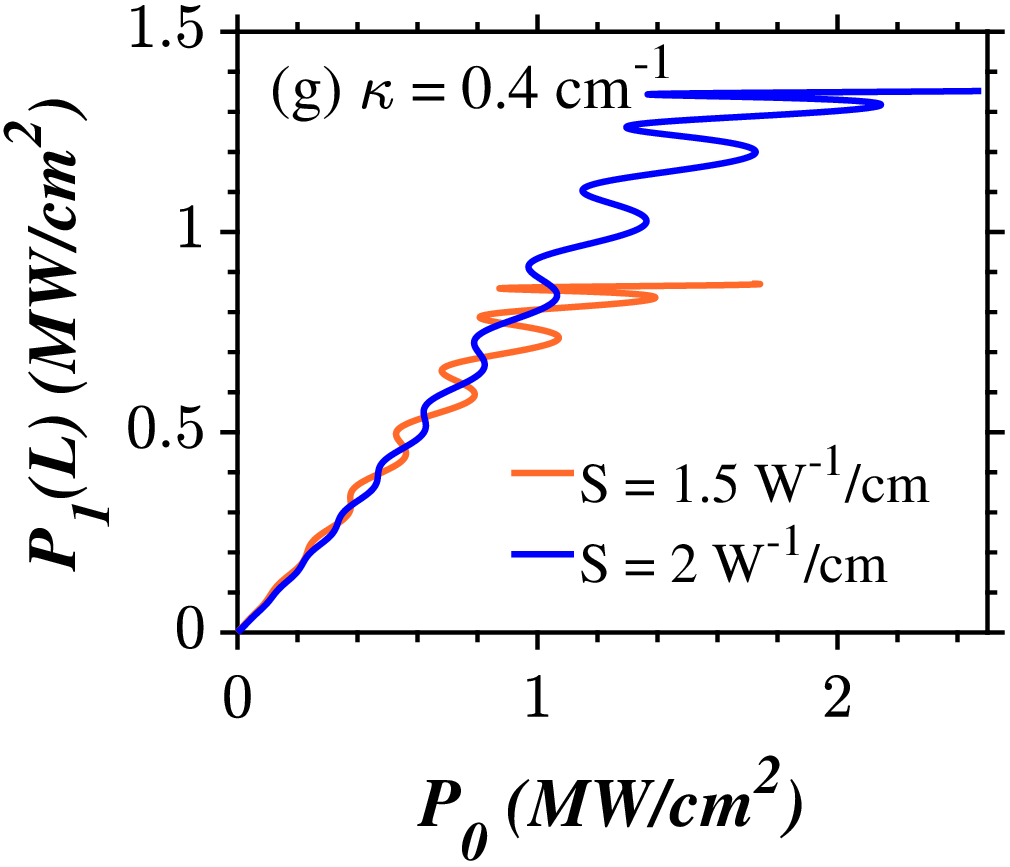}\includegraphics[width=0.5\linewidth]{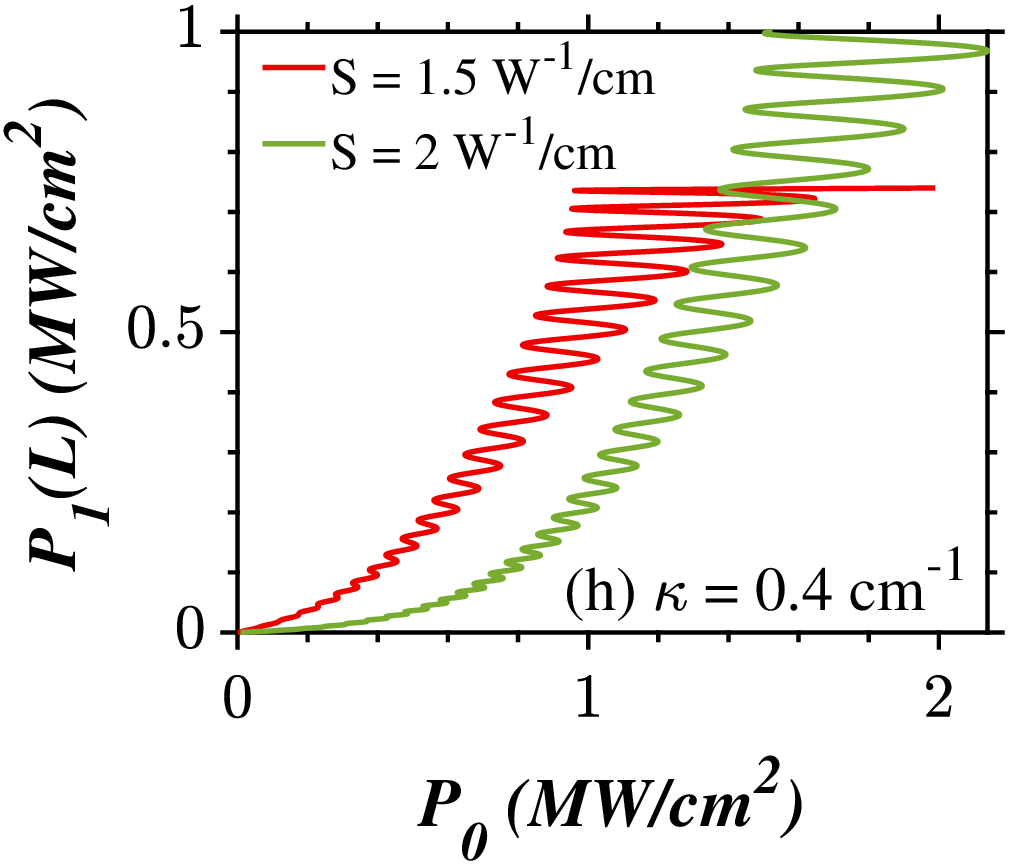}
		\caption{OB (OM) curves exhibited by a conventional FBG with SNL at $\delta$ = 0 $cm^{-1}$. The device length assumes a value of $L = 20$ and 70 $cm$ in the left and right panels, respectively. Role of (a)--(d) coupling coefficient ($\kappa$) and (c) -- (h) NL on OB (OM) curves.}
		\label{fig1}
	\end{figure}
	
	As the input intensity varies from zero, the output intensities increase sharply, leading to the formation of a ramp-like first stable state, as shown in Fig. \ref{fig1}.  The output of the system jumps from the first to the second stable state at the switch-up intensity and remains in it for a finite increase in input intensities as delineated in Fig. \ref{fig1}(a). Since the curve in Fig. \ref{fig1}(a) features a ramp-like first stable state, it is called a ramp-like OB curve to differentiate it from the conventional S-shaped OB curve that features a gradual variation in the output against the input intensities. Along these lines, if multiple stable branches appear on the top of a ramp-like first stable state, they can be referred to as the ramp-like OM curves, provided that the variations in the output against the input intensities are sharp (abrupt) in all the stable branches. 
	
	As we tune the nonlinearity parameter gradually, a transition from the ramp-like OB ($S < 1$ $W^{-1}/cm$) to the ramp-like OM  ($S \ge 1$ $W^{-1}/cm$) is visible in Figs. \ref{fig1} (c) and (e). The number of stable states in the ramp-like OM curve increases with an increase in the NL.  For an easier understanding, we  name the stable branches in a specific order, namely following the first stable branch in Fig. \ref{fig1}, they are named as $2^{nd}$, $3^{rd}$, $\dots$, $n-1$, and $n$. Let $P_{th}^{\uparrow (1)}$, $P_{th}^{\uparrow (2)}$, $\dots$, and $P_{th}^{\uparrow (n)}$ be the input intensities required to switch-up from first to second, second to third, $\dots$ and $n-1$ to $n^{th}$ branch, respectively. The corresponding switch-down values are given by $P_{th}^{\downarrow (1)}$, $P_{th}^{\downarrow (2)}$, $\dots$, and $P_{th}^{\downarrow (n)}$. Once the system switches to the $n^{th}$ stable branch, no further switching in the form of OB or OM is supported by the system for any increase in input intensity ($P_0 > P_{th}^n$), as confirmed by the numerical simulations. 
	
	 From Figs. \ref{fig1}(a), (c), (e) and (g), we infer that the width and switching intensities of the first hysteresis curve are low. However, when the input intensity increases, the width of the successive stable branches increases in a ramp-like OM curve and their switch-up and down intensities decrease with an increase in the coupling parameter ($\kappa$), as shown in Figs. \ref{fig1}(a) -- (d).  Additionally, the output intensities fall steeply with an increase in the coupling coefficient.    An increase in the NL parameter ($S$) leads to an increase in the output, switch-up, and down intensities of all the stable branches.  In the left panel of Fig. \ref{fig1}, we find that the number of stable states supported by the system for smaller device lengths is less.  We can generate ramp-like OM curves with more stable states by tuning the device length in the simulations, as shown in the right panel of Fig. \ref{fig1}. In other words, the higher the value of $L$, the higher the number of stable states.    Before switching, the input-output curves admit a ramp-like first stable state, as shown in Figs. \ref{fig1}(b), (d), (f) and (h).  
	
	Generally, nonlinear FBGs display an S-shaped hysteresis curve in their input-output characteristics \cite{winful1979theory,karimi2012all,lee2003nonlinear,ping2005bistability,radic1994optical,radic1995analysis,radic1995theory}. On the contrary, the input-output characteristics of the proposed system display ramp-like OB and OM curves. In the literature, we can find these kinds of OM curves in coupled active ring resonators \cite{zhang2012novel} and photonic metamaterial multilayers with graphene sheets \cite{wen2020tunable}.   What differentiates the ramp-like OM curves in Refs. \cite{raja2019multifaceted, PhysRevA.100.053806} from the ramp-like OM curves admitted by the proposed system is as follows: The width of the successive stable branches decreases with an increase in the input intensities in Refs. \cite{raja2019multifaceted, PhysRevA.100.053806}. On the other hand, in the proposed system of FBG with SNL, the converse effect occurs, i.e., the width of the successive stable branches increases with an increase in the input intensities. 
	
	\section{OB in the unbroken $\mathcal{PT}$-symmetric regime: Left incidence}
	\label{Sec:4}
	
	\subsection{Controlling the switching dynamics at Bragg condition with $\mathcal{PT}$-symmetry}
	\label{Sec:4A}
	\begin{figure}[h!]
		\centering	\includegraphics[width=0.5\linewidth]{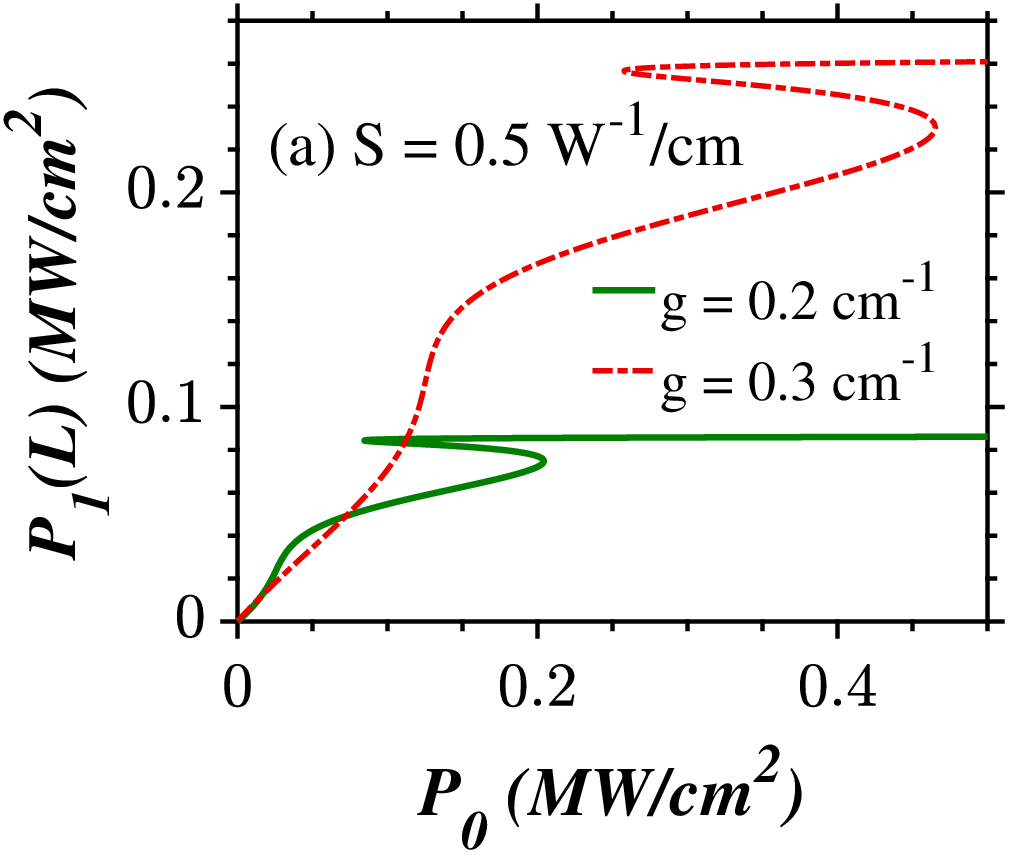}\includegraphics[width=0.5\linewidth]{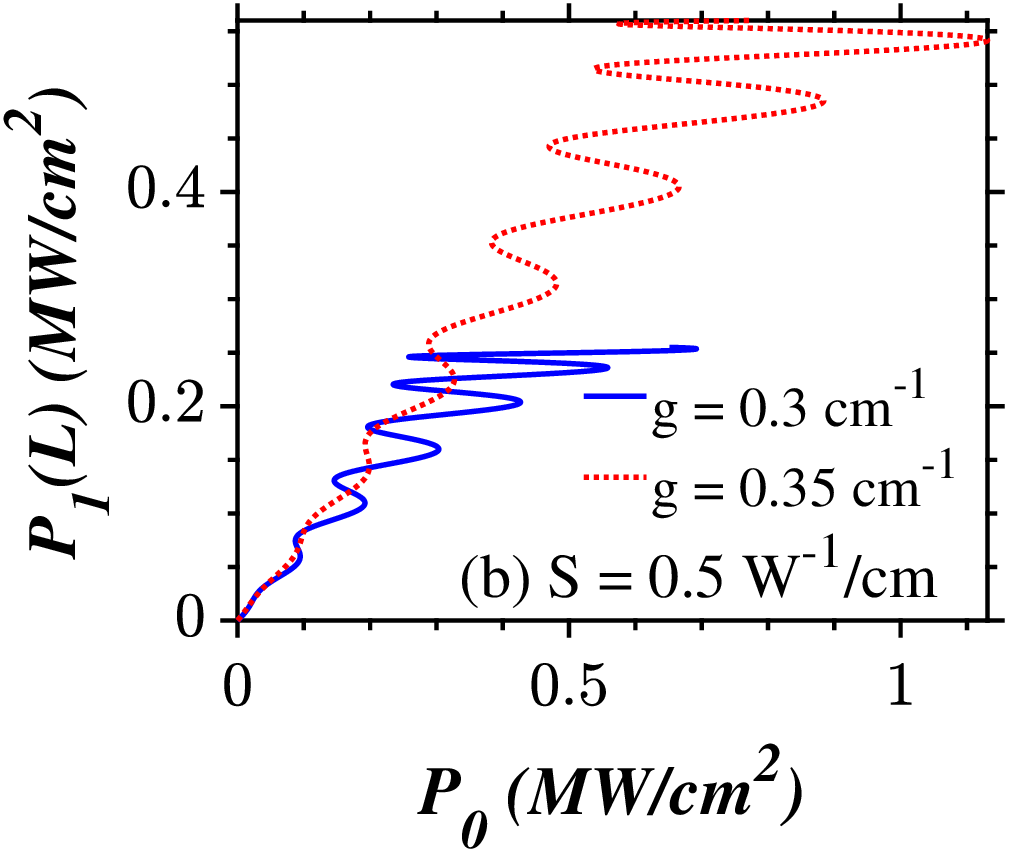}\\\includegraphics[width=0.5\linewidth]{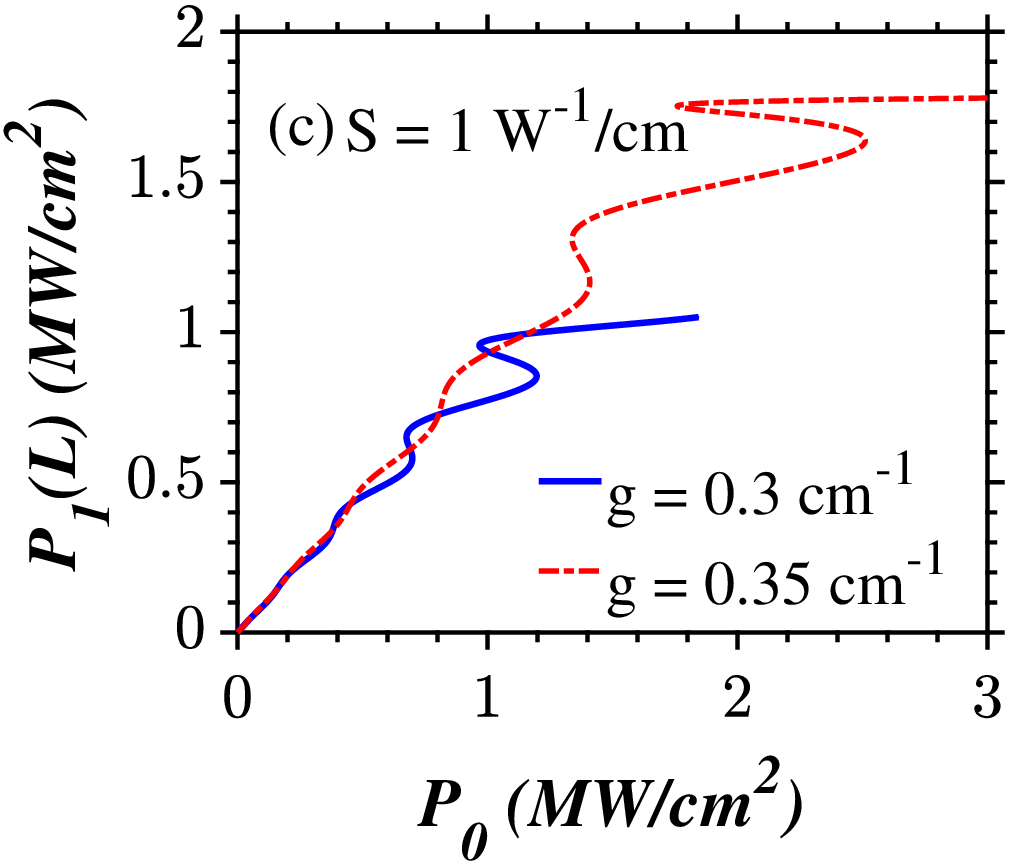}\includegraphics[width=0.5\linewidth]{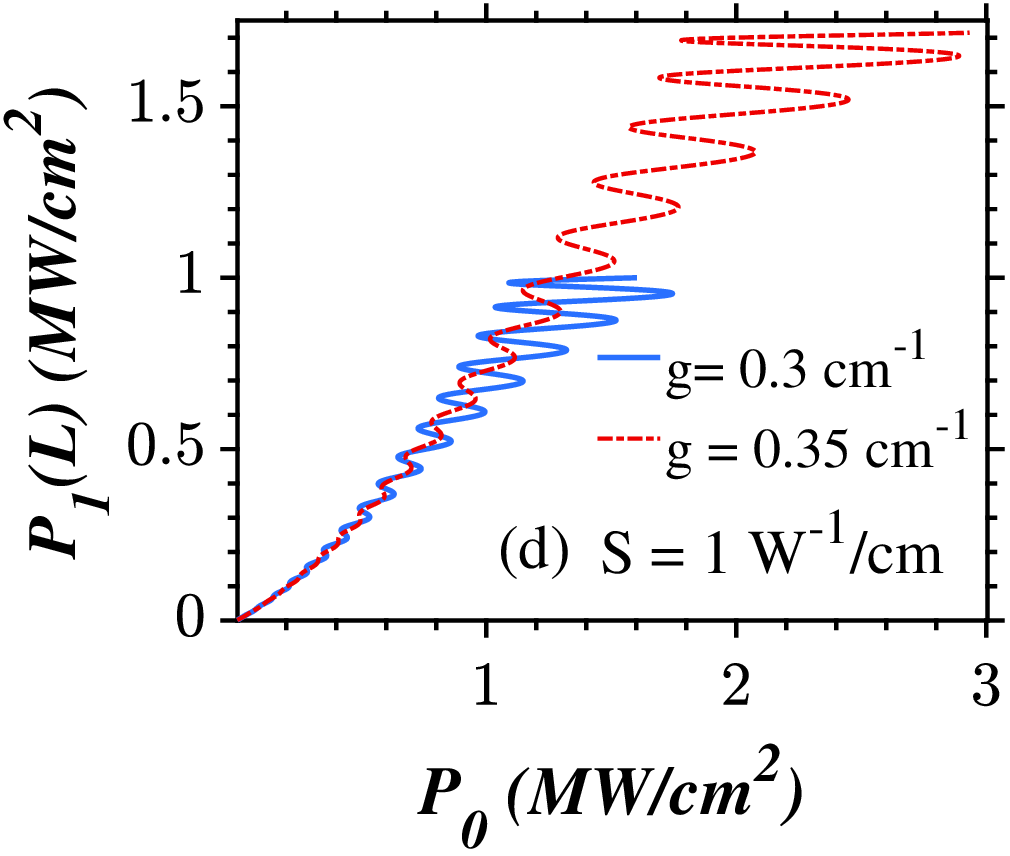}\\\includegraphics[width=0.5\linewidth]{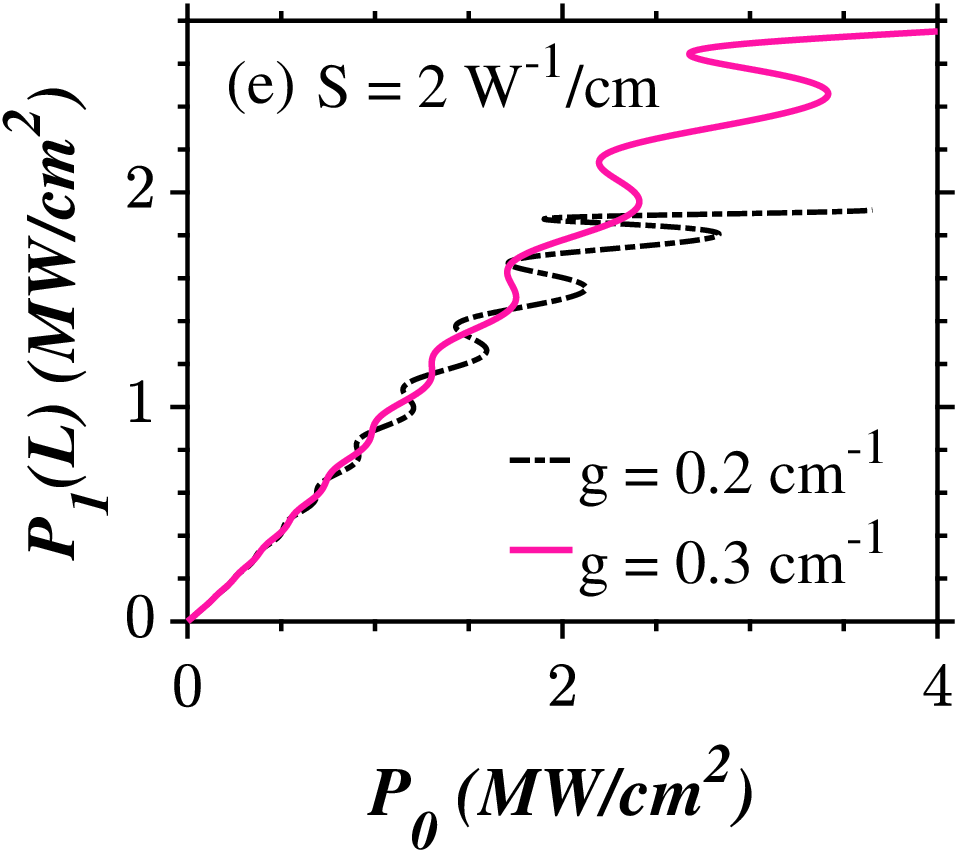}\includegraphics[width=0.5\linewidth]{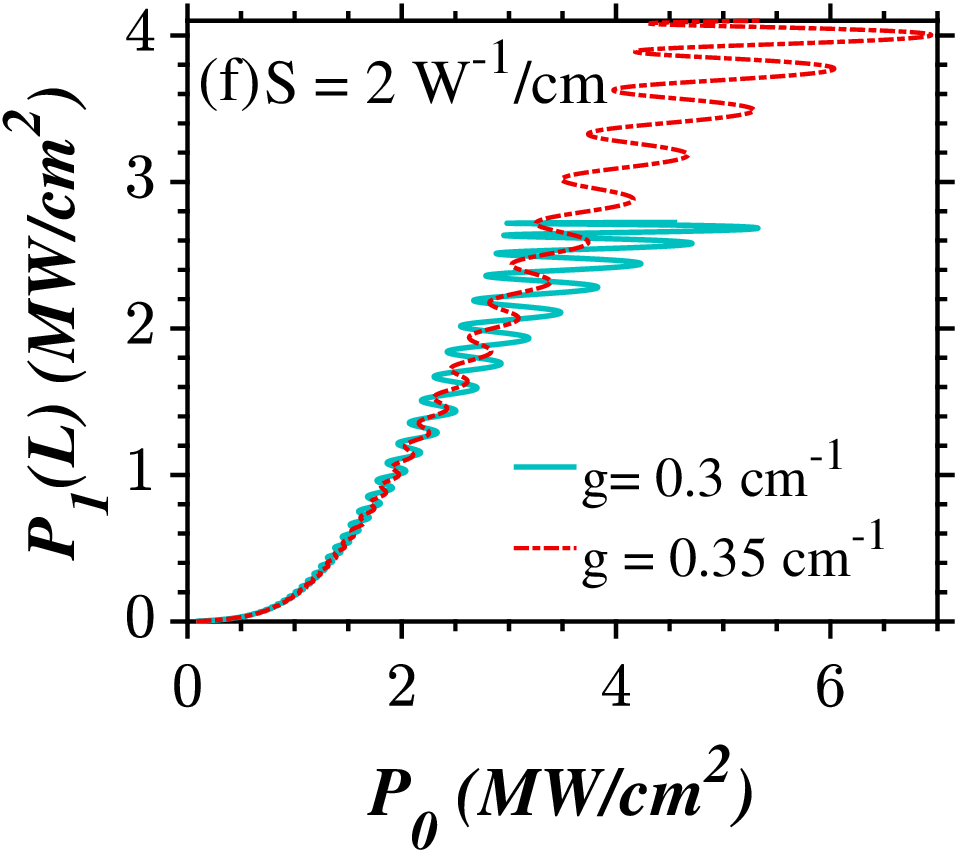}\\\includegraphics[width=0.5\linewidth]{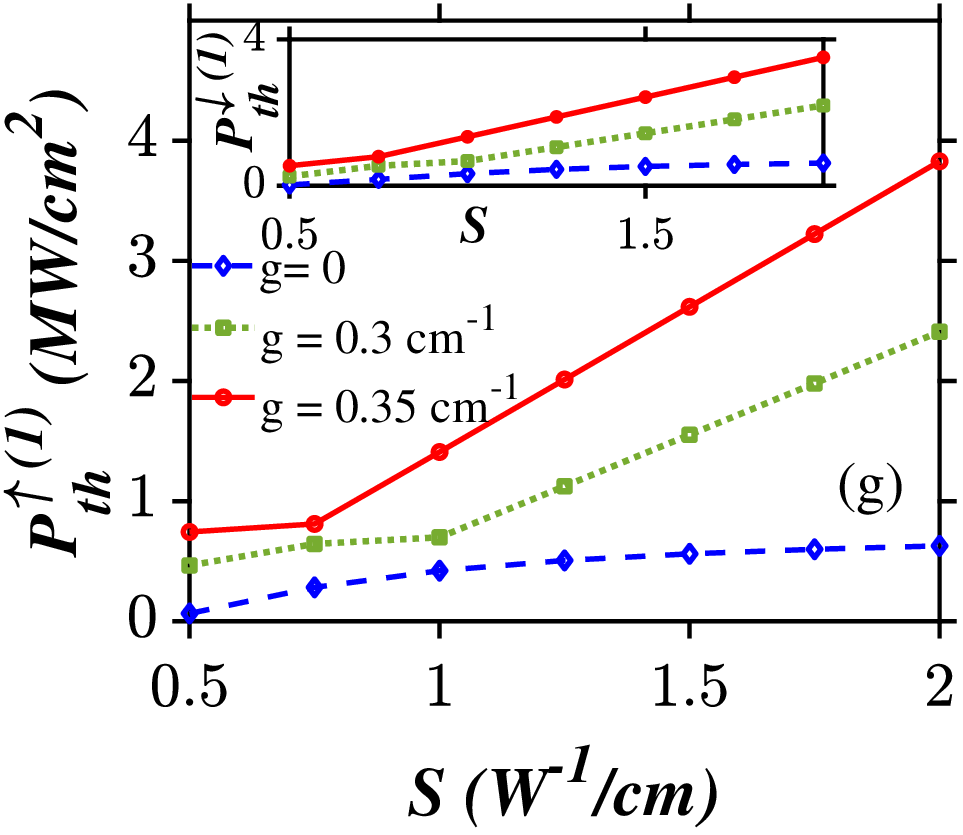}\includegraphics[width=0.5\linewidth]{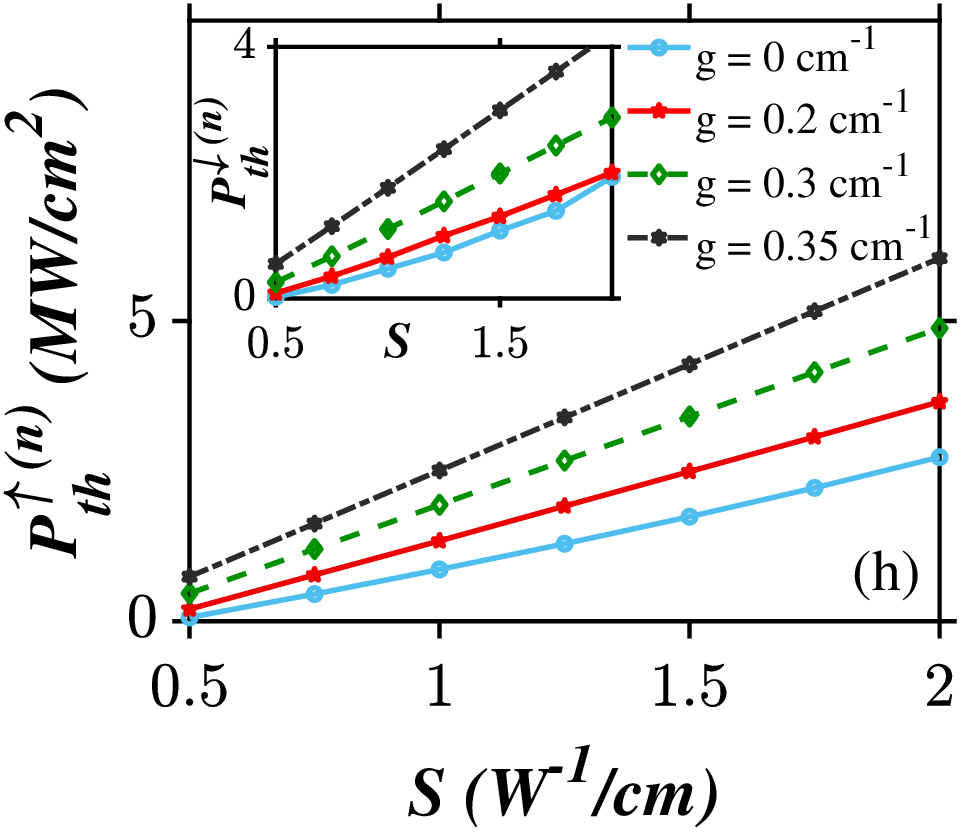}
		\caption{(a) -- (f) Variations in the ramp-like OB (OM) curves against the gain and loss parameter ($g$) at $\kappa = 0.4$ $cm^{-1}$.  (g) and (h) Continuous variation of first and $n^{th}$ switch-up and down intensities, respectively. The device length assumes a value of $L = 20$ and 70 $cm$ in the left and right panels, respectively. }
		\label{fig2}
	\end{figure}

	In the previous section, we have investigated the OB (OM) behavior in a conventional FBG with SNL. In this section, we present the results pertaining to the OB (OM) behavior induced by the impact of $\mathcal{PT}$-symmetry in the unbroken regime. As we tune the input intensities, the system's output varies sharply along the ramp-like first stable branch. As a consequence of the increase in the NL parameter, the ramp-like OB ($S < 1$ $W^{-1}/cm$)  curves transform into ramp-like OM ($S \ge 1$ $W^{-1}/cm$), as shown in Figs. \ref{fig2}(a), (c), and (e). An increase in the device length increases the number of stable states for a given range of input intensities, as shown in the right panel of Fig. \ref{fig2}. The width of each hysteresis curve is broader than its former in these figures. These results are similar to the observations made from Figs. \ref{fig1}(a), (c), and (e).  Compared to the conventional case discussed in Fig. \ref{fig1}, the switch-up, down and output intensities of different stable branches of the ramp-like OM curves increase with an increase in the gain and loss parameter in the unbroken regime for all the values of $S$, as depicted in Figs. \ref{fig2}(a), (c),(e), and (g). 
	\subsection{Retrieving S-shaped OB via frequency detuning}
	\label{Sec:4B}
	\begin{figure*}
		\centering	\includegraphics[width=0.25\linewidth]{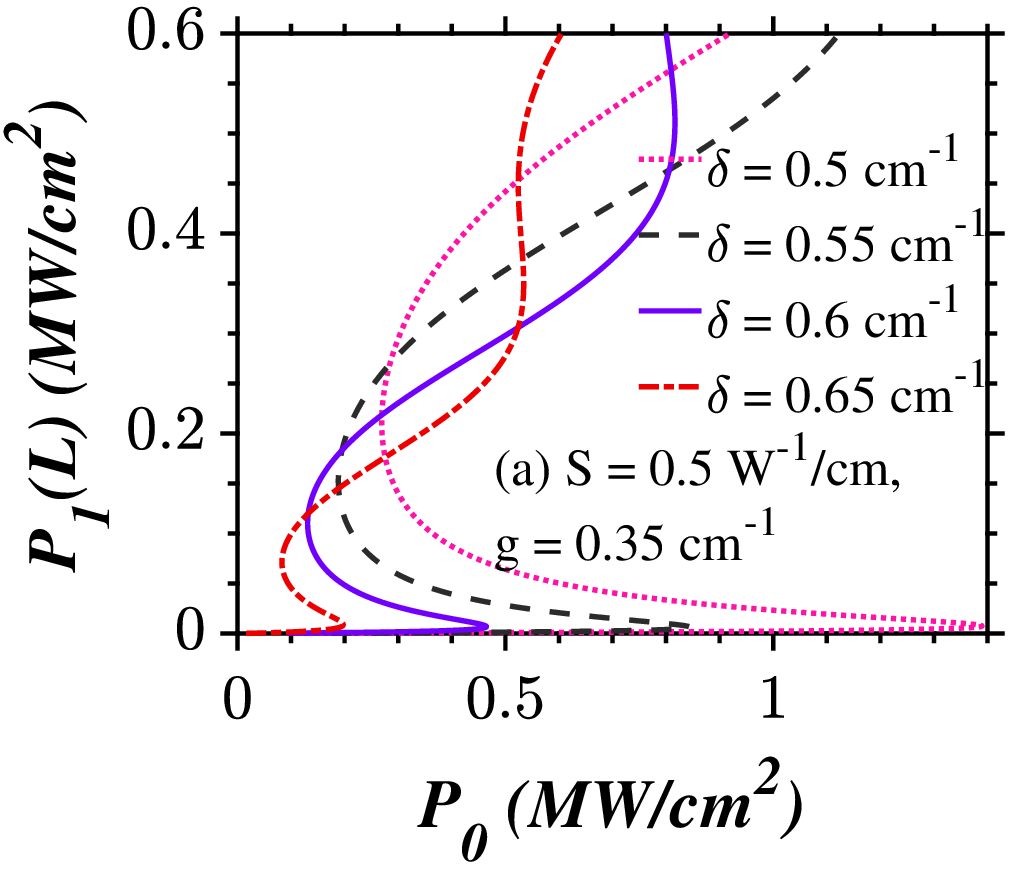}\includegraphics[width=0.25\linewidth]{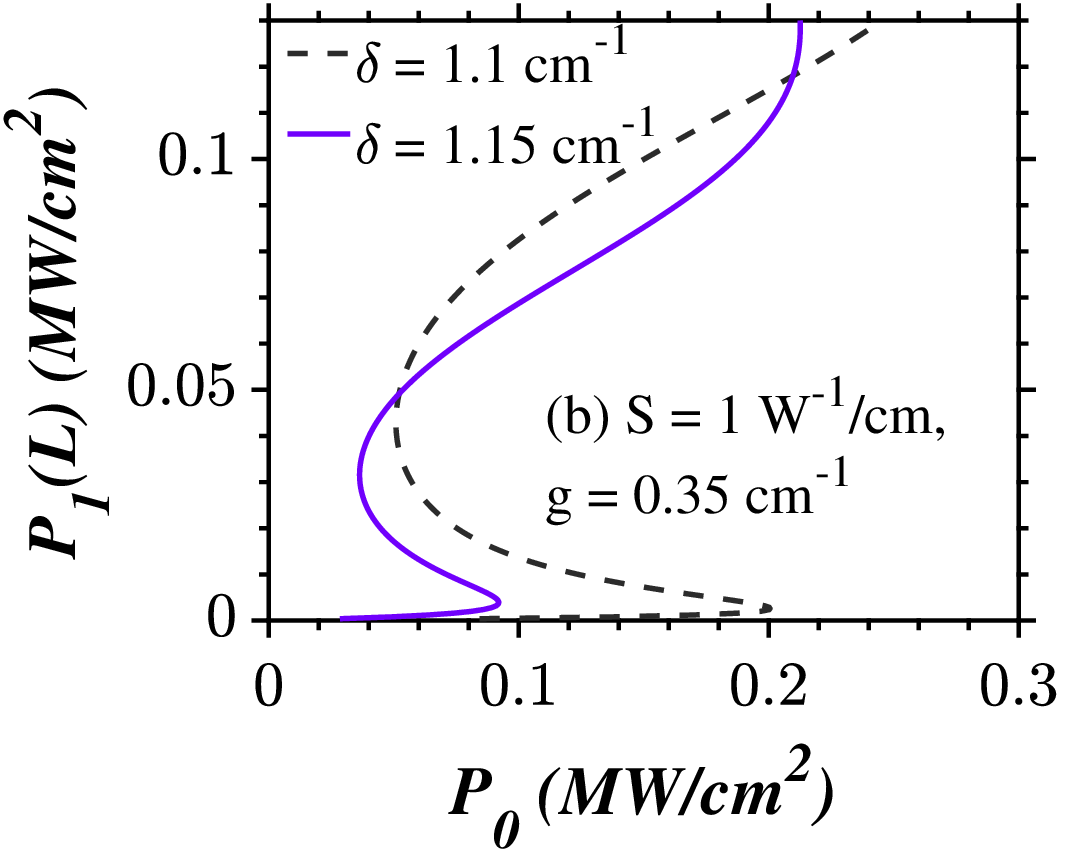}\includegraphics[width=0.25\linewidth]{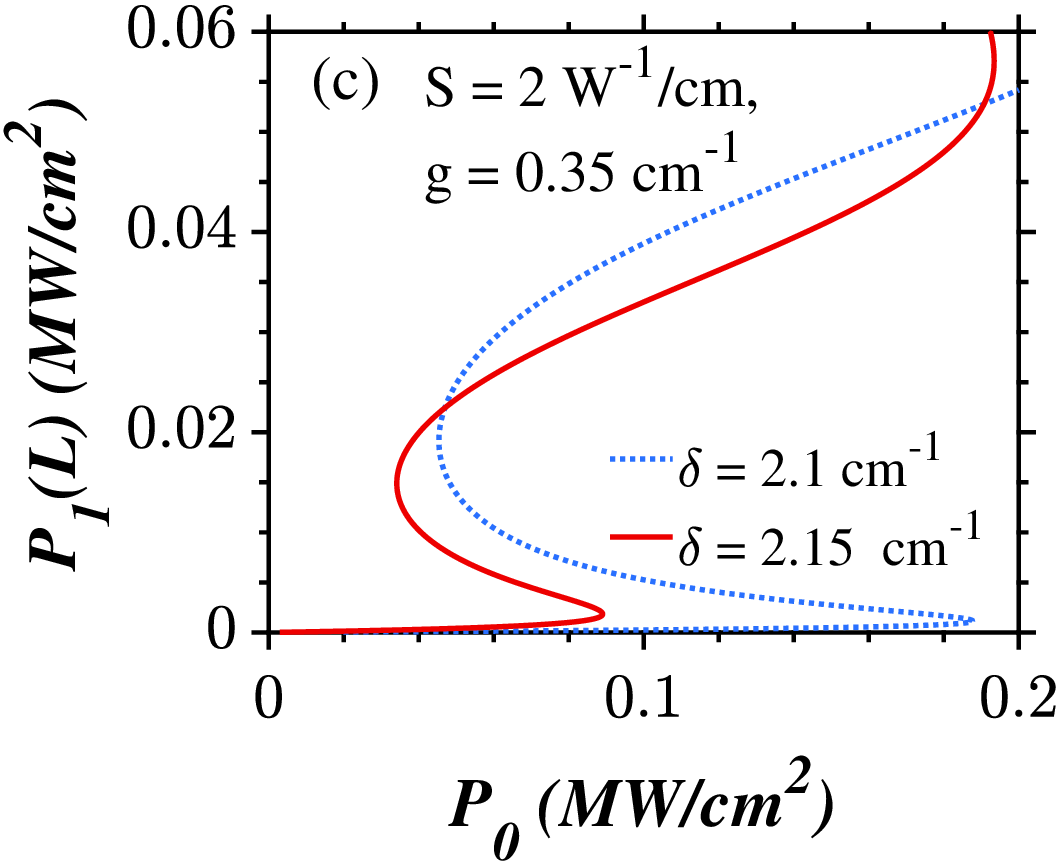}\includegraphics[width=0.25\linewidth]{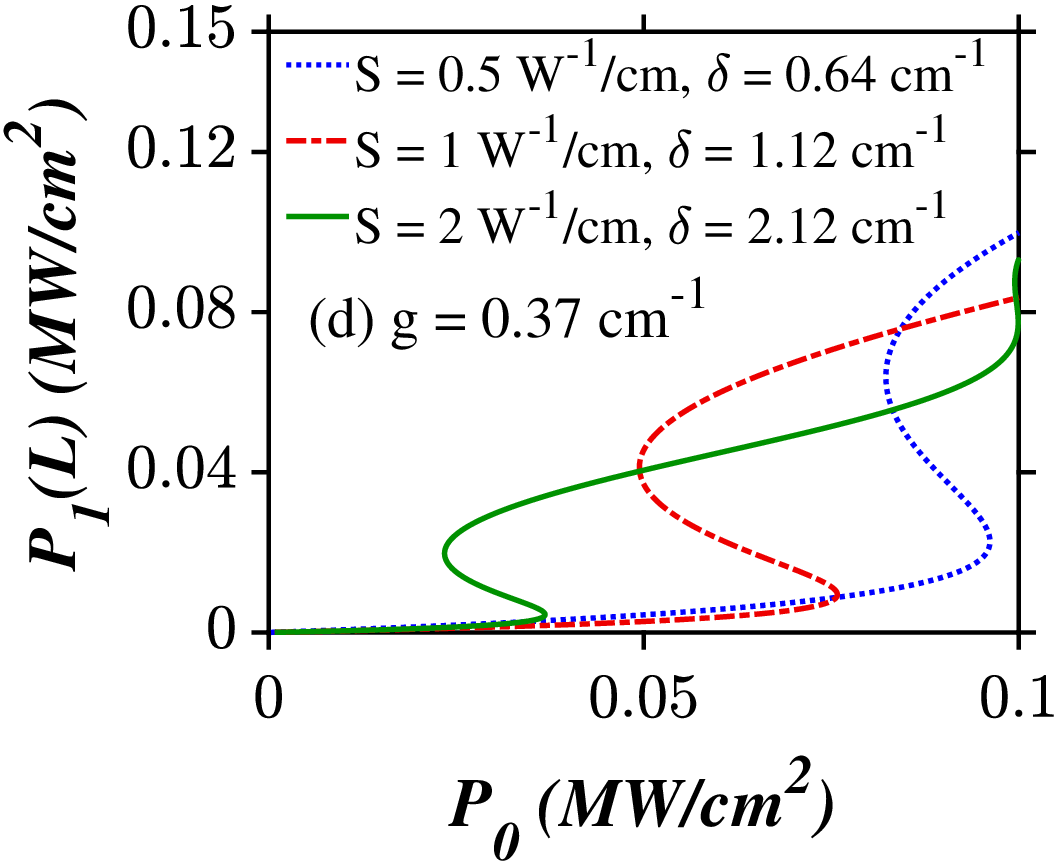}\\\includegraphics[width=0.25\linewidth]{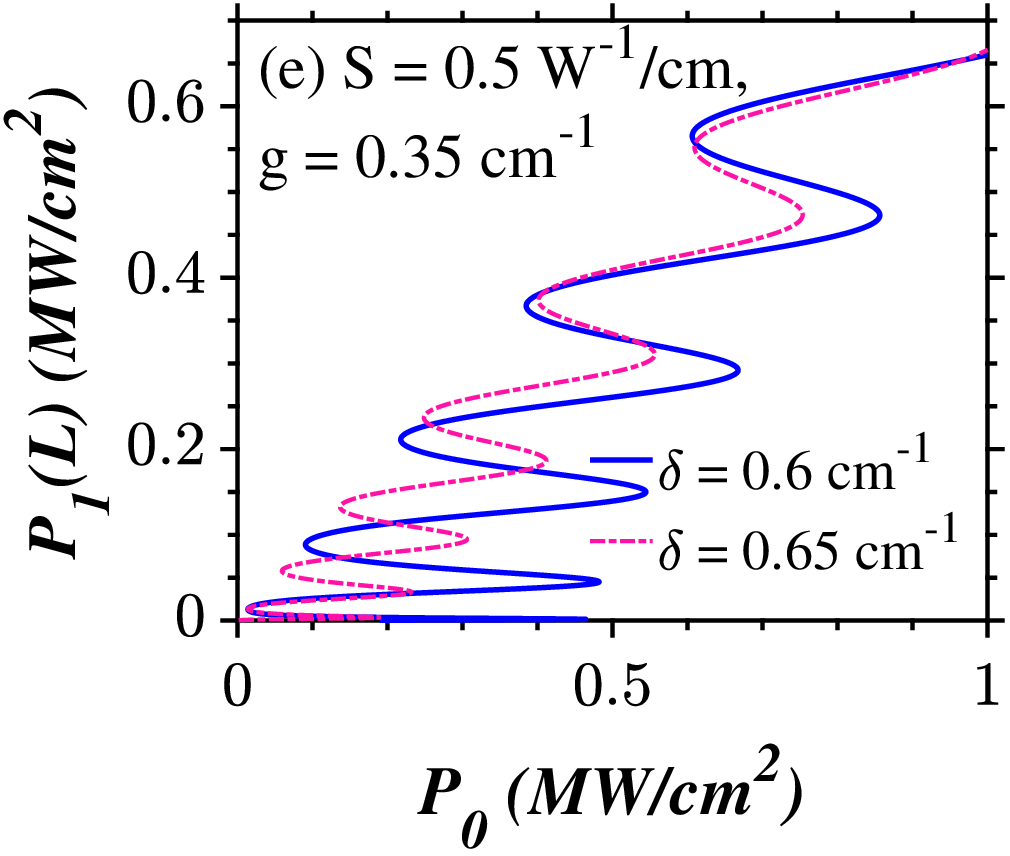}\includegraphics[width=0.25\linewidth]{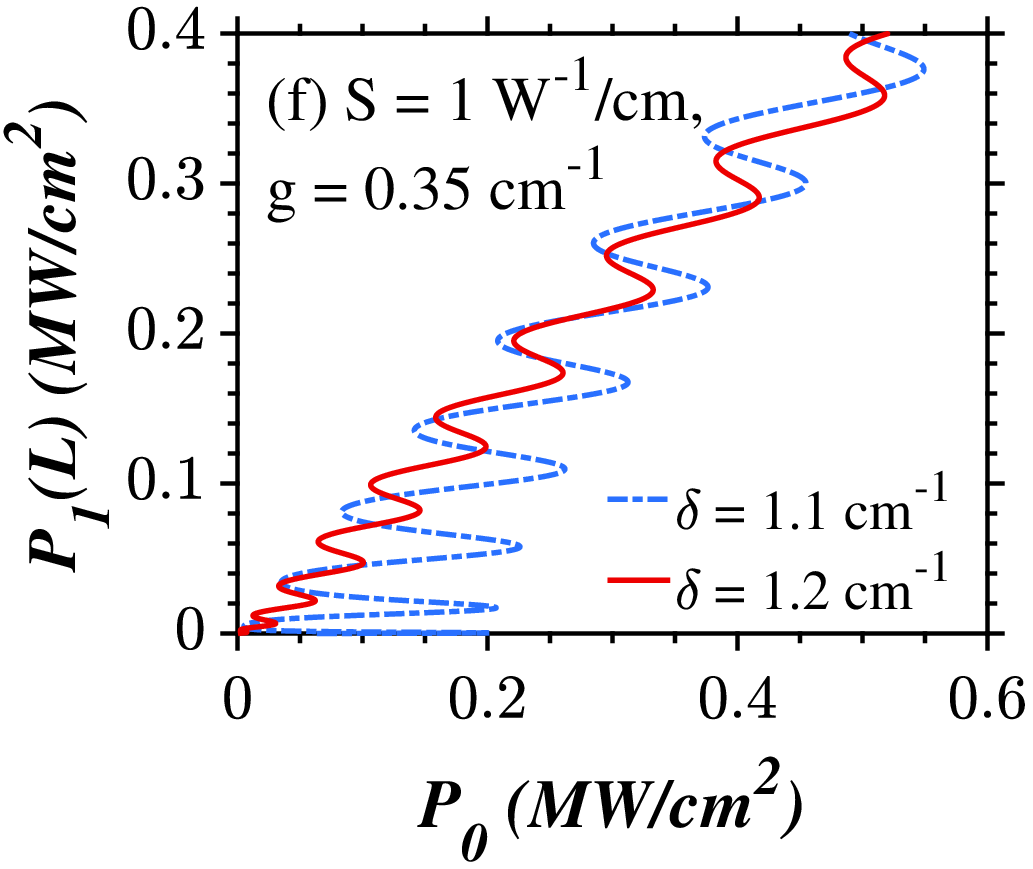}\includegraphics[width=0.25\linewidth]{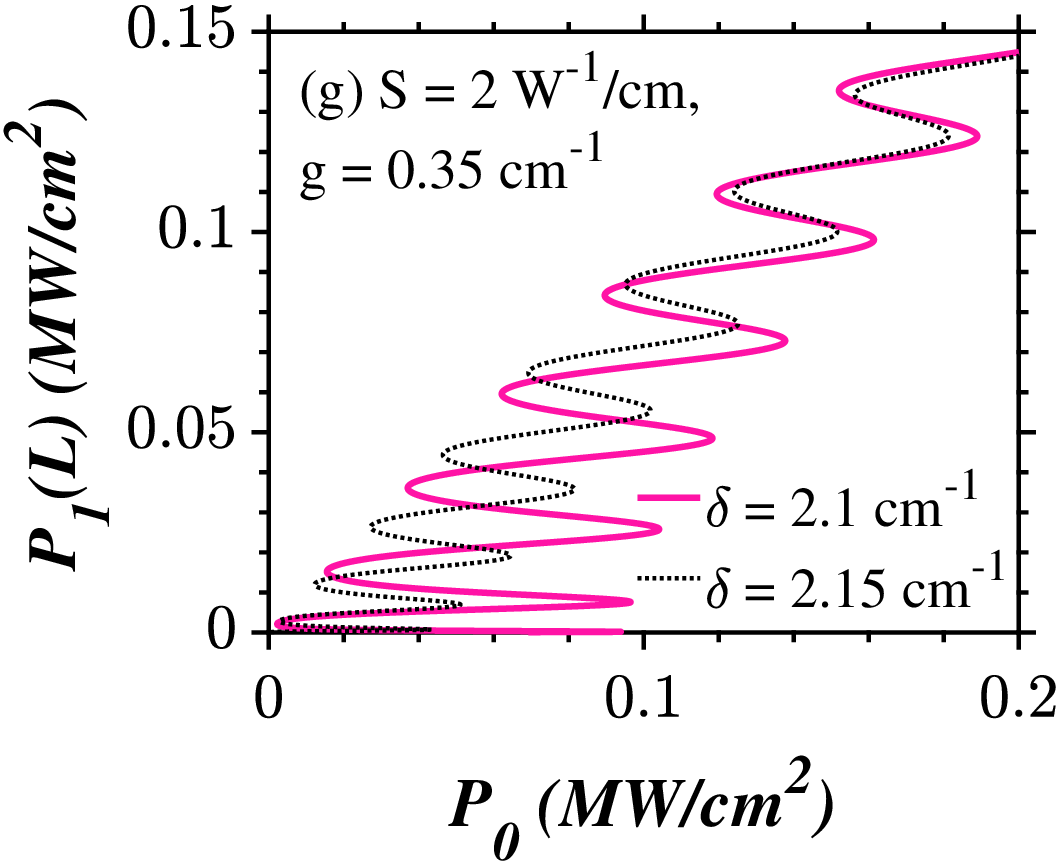}\includegraphics[width=0.25\linewidth]{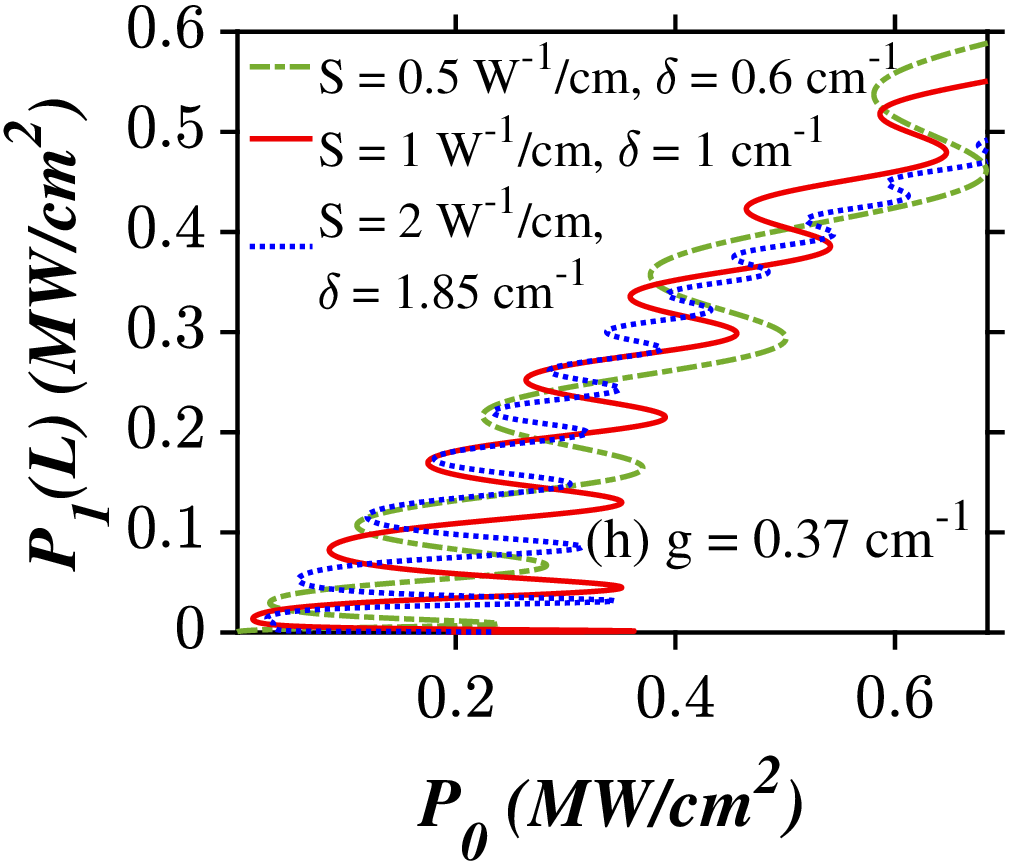}\\\includegraphics[width=0.25\linewidth]{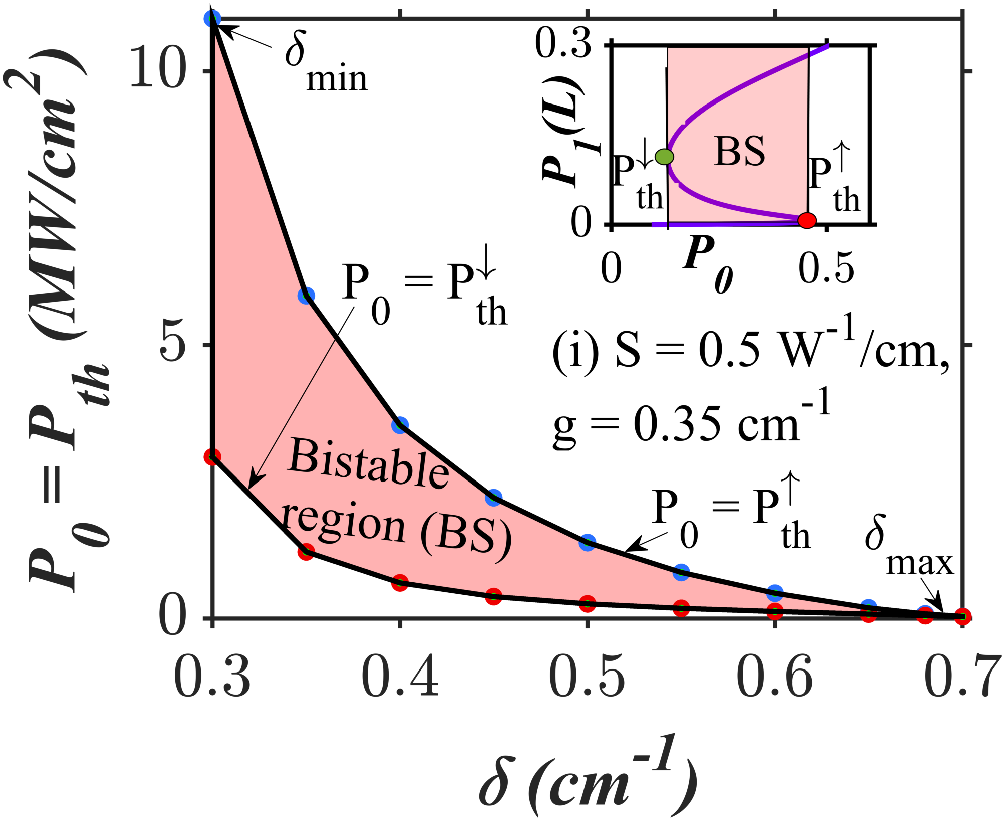}\includegraphics[width=0.25\linewidth]{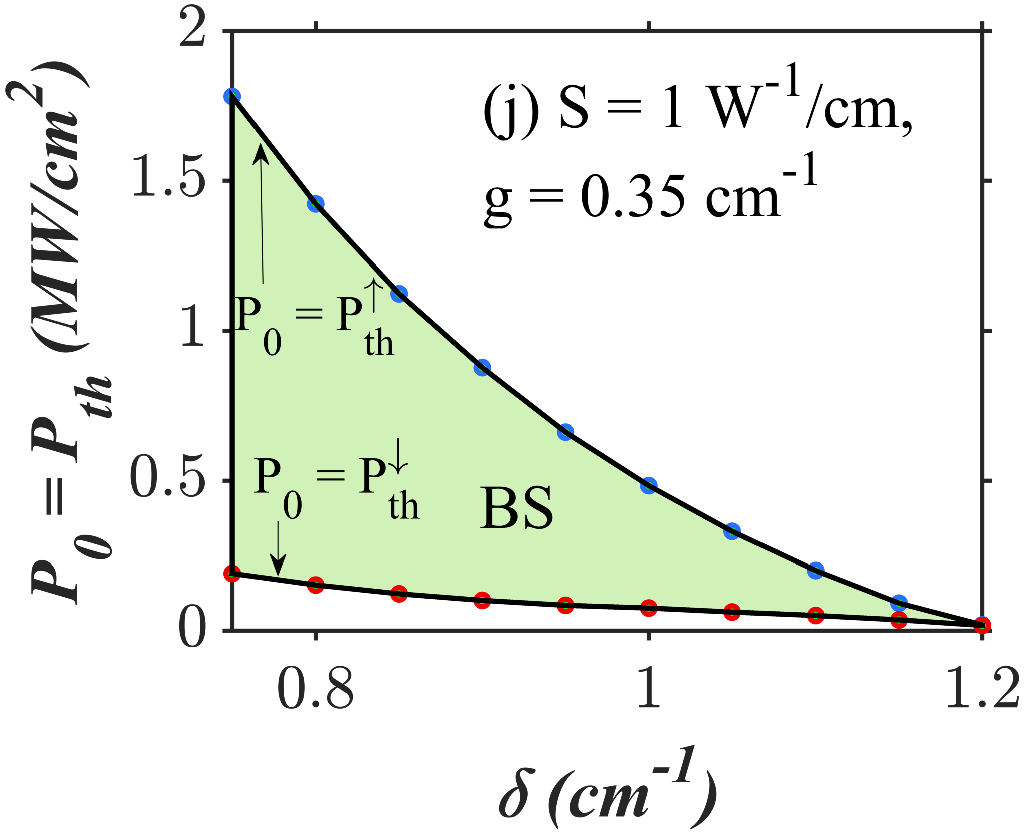}\includegraphics[width=0.25\linewidth]{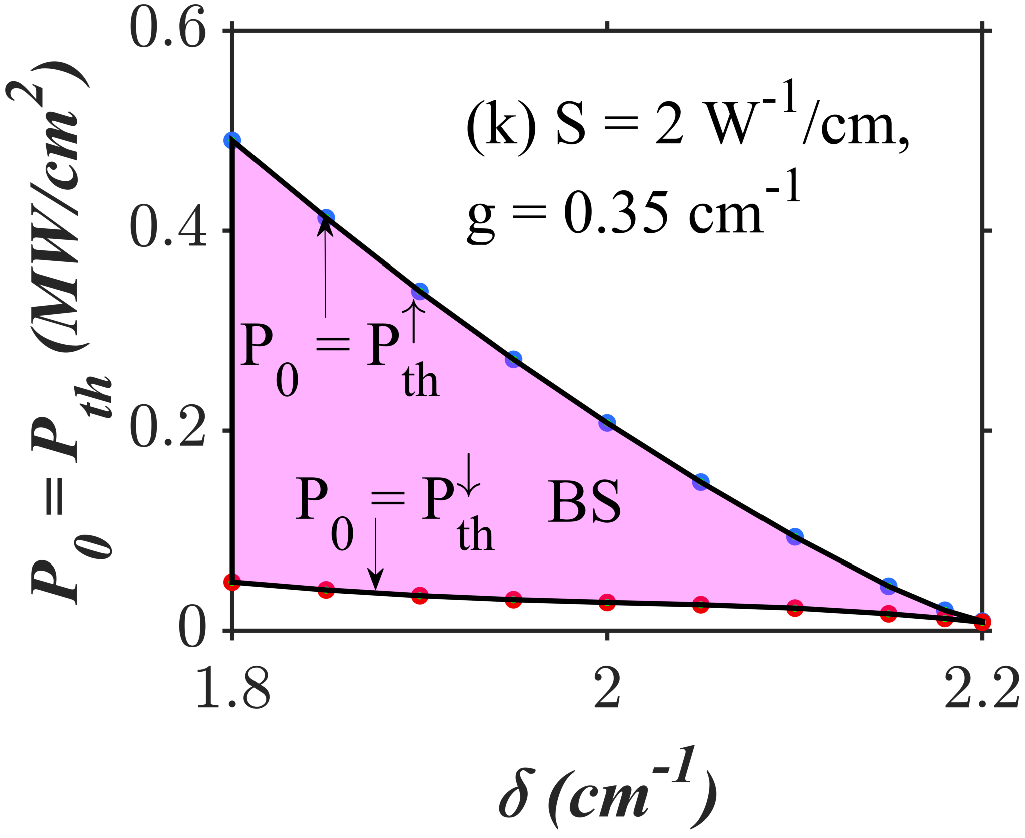}\includegraphics[width=0.25\linewidth]{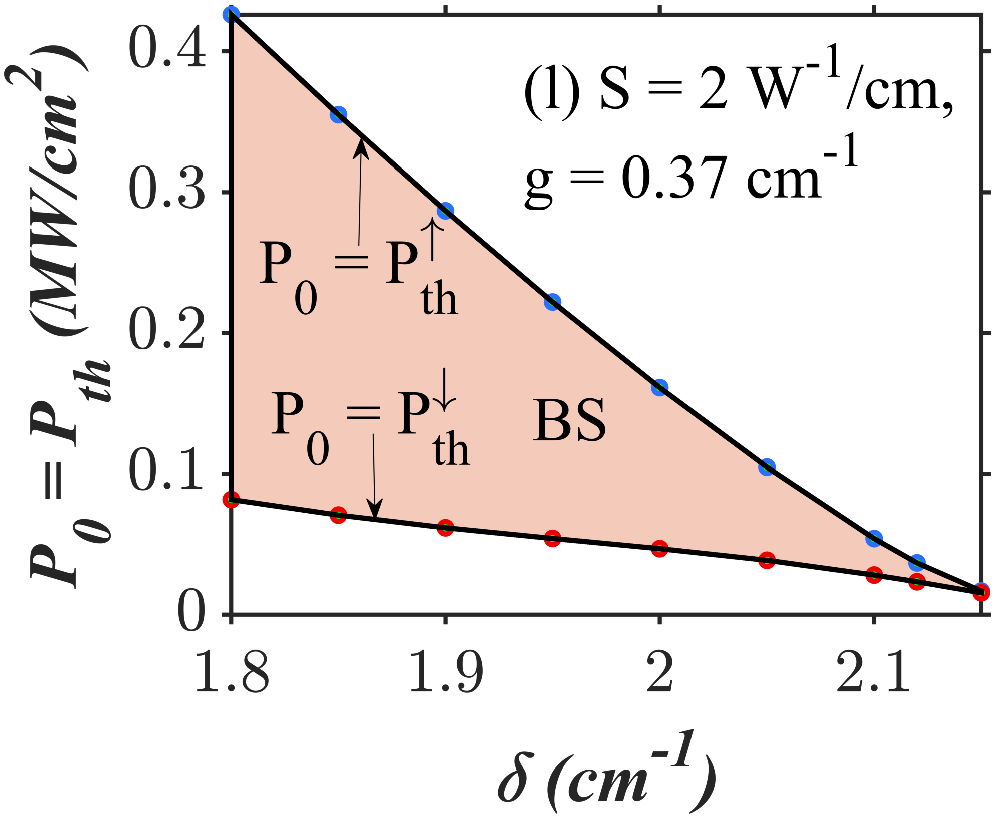}
		\caption{Frequency detuning induced S-shaped OB (OM) curves in an unbroken PTFBG ($g = 0.35$ $cm^{-1}$) with SNL at $\kappa = 0.4$ $cm^{-1}$. The light launching direction is left. The device length is $L = 20$ and 70 $cm$ in the top and middle panels, respectively.  The plots on the rightmost panel depict the decrease in switching intensities with an increase in the gain and loss parameter ($g = 0.37$ $cm^{-1}$). The bottom panel illustrates the variation in the values of detuning parameter or the spectral span ($\delta_{min} < \delta < \delta_{max}$) for which S-shaped OB curves occur. The switching intensities are high and low at $\delta_{min}$ and $\delta_{max}$, respectively. The OB curves with the broadest and narrowest hysteresis width appear at $\delta_{min}$ and $\delta_{max}$, respectively. }
		\label{fig3}
	\end{figure*}
	
	We have observed that the conventional FBGs and unbroken PTFBGs with SNL give rise to unconventional OB (OM) curves in their nonlinear transmission characteristics at the synchronous wavelength ($\delta = 0$ $cm^{-1}$).  Recall that the detuning parameter aids in lowering the input intensity required for switching, provided that its sign matches with the type of nonlinearity \cite{raja2019multifaceted,PhysRevA.100.053806}. Keeping this fact in mind, we present the results pertaining to the operation of PTFBG with SNL for positive values of the detuning parameter ($\delta > 0$ $cm^{-1}$). Interestingly,  detuning the system far away from the synchronous wavelength favors the formation of the typical S-shaped hysteresis curves, as shown in Figs. \ref{fig3}(a) -- \ref{fig3}(c).  The nonlinear parameter controls the steering dynamics in two different ways: First, it serves as an additional degree of freedom to control the switching intensities, as shown in the top panel of Fig. \ref{fig3}.  Second, it alters the range of the detuning parameters or the spectral span ($\delta^{S}$) at which the S-shaped OB (OM) curves occur, as shown in the bottom panels of Fig. \ref{fig3}. Let the minimum and maximum values of the detuning parameter at which the S-shaped OB curves appear be $\delta_{min}$ and $\delta_{max}$, respectively. For the values of the detuning parameter below $\delta_{min}$, the system exhibits a hysteresis curve with a different envelope which will be discussed separately in the next section. 
	
	The switch-up and down ($P_{th}^{\uparrow}$ and $P_{th}^{\downarrow}$) intensities and hysteresis width of the S-shaped OB curve decrease with an increase in the value of the detuning parameter, as shown in the bottom panels of Fig. \ref{fig3}. For instance, the bistable curves with the broadest and narrowest width occur at $\delta_{min}$ and $\delta_{max}$, respectively. At this juncture, one may wish to reduce the switch-up and down intensities by increasing the detuning parameter further. However, the value of the detuning parameter cannot be greater than $\delta_{max}$ because it would result in insufficient feedback to create any desirable bistable feature. Thus, the switch-up and down intensities corresponding to the S-shaped hysteresis curves are lowest at $\delta_{max}$ and highest at $\delta_{min}$ for a given set of system parameters. A further reduction in them is feasible via an increase in the gain and loss parameter, as shown in Figs. \ref{fig3}(d) and (l). Specifically, an increase in $g$ leads to a significant decrease in the switch-up and down intensities for fixed values of $S$ and $\delta$. 
	
	 As we tune the input intensity further, the OB curves transform into the OM curves with an increase in the device length ($L = 70$ $cm$),  provided that the other system parameters remain unchanged. For some values of input intensities, the output intensity even experiences three (tristability) or four (tetrastability) stable states. In contrast to the OM curves shown in Fig. \ref{fig1}, the width of the successive stable branches decreases with an increase in the input intensity.  There is a significant reduction in the switching intensities (corresponding to the different stable branches) with an increase in the detuning, nonlinearity, and gain-loss parameters, as shown in Figs. \ref{fig3}(f), (g), and (h), respectively.  
	
	\subsection{Mixed OM curves}
	\label{Sec:4C}
	\begin{figure*}
		\centering	\includegraphics[width=0.29\linewidth]{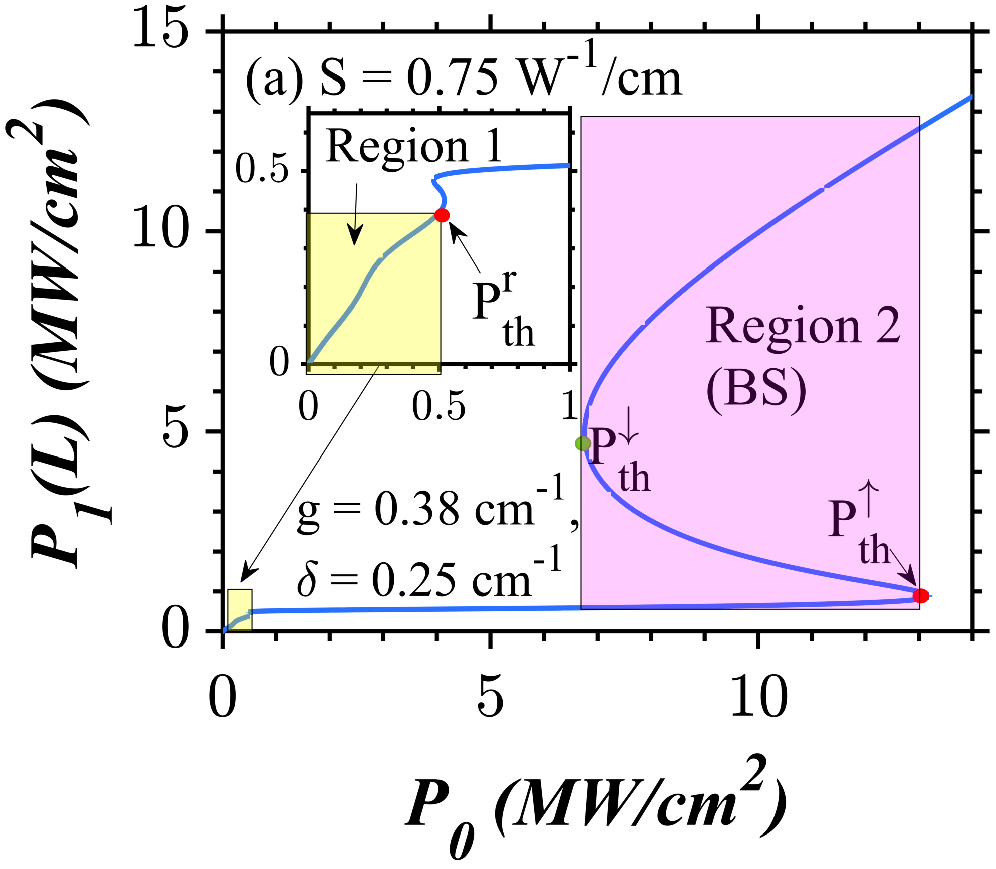}\includegraphics[width=0.28\linewidth]{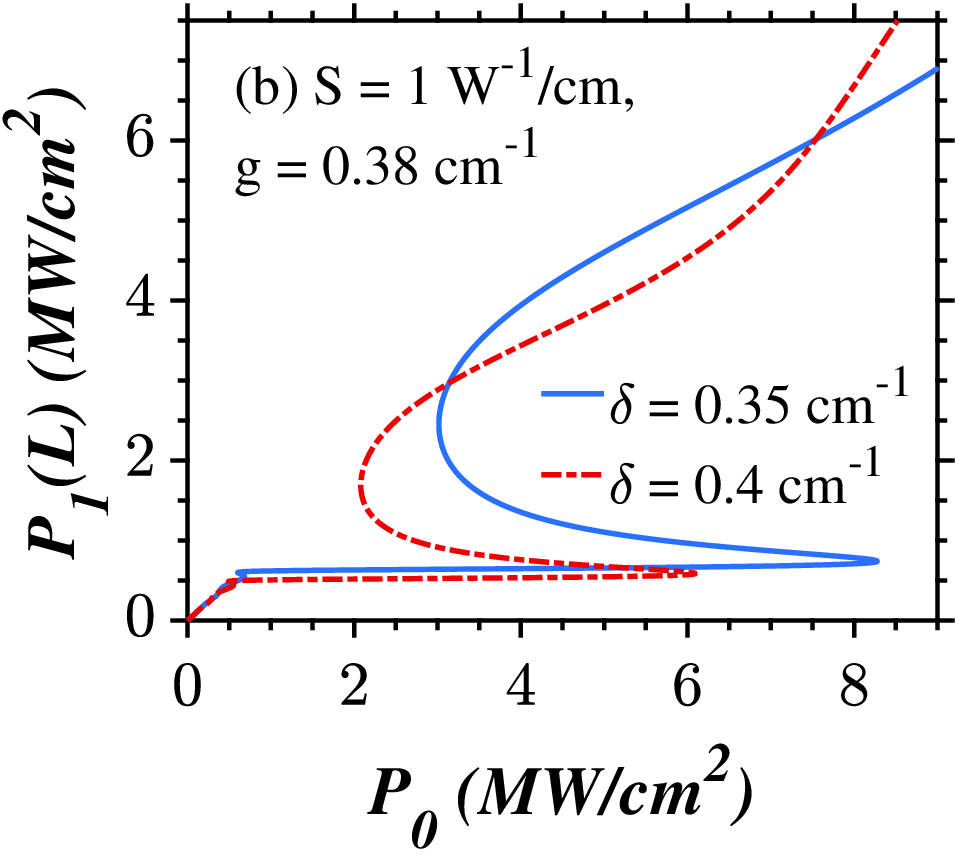}\includegraphics[width=0.28\linewidth]{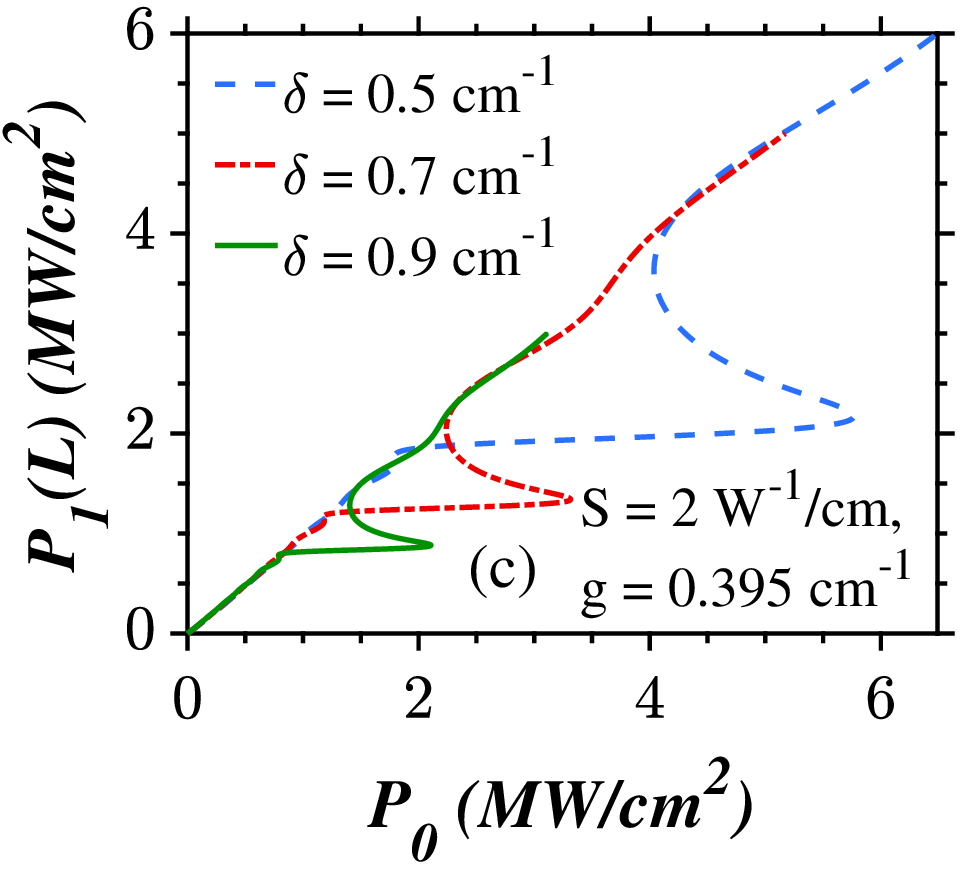}\\\includegraphics[width=0.28\linewidth]{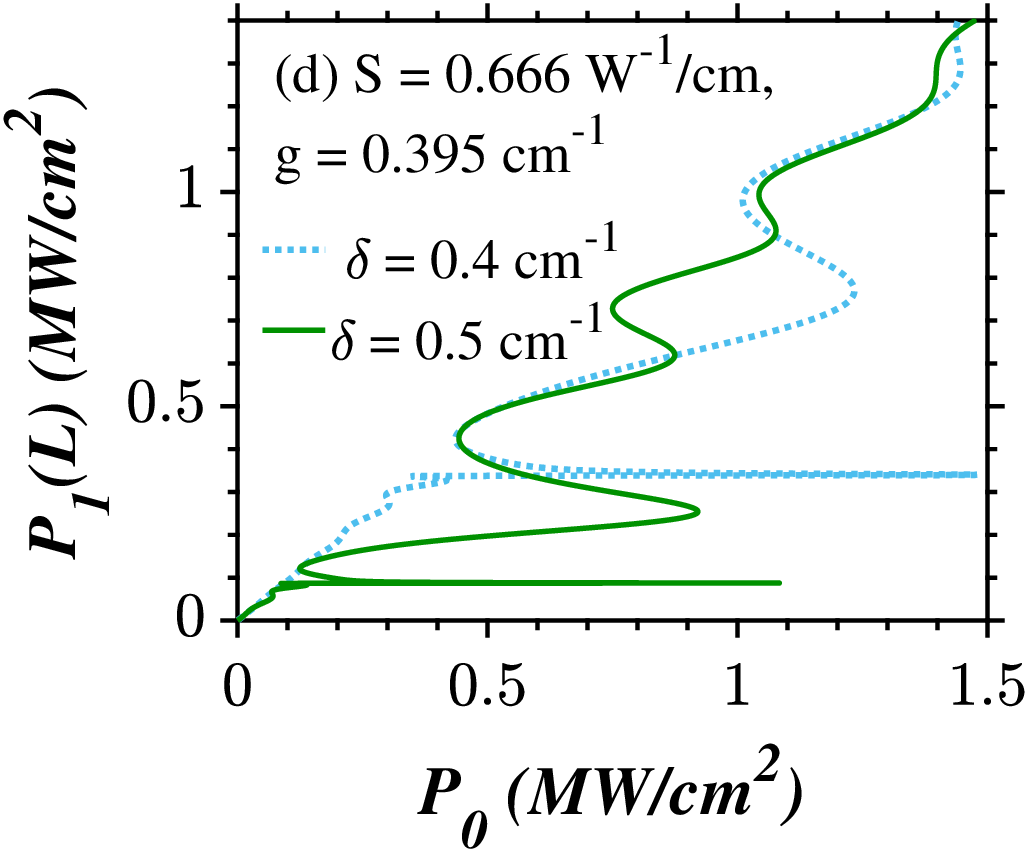}\includegraphics[width=0.28\linewidth]{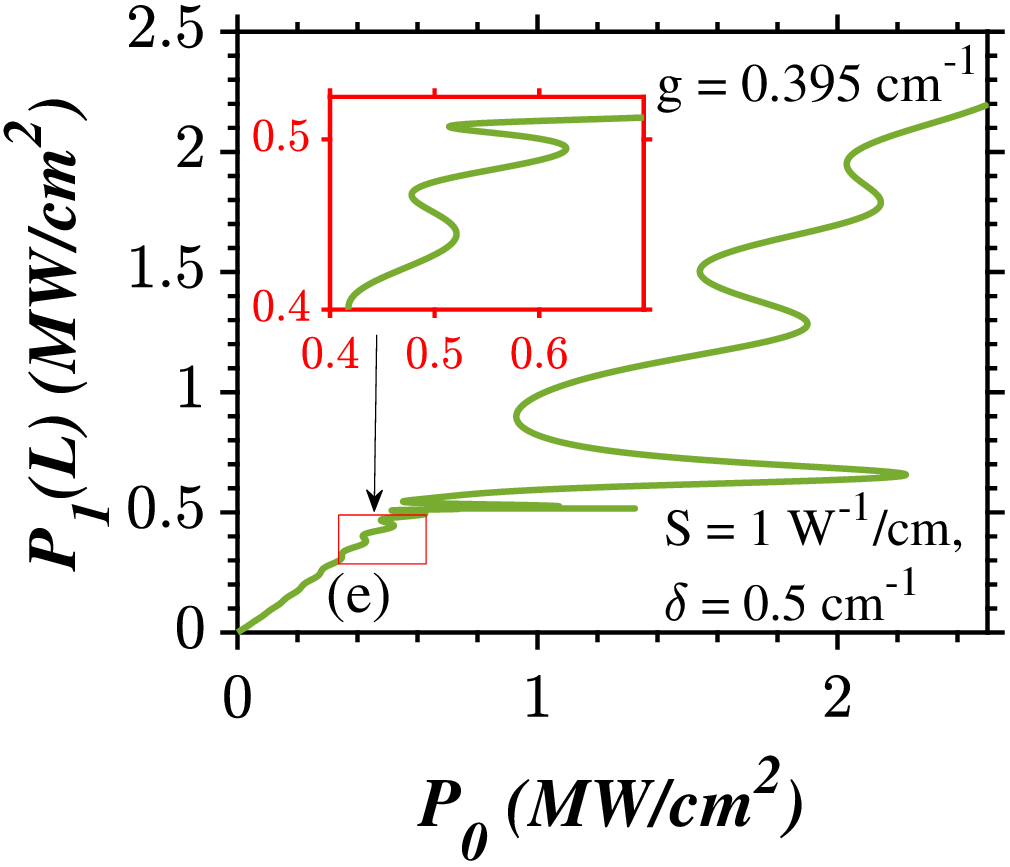}\includegraphics[width=0.28\linewidth]{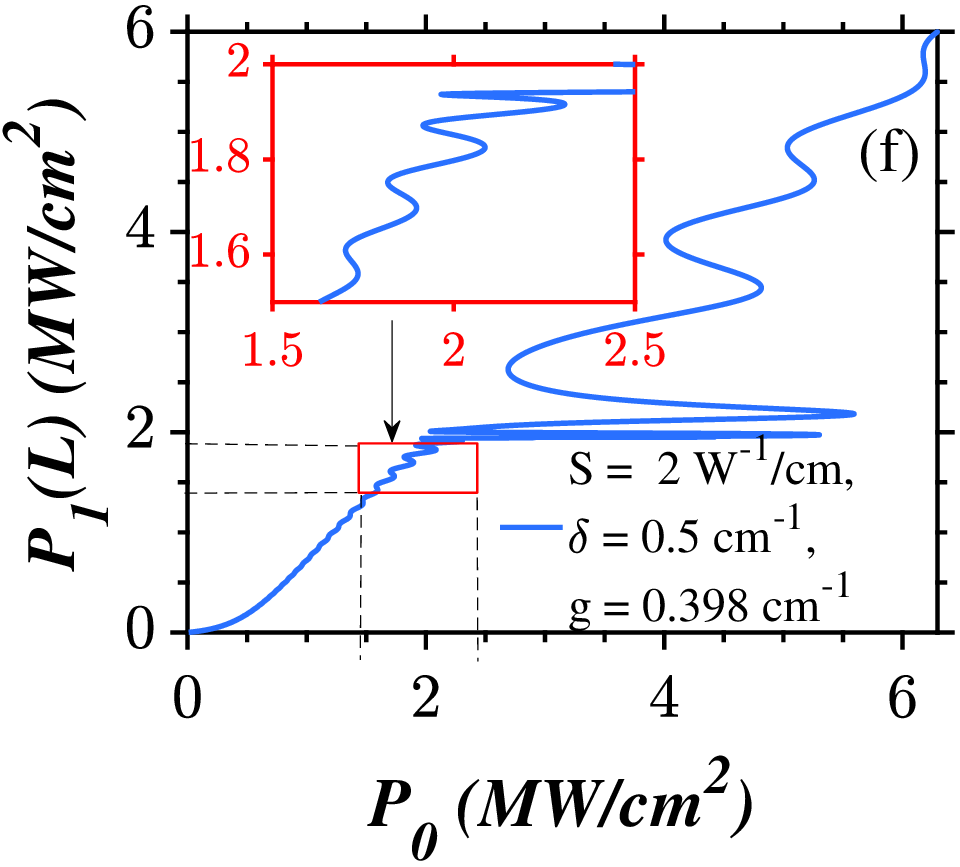}\\\includegraphics[width=0.28\linewidth]{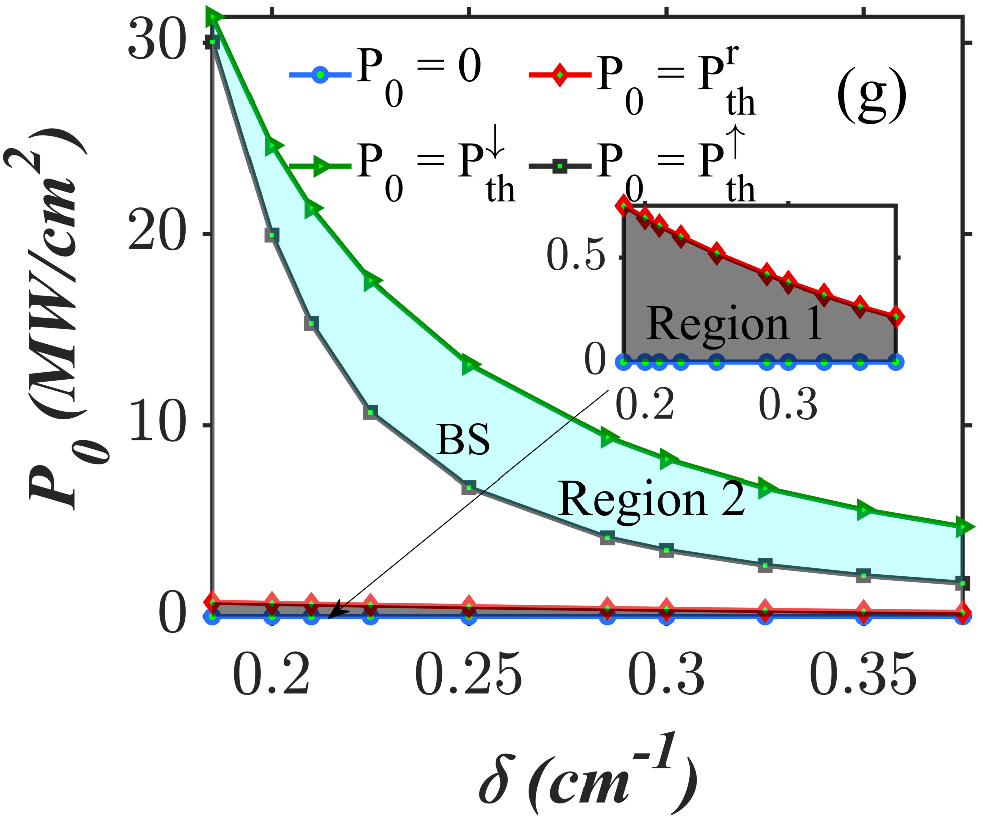}\includegraphics[width=0.28\linewidth]{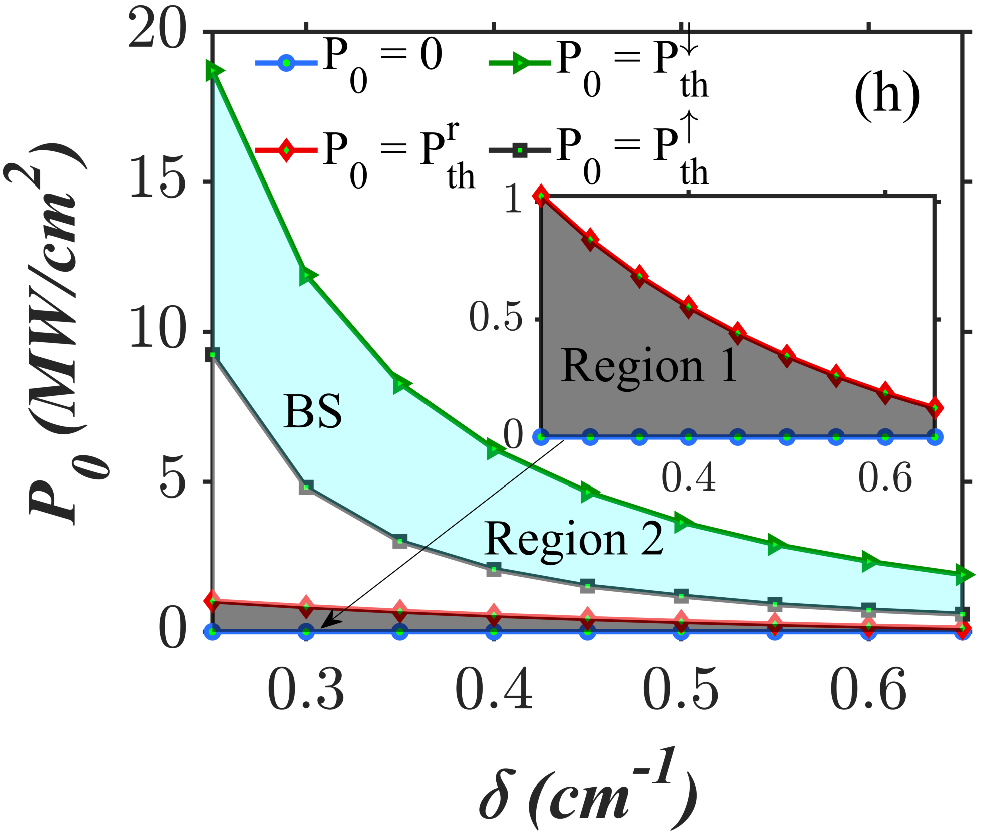}\includegraphics[width=0.28\linewidth]{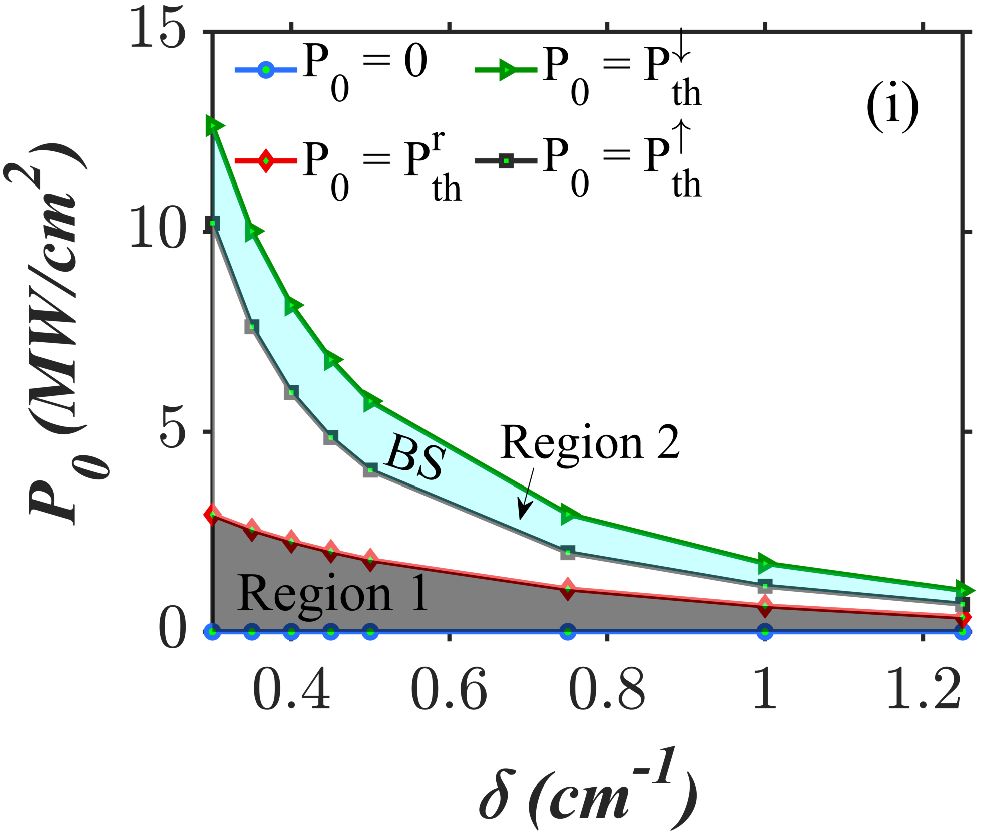}
	\caption{Mixed OB (OM) curves in an unbroken PTFBG with SNL at $\kappa = 0.4$ $cm^{-1}$. The light launching direction is left. The device lengths are $L = 20$ and 70 $cm$ in the top and middle panels, respectively.  (g), (h) and (i) Variations in the range of input intensities at which ramp-like stable state appear, and the bistable region (BS) region in (a), (b), and (c), respectively against detuning.}
		\label{fig4}
	\end{figure*}
	
	  The bottom line of the investigations carried out in the previous sections is as follows: FBGs and PTFBGs with SNL do not admit a typical S-shaped hysteresis curve in their input-output characteristics at $\delta = 0$ $cm^{-1}$. The concept of frequency detuning makes it feasible to retrieve the S-shaped OB (OM) curves in the presence of $\mathcal{PT}$-symmetry. It should be noted that the values of the detuning parameter used in Fig. \ref{fig3} are considerably large, which means that the system is operated far away from the Bragg wavelength. A natural question that comes to mind at this point is that what happens to the characteristics of the OB (OM) curves for smaller values of detuning parameter, i.e., operating wavelengths of incident light close to the synchronous wavelength. To address this query, we choose the values of the detuning parameter in the range of $0<\delta<\delta_{min}$ and investigate the nonlinear response of the proposed system via numerical simulations. Recall that $\delta_{min}$ signifies the minimum  value of the detuning parameter required to generate the S-shaped OB (OM) curves in the nonlinear regime.  
	  
	  In the left panels of Fig. \ref{fig4} ($L = 20$ $cm$), we observe two distinct regions in the input-output characteristics curves. In region 1, increasing the input intensities induce sharp variations in the output intensities leading to the formation of ramp-like first stable states ($0<P_0<P_{th}^{r}$), as shown in Fig. \ref{fig4}(a).  As we tune the input intensity further, the system's output jumps to the second stable state branch. There exists a bistable region with a narrow hysteresis width between the first and the second stable branch. The output intensities show gradual variations against the increasing input intensities ($P_{th}^{r}<P_0<P_{th}^{\uparrow}$). If the input intensity is tuned further ($P_0>P_{th}^{\uparrow}$), the output jumps to the next stable branch. The output intensities vary gradually in it for a given range of input intensities ($P_0>P_{th}^{\uparrow}$) at $L = 20$ $cm$. When the input intensities decrease, the system returns to the second stable branch at $P_{th}^{\downarrow}$.  Region 2 represents the values of input intensities for which the system's output is bistable ($P_{th}^{\downarrow}<P_0<P_{th}^{\uparrow}$).  Thus, the overall shape of the curves looks like a mix of ramp-like and S-shaped OB curves. These kinds of OB curves were previously reported in graphene-based quantum systems \cite{aichun2022optical}. 
	  
	  An increase in the value of the detuning parameter leads to a reduction in the switch-up and down intensities ($P_{th}^{\downarrow}$ and $P_{th}^{\uparrow}$) of the different stable branches, as shown in Figs. \ref{fig4}(b), (c), (h) and (i). The mixed OB (OM) curves appear only for a set of values of the detuning parameter ($\delta^{mix}$), as shown in the continuous variation curves in the bottom panel of Fig. \ref{fig4}.   The spectral span or the range of detuning parameters for which mixed OM curves ($\delta^{mix}$) occur varies according to the values of the NL parameter, as shown in Figs. \ref{fig4}(h) and (i). 
	
	Increasing the device length  ($L = 70$ $cm$) brings significant variations in the characteristics of the mixed OM curves, as shown in Figs. \ref{fig4}(d), (e), and (f). As the value of input intensity increases, the system exhibits a ramp-like first stable state followed by a series of ramp-like hysteresis curves in which the variations in the output intensities against the input are sharp.  Also, the width of the successive stable branches increases in region 1.  The device generates S-shaped OM curves in region 2 for higher intensities. In this regime, the width of the successive stable branches decreases with an increase in the input intensities. In the literature  similar OM curves that feature a mix of ramp-like OM and S-shaped OM curves have already been reported in graphene nanodisk–quantum dot hybrid systems \cite{tohari2020optical}. The interplay between the detuning and nonlinearity parameters decreases the switch-up and down intensities of the different stable branches considerably.

	\section{OB in the unbroken $\mathcal{PT}$-symmetric regime: Right incidence}
	\label{Sec:5}
	In the context of linear FBGs, Kulishov \emph{et al.} have first conceptualized the launching of the incident light from the rear end \cite{kulishov2005nonreciprocal}. Later, Komissarova \emph{et al.} and Raja \emph{et al.} have extended this idea to the nonlinear domain \cite{komissarova2019pt,raja2019multifaceted,PhysRevA.100.053806}. It remains the simplest yet the most effective method to realize low-power AOSs \cite{sudhakar2022low,sudhakar2022inhomogeneous}. 
	\subsection{Low power ramp-like OB (OM) curves }
	\label{Sec:5A}
	\begin{figure}
		\centering	\includegraphics[width=0.5\linewidth]{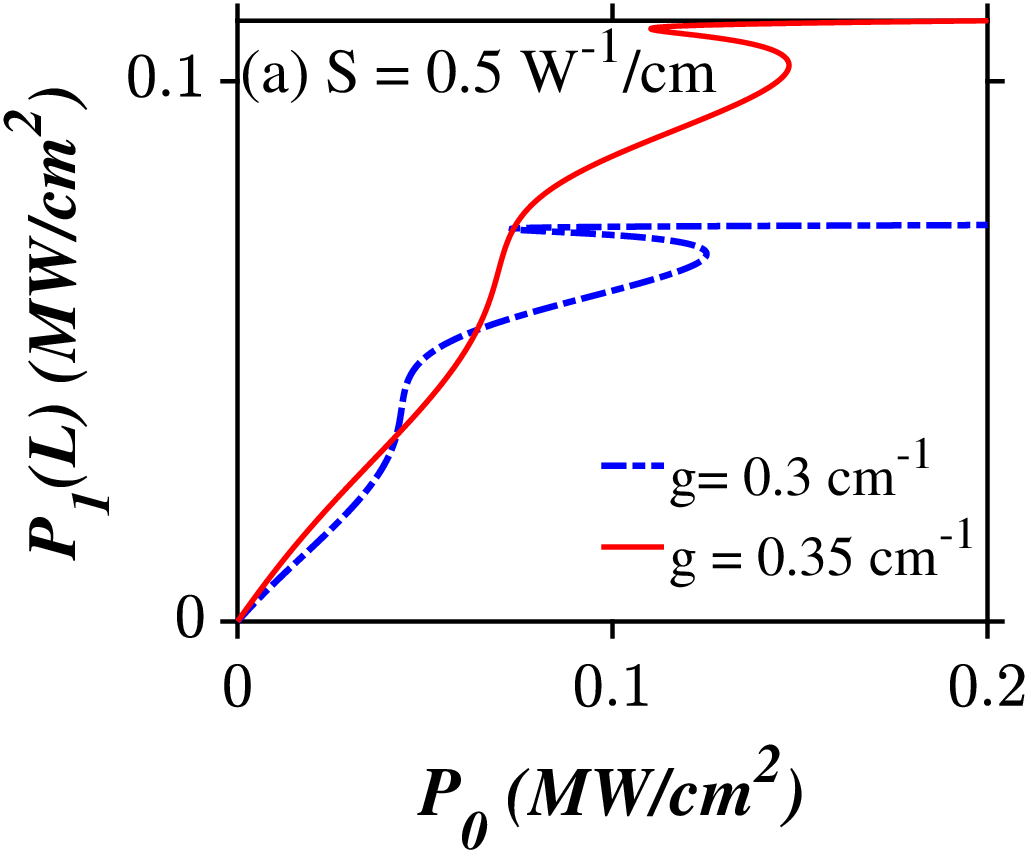}\includegraphics[width=0.5\linewidth]{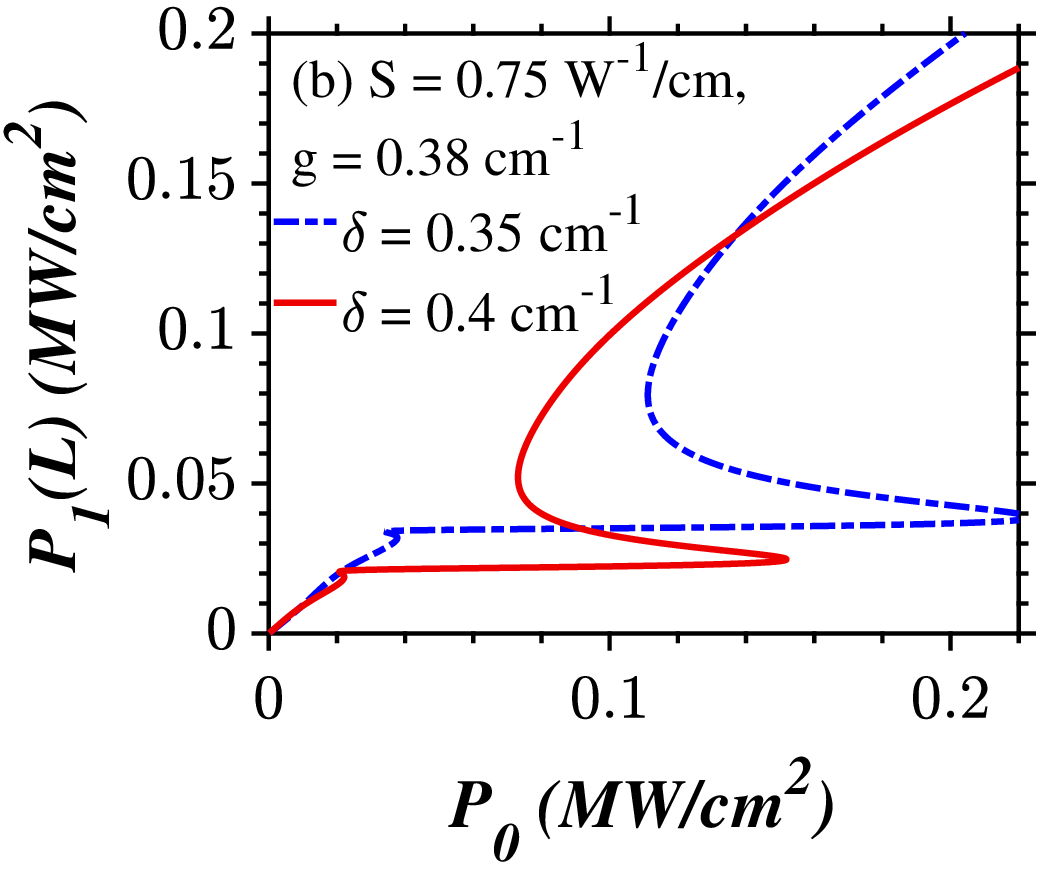}\\\includegraphics[width=0.5\linewidth]{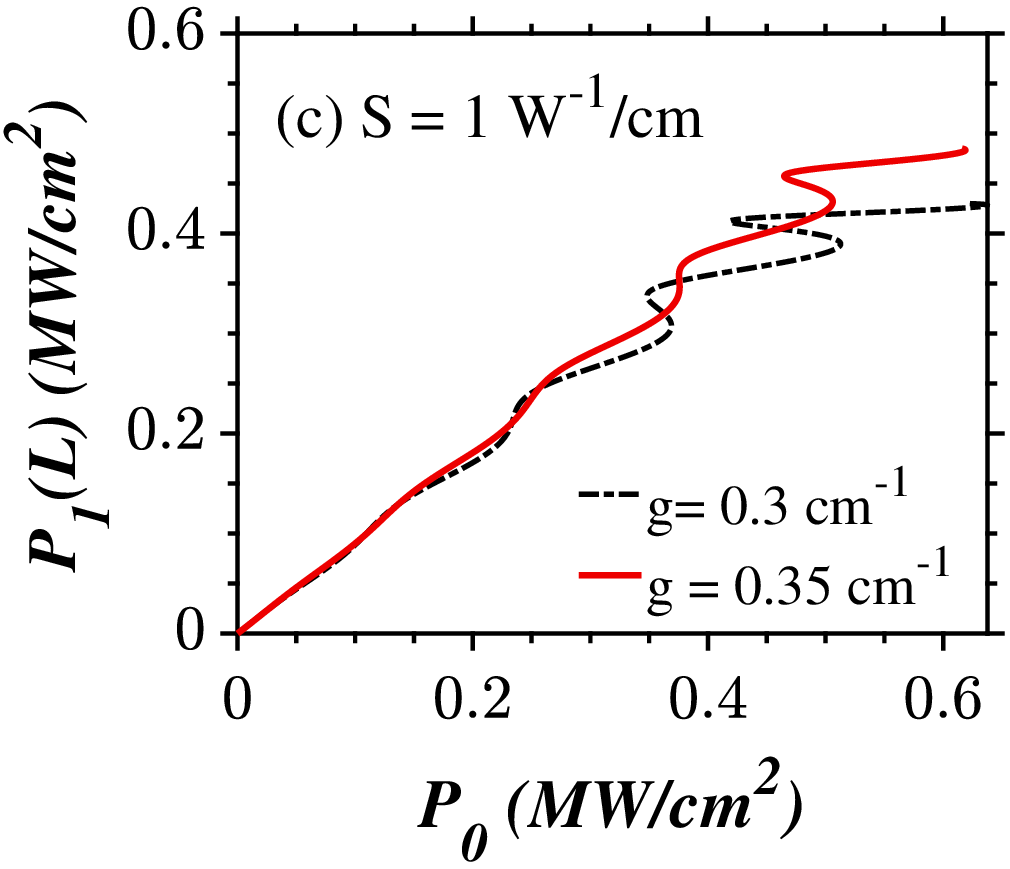}\includegraphics[width=0.5\linewidth]{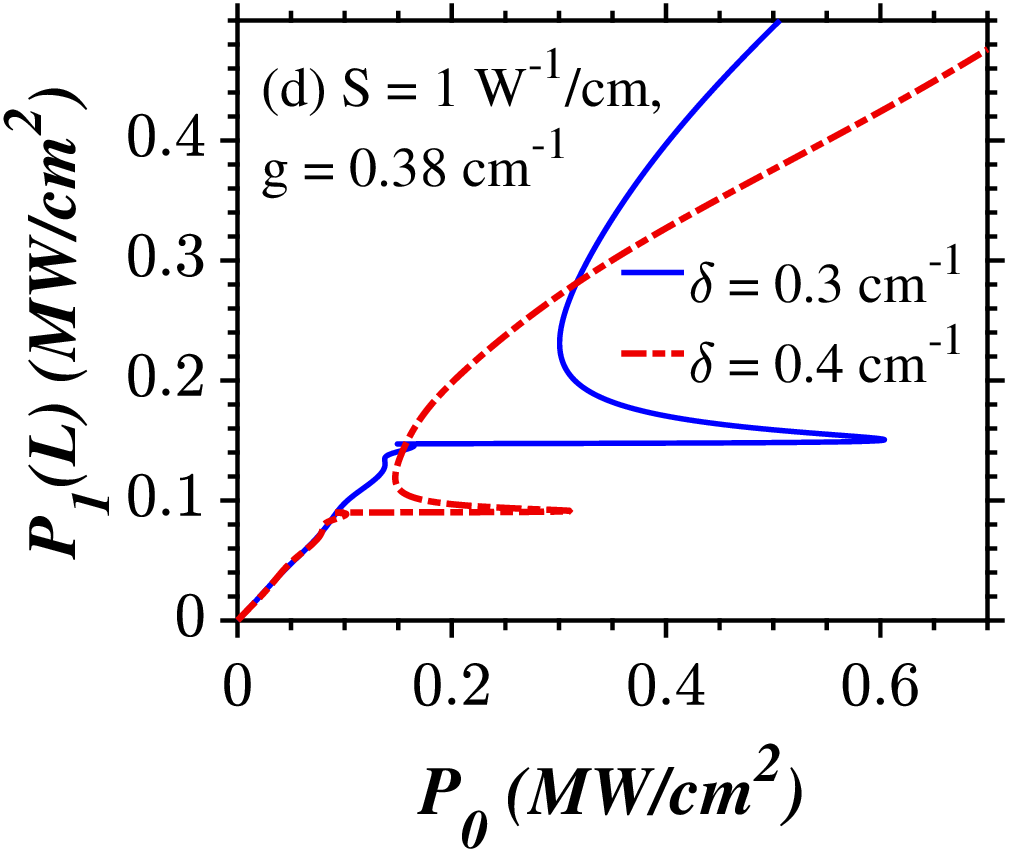}\\\includegraphics[width=0.5\linewidth]{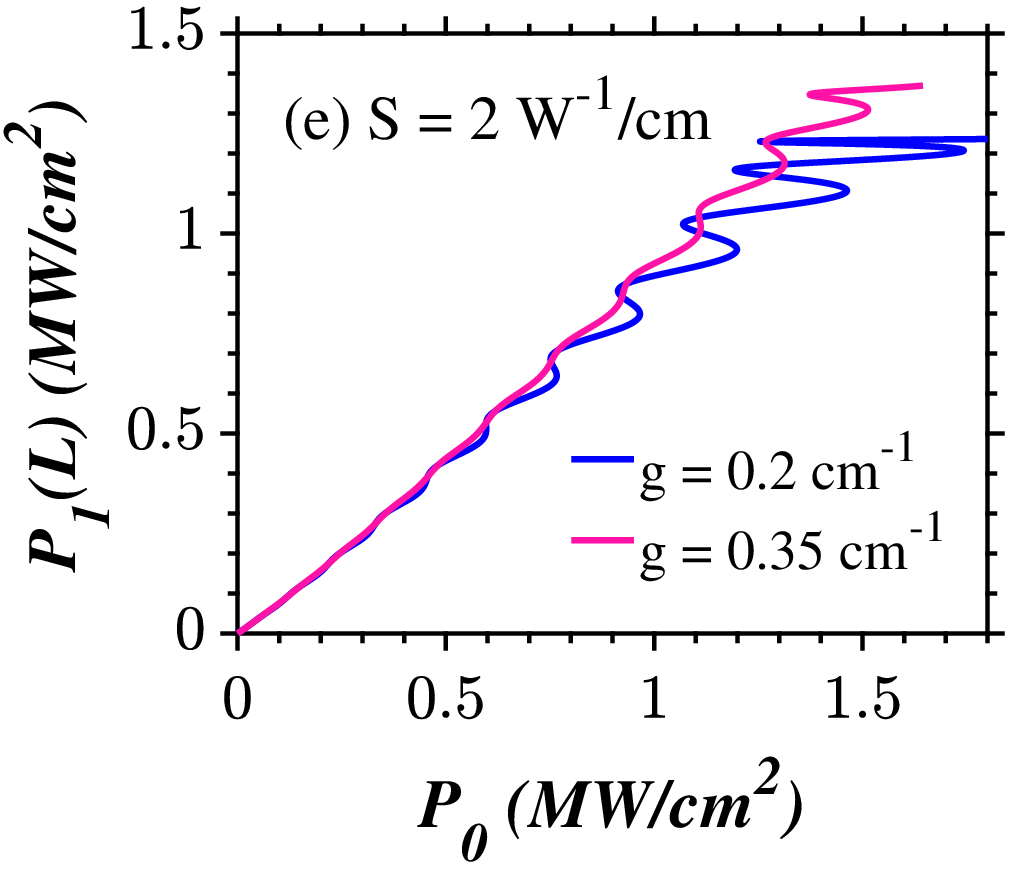}\includegraphics[width=0.5\linewidth]{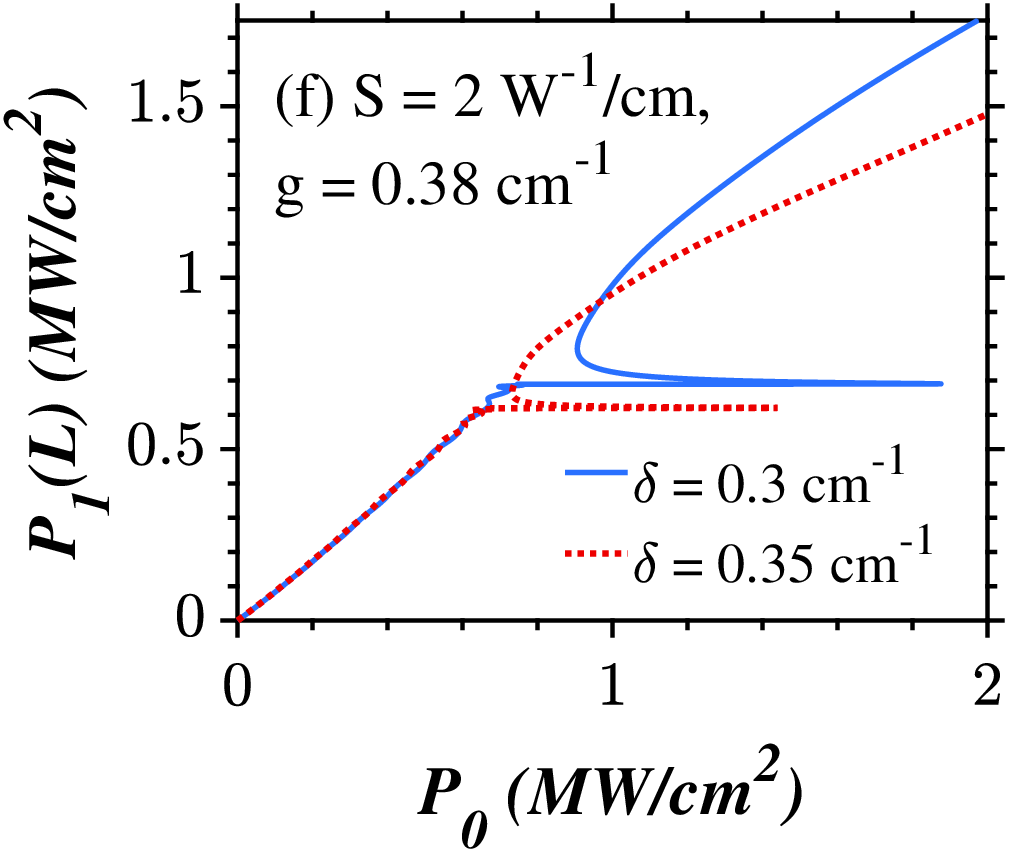}\\\includegraphics[width=0.5\linewidth]{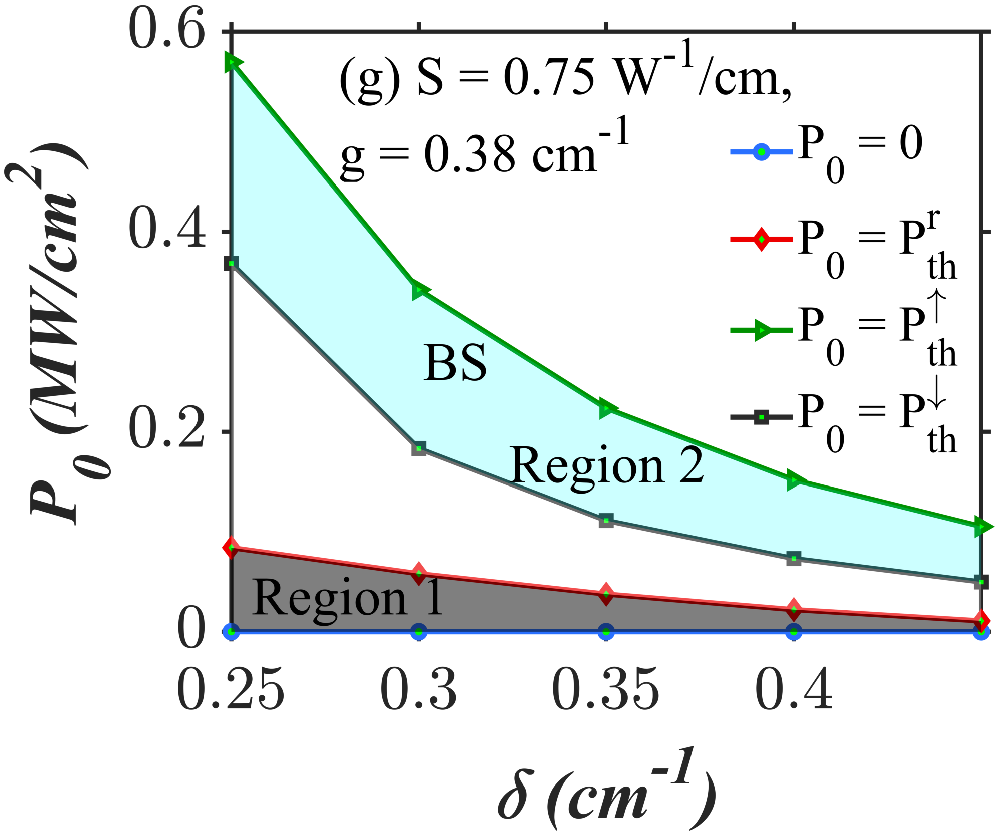}\includegraphics[width=0.5\linewidth]{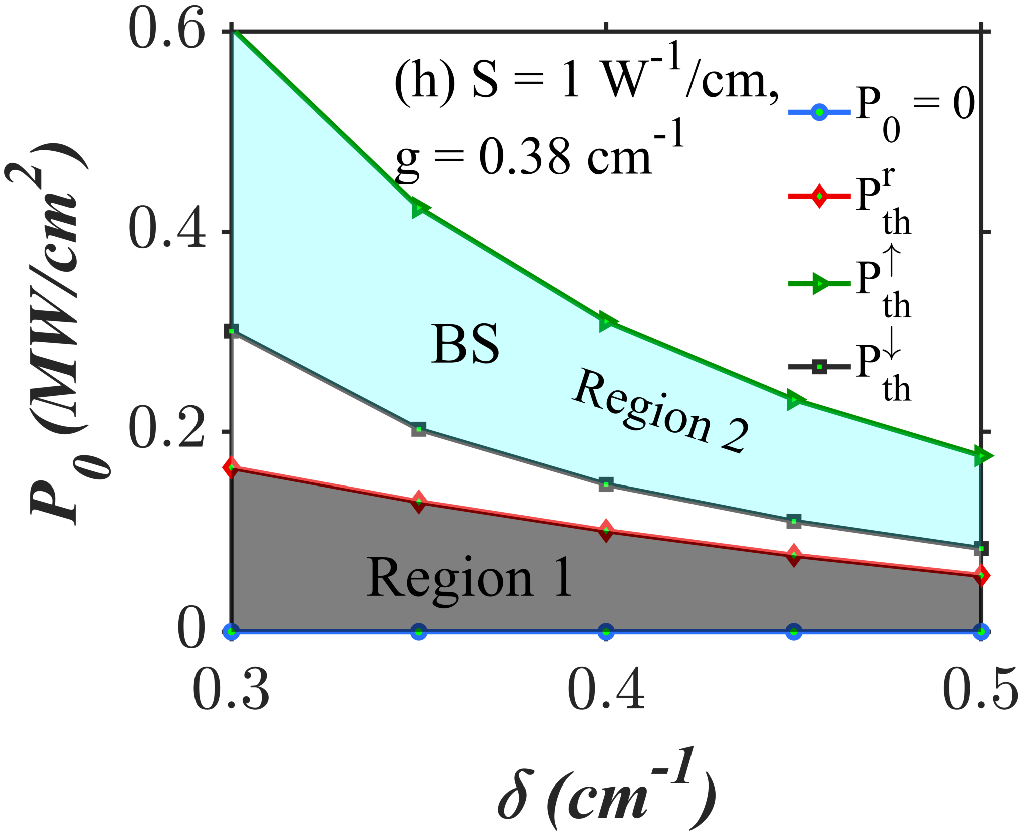}
	\caption{Low-power (a)  ramp-like OB (b), (d), and (f) Mixed OM, (c) and (e) ramp-like OM curves exhibited by an unbroken PTFBG with SNL at  $L = 20$ $cm$ and $\kappa = 0.4$ $cm^{-1}$. (f) and (g) Continuous variations in different regions of mixed OM curves.   The direction of light incidence is right.  }
	\label{fig5}
	\end{figure}
	
	Except for a reversal in the light launching direction, Figs. \ref{fig5}(a), (c), and (e) depict the same light-guiding dynamics with the same system parameters as in Figs. \ref{fig2}(a), (c), and (e), respectively. We observe a low-power ramp-like OB curve at  $S = 0.5$ $W^{-1}/cm$ which transforms to a ramp-like OM curve for $S \ge 1$ $W^{-1}/cm$, at $\delta = 0$ $cm^{-1}$, which confirms that the transition in the OB (OM) curves with an increase in the NL is independent of the direction of light incidence.

	 Comparing the plots in the left panels of Figs. \ref{fig5} and \ref{fig2}, we can justify that the reversal in the light incidence direction results in a significant reduction in the switching intensities of various stable states of a ramp-like OM curve.    Similar to Fig. \ref{fig2}, an increase in the value of $g$ or $S$ increases the switch-up and down intensities corresponding to the  different stable branches of the ramp-like OB (OM) curve in Fig. \ref{fig5}(a), (c) and (e).

	\subsection{ Mixed OM curves}
	\label{Sec:5B}
	When the unbroken PTFBG operates at wavelengths closer to the synchronous wavelength, a mix of ramp-like and S-shaped stable states appear in the input-output characteristics of the system, as shown in Figs. \ref{fig5}(b), (d), and (e). The mixed OB curves have two distinct regions, as shown in the schematic in Fig. \ref{fig4}(a). In region 1, as we tune the value of input intensity from zero, the output intensities vary sharply in the first stable branch for $0< P_0< P_{th}^r$. An increase in input intensities above $P_{th}^r$ allows the output intensities to jump from the ramp-like first stable branch to the second stable branch, indicating the first switching scenario. The output intensities do not show significant variations against the input for a wide range of input intensities $P_{th}^r<P_0<P_{th}^{\uparrow}$. Further increase in the input intensities above $P_{th}^{\uparrow}$ leads to a sudden jump in the output representing the second switching scenario. The system does not return to the second stable branch at $P_{th}^{\uparrow}$ under a decrement in the input intensities. Rather, it occurs only at $P_{th}^{\downarrow}$. For the values of input intensities between $P_{th}^{\downarrow}<P_0<P_{th}^{\uparrow}$, the input-output curve features two stable states for any given input intensity value, as shown in Figs. \ref{fig5}(g) and (h). The switch-up and down intensities required to jump from (to) the ramp-like first stable branch to (from) the second stable branch of the mixed OM curve reduce dramatically, thanks to the concept of reversal in the light incidence direction, as confirmed by Figs. \ref{fig5}(d) and \ref{fig5}(f). The switch-up and down intensities of the S-shaped hysteresis curve in region 2 also decrease to a large extent under a reversal in the direction of light incidence condition.  The values of the detuning parameter at which the mixed OB occurs ($\delta^{mix}$) depend on the NL, as shown in Figs. \ref{fig5}(g) and (h). 
	
		\subsection{OM curves at $L = 70$ $cm$}
	\subsubsection{Low-power ramp-like OM curves}
	\begin{figure}
		\centering	\includegraphics[width=0.5\linewidth]{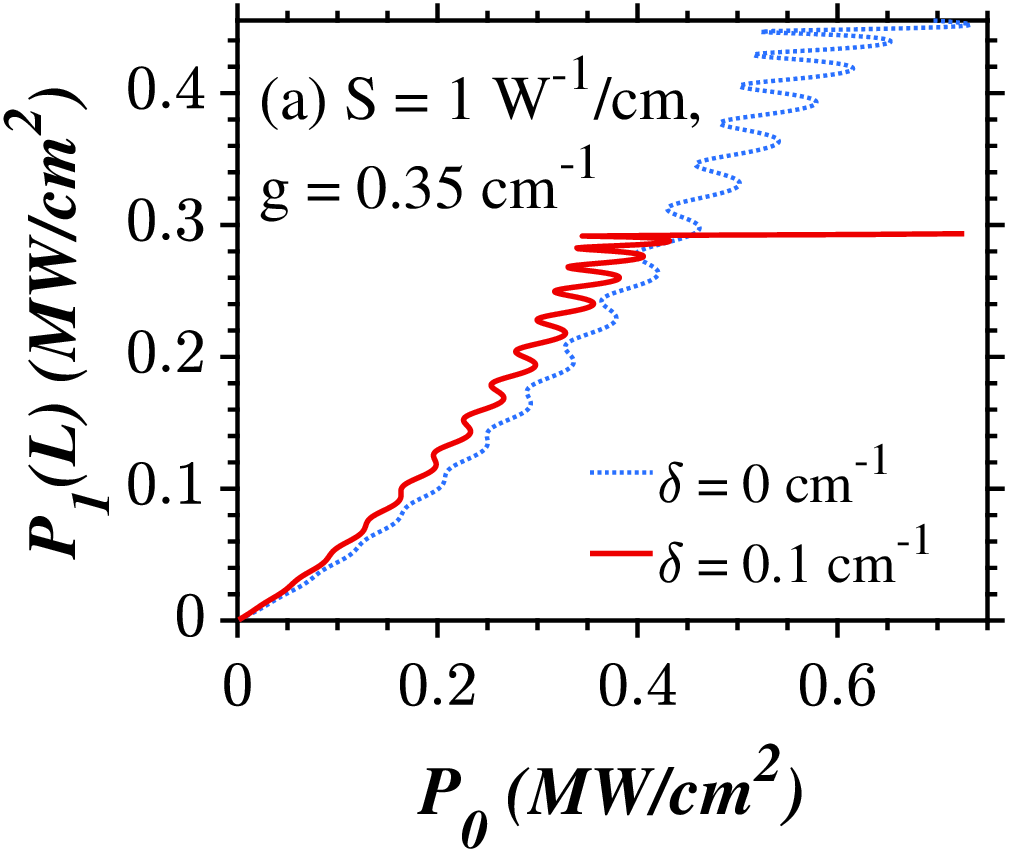}\includegraphics[width=0.5\linewidth]{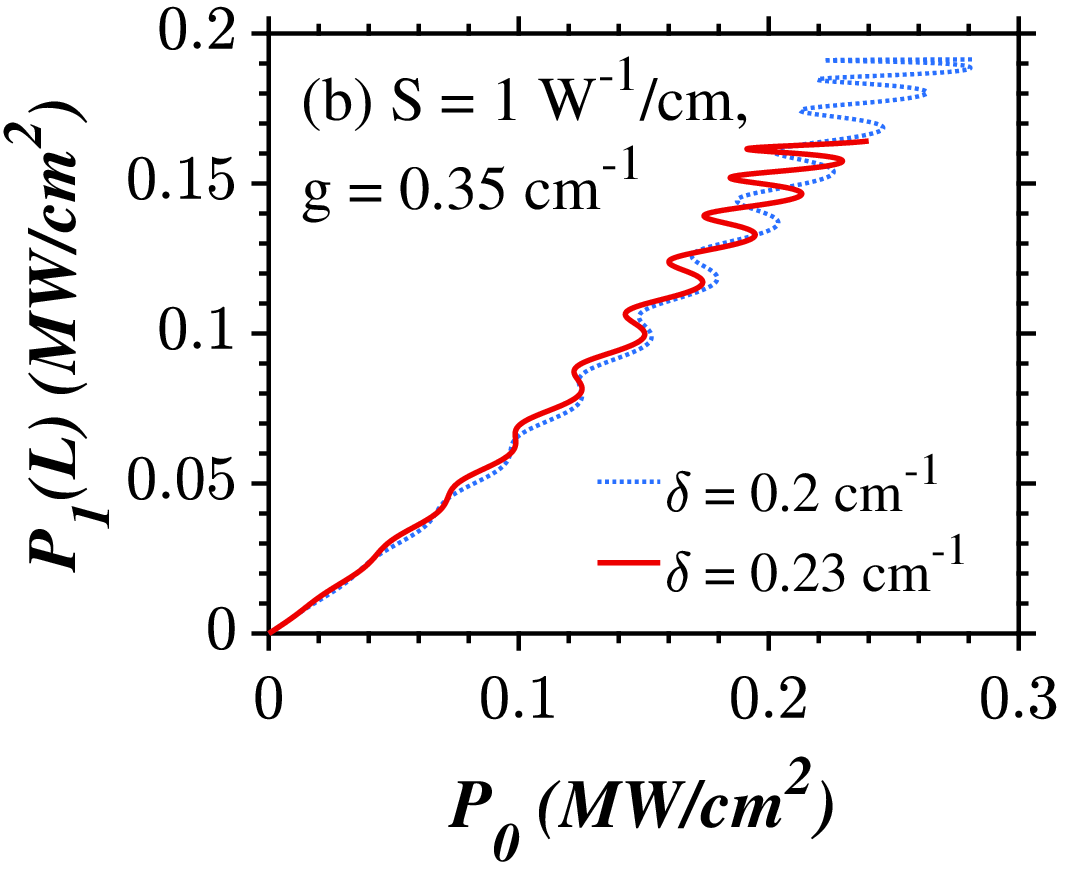}\\\includegraphics[width=1\linewidth]{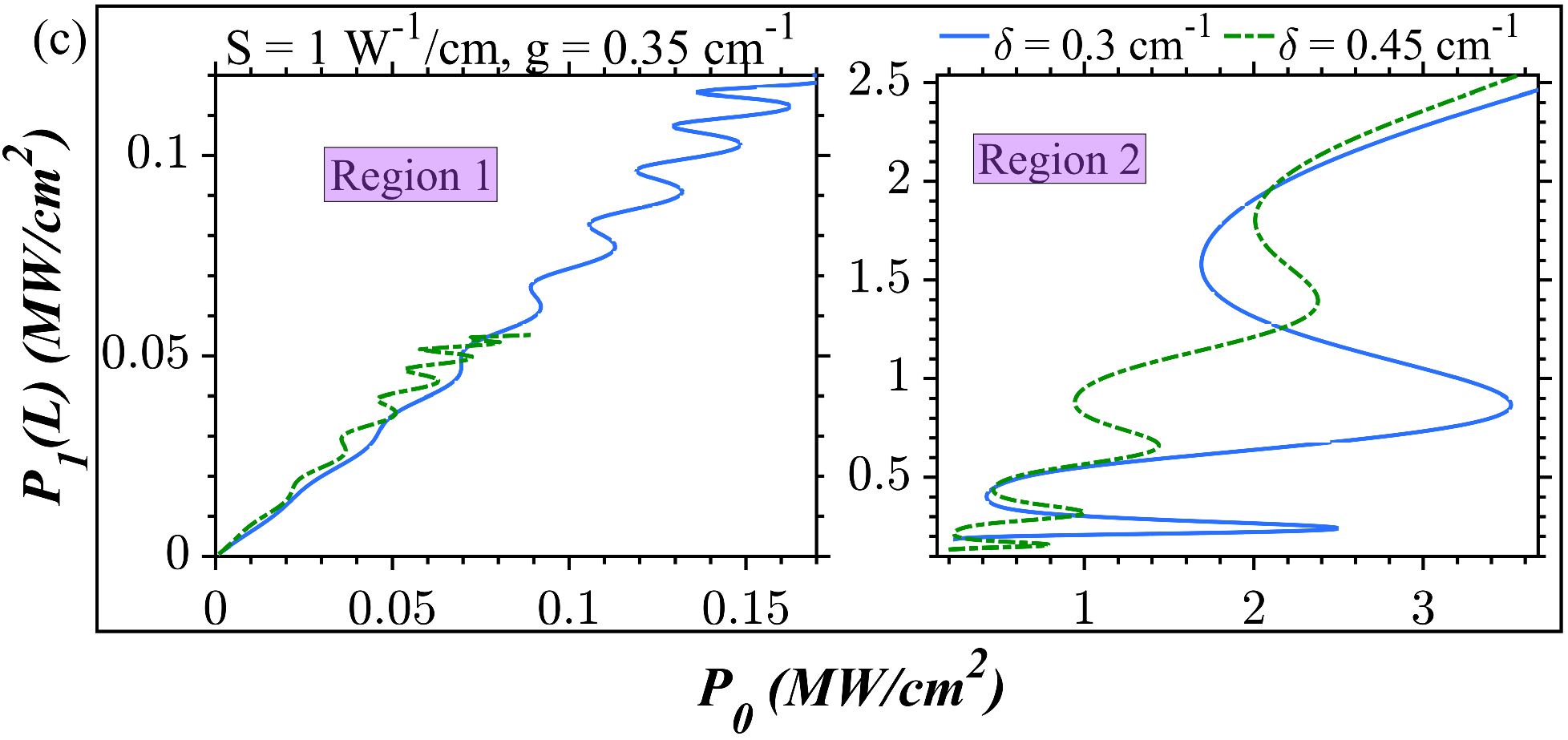}\\\includegraphics[width=1\linewidth]{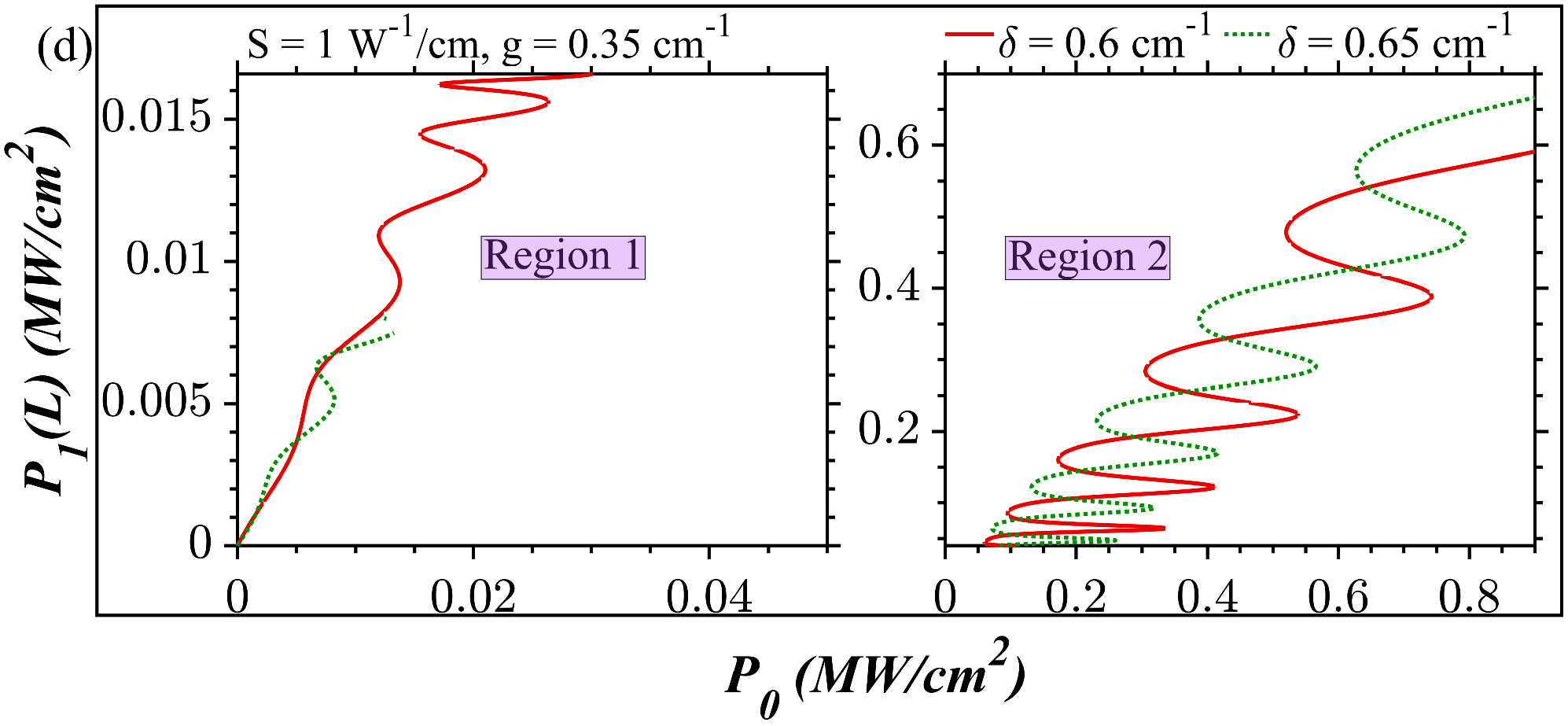}
		\caption{Low-power (a), (b) Ramp-like OM, (c) mixed OM curve with more number of ramp-like stable branches (d) Mixed OM curve with more number of S-shaped stable branches at $L = 70$ $cm$, $\kappa = 0.4$ $cm^{-1}$ and $S = 1$ $W^{-1}/cm$. The direction of light incidence is right.  }
	\label{fig6}
	\end{figure}

\begin{figure}
	\centering	\includegraphics[width=0.5\linewidth]{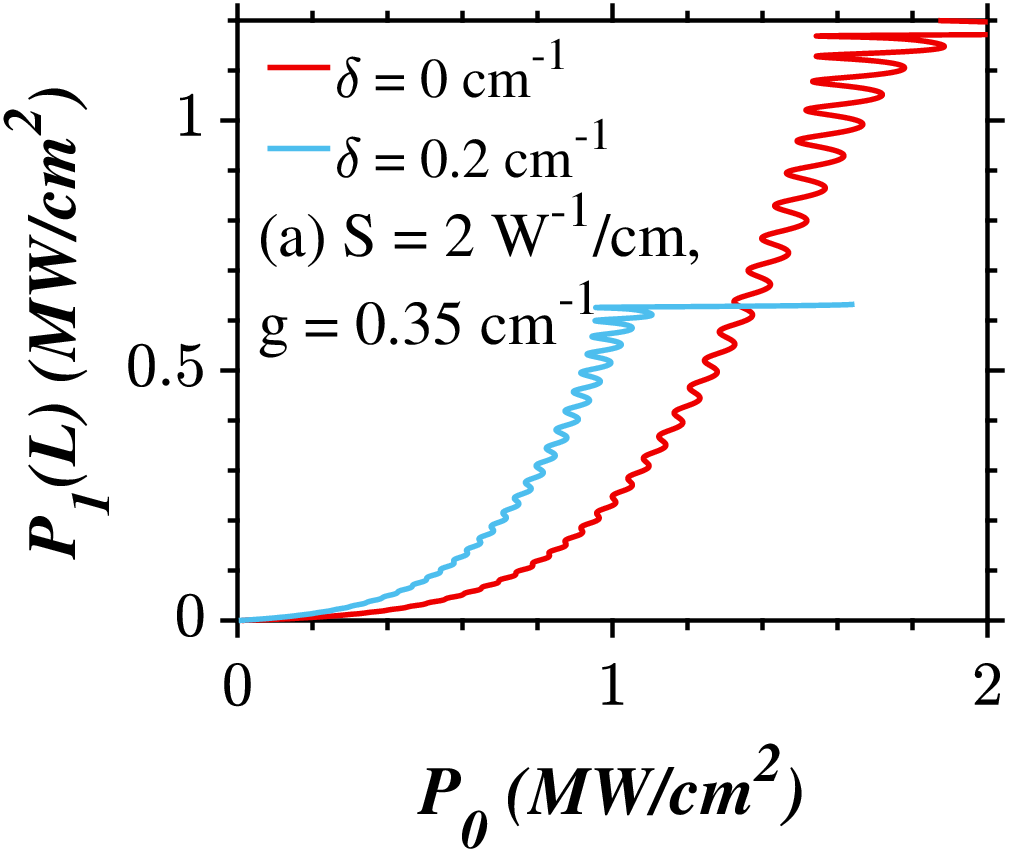}\includegraphics[width=0.5\linewidth]{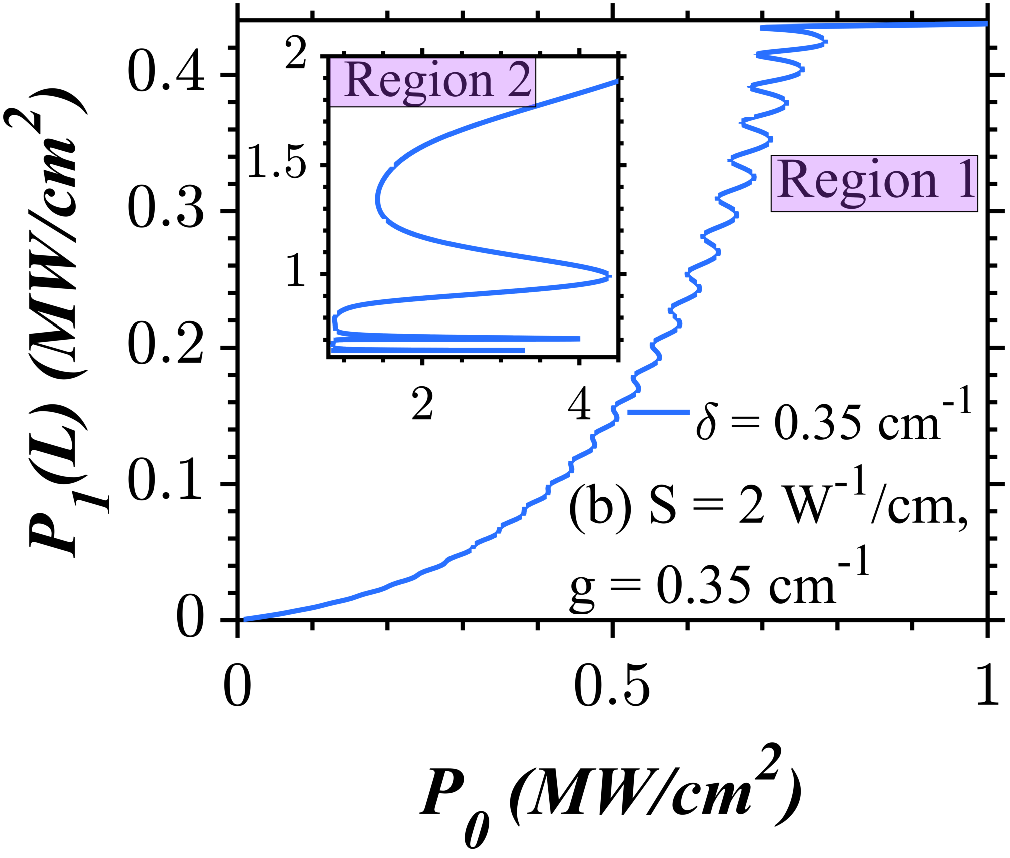}\\\includegraphics[width=1\linewidth]{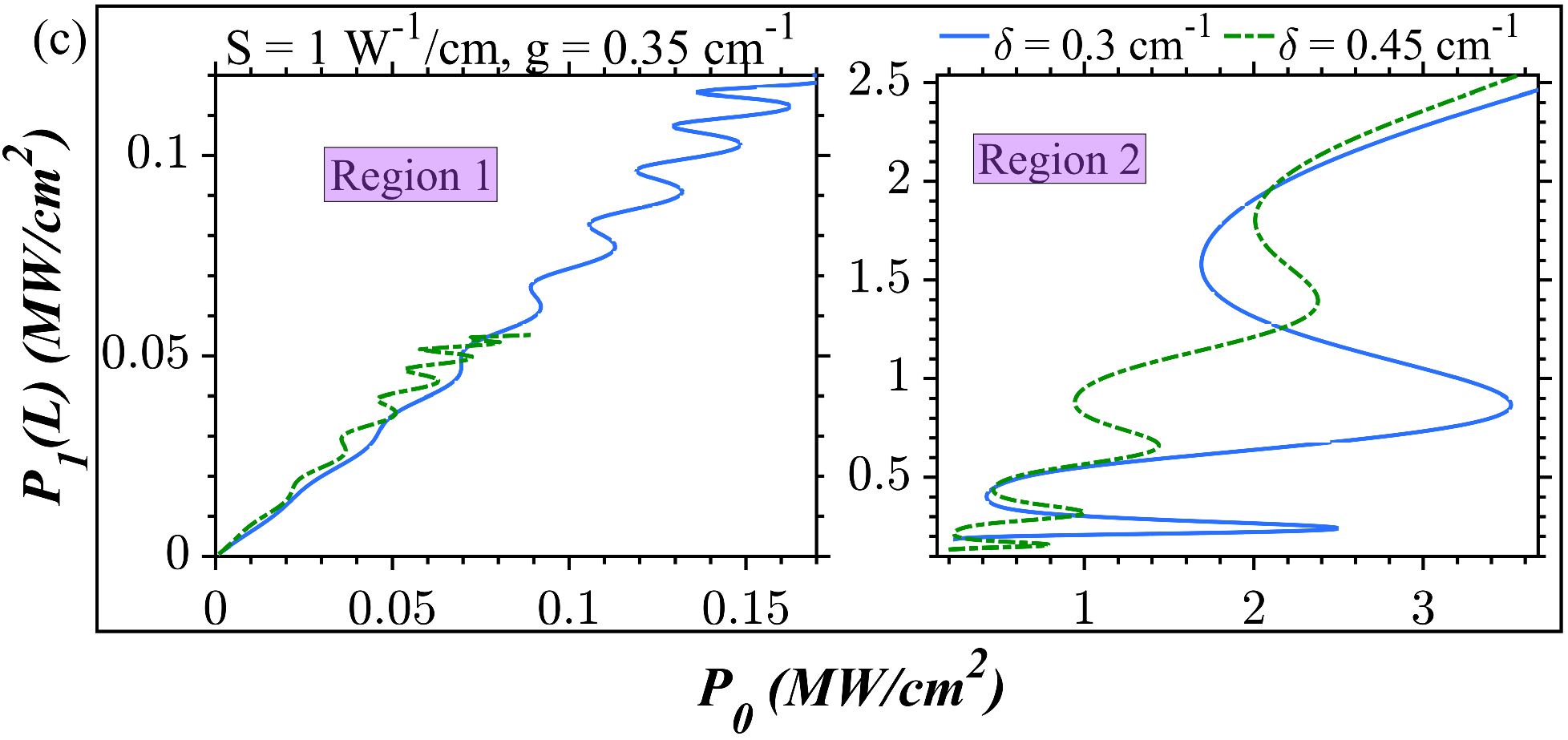}\\\includegraphics[width=1\linewidth]{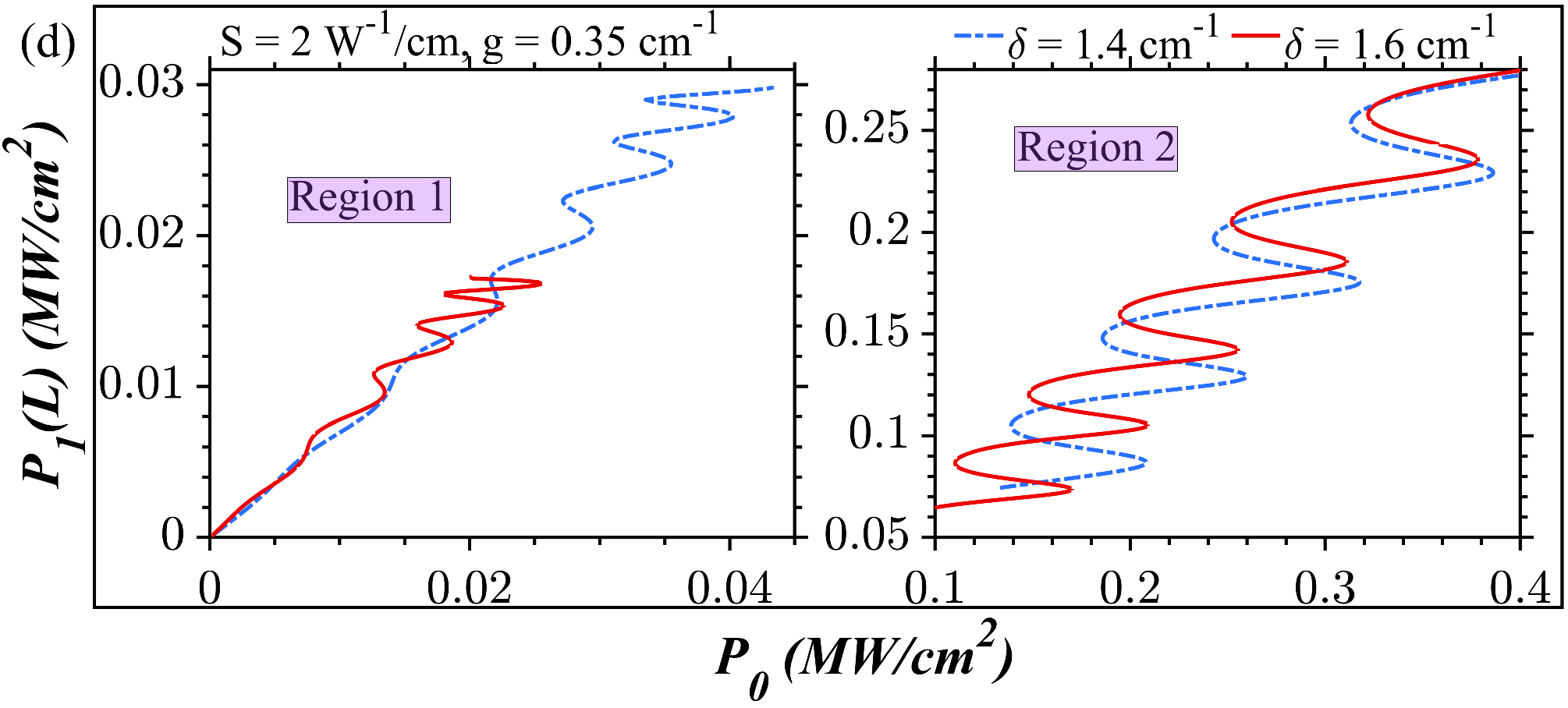}
	\caption{Low-power (a), (b) Ramp-like OM, (c) mixed OM curve with more number of ramp-like stable branches (d) Mixed OM curve with more number of S-shaped stable branches at $L = 70$ $cm$, $\kappa = 0.4$ $cm^{-1}$ and $S = 2$ $W^{-1}/cm$. The direction of light incidence is right.  }
	\label{fig7}
\end{figure}

An increase in the device length from $L = 20$ to 70 $cm$ increases the number of stable states for the given values of the input intensity, as shown in Fig. \ref{fig6}. The characteristics of the OM curve varies accordingly with the value of the detuning parameter. For $\delta < 0.3$ $cm^{-1}$, the system generates ramp-like OM curves in its input-output characteristics, as shown in Figs. \ref{fig6}(a) and (b).  The first and the successive stable branches show ramp-like or sharp variations in the output with the input intensities. The width of the successive stable branches increases with an increase in the input intensities in the ramp-like OM curves. An increase in the detuning parameter leads to a decrease in the switch-up and down intensities.

	\subsubsection{Low power mixed OM curves at  $S = 1$ $W^{-1}/cm$}
For some values of the detuning parameter, the system generates mixed OM curves that feature a fusion between ramp-like and S-shaped OM curves, as shown in Figs. \ref{fig6}(c) and (d). Earlier, we have presented a schematic of the mixed OB in Fig. \ref{fig4}(a), which holds good for all device lengths.  The system features diverse forms of mixed OM curves. For $0.25$ $cm^{-1}$ $< \delta <$ $0.4$ $cm^{-1}$, the number of ramp-like hysteresis curves in region 1 is more than the S-shaped ones in region 2, as shown in Fig. \ref{fig6}(c). An increase in the detuning parameter leads to a decrease in the switch-up and down intensities of various stable branches. Any further increase in the detuning parameter leads to the formation of asymmetric mixed OM curves in which the number of S-shaped hysteresis curves is more than that of the ramp-like states. For instance, the plot features only one ramp-like OB curve in region 1 at $\delta = 0.65$ $cm^{-1}$, as shown in  Fig. \ref{fig6}(d). 

\subsubsection{Low power ramp-like and mixed OM curves when $S = 2$ $W^{-1}/cm$}
The range of the detuning parameters at which low-power mixed OM curves appears at $S = 1$ $W^{-1}/cm$ is limited. The spectral span of the low-power mixed OM ($\delta^{mix}$) curves increases with an increase in the NL parameter as shown in Fig. \ref{fig7}. For $\delta < 0.2$ $cm^{-1}$, the system exhibits low-power ramp-like OM behavior for a given value of input intensities, as shown in Fig. \ref{fig7}(a). For $\delta > 0.2$ $cm^{-1}$, the mixed OM curves appear in the plots, as shown in Fig. \ref{fig7}(b).   An increase in the detuning parameter decreases the output, switch-up and down intensities of the different stable branches of the mixed OM curve. Also, it leads to a reduction in the number of ramp-like stable branches in region 1 and an increase in the number of S-shaped stable branches in region 2. 

	\subsection{Ultra-low power S-shaped OB (OM) curves}
	\label{Sec:5D}
	\begin{figure*}
		\centering	\includegraphics[width=0.25\linewidth]{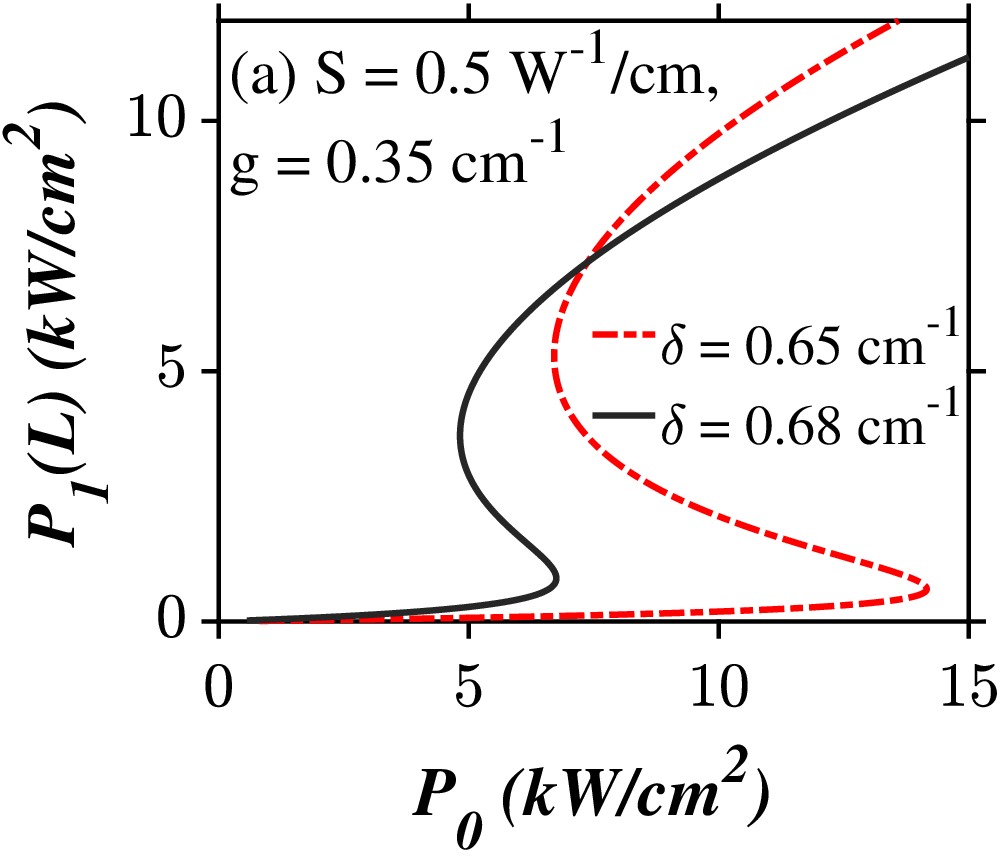}\includegraphics[width=0.25\linewidth]{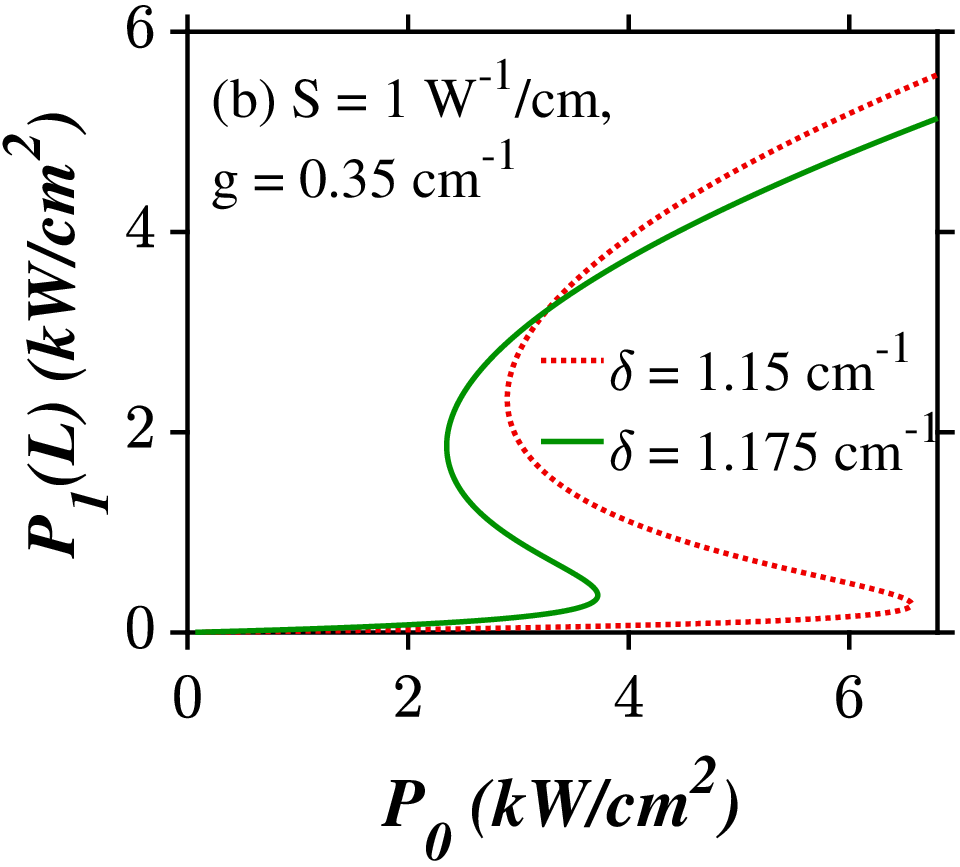}\includegraphics[width=0.25\linewidth]{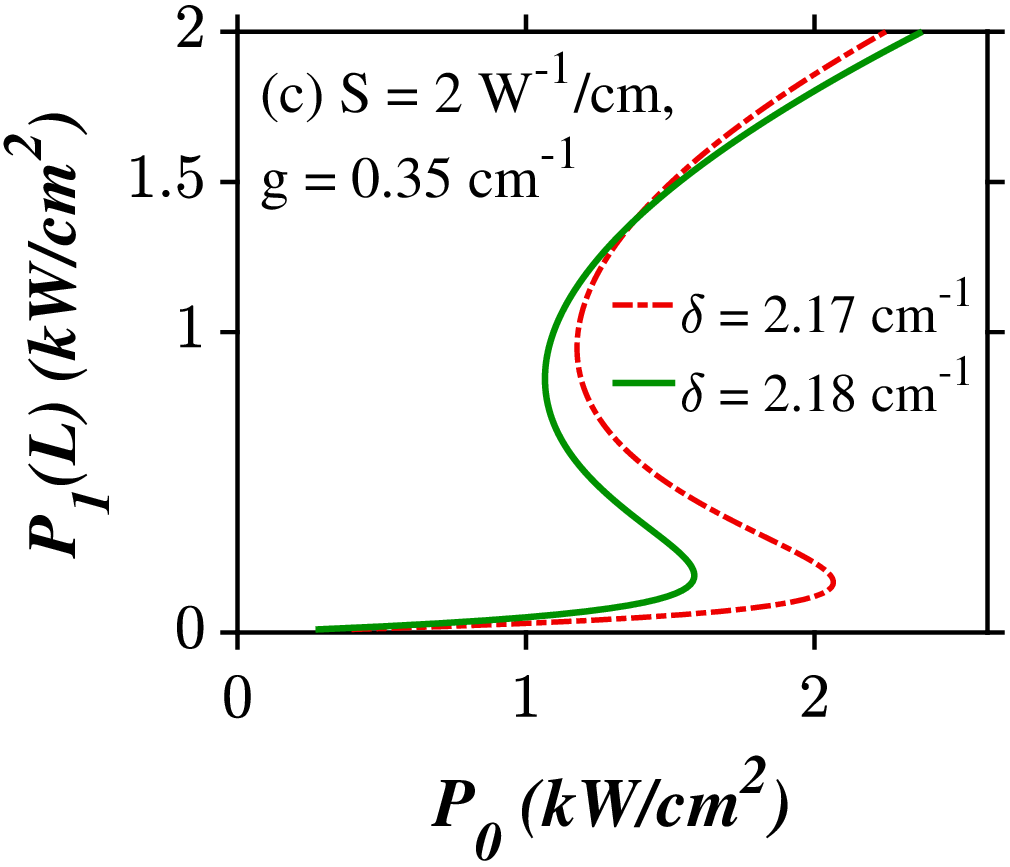}\includegraphics[width=0.25\linewidth]{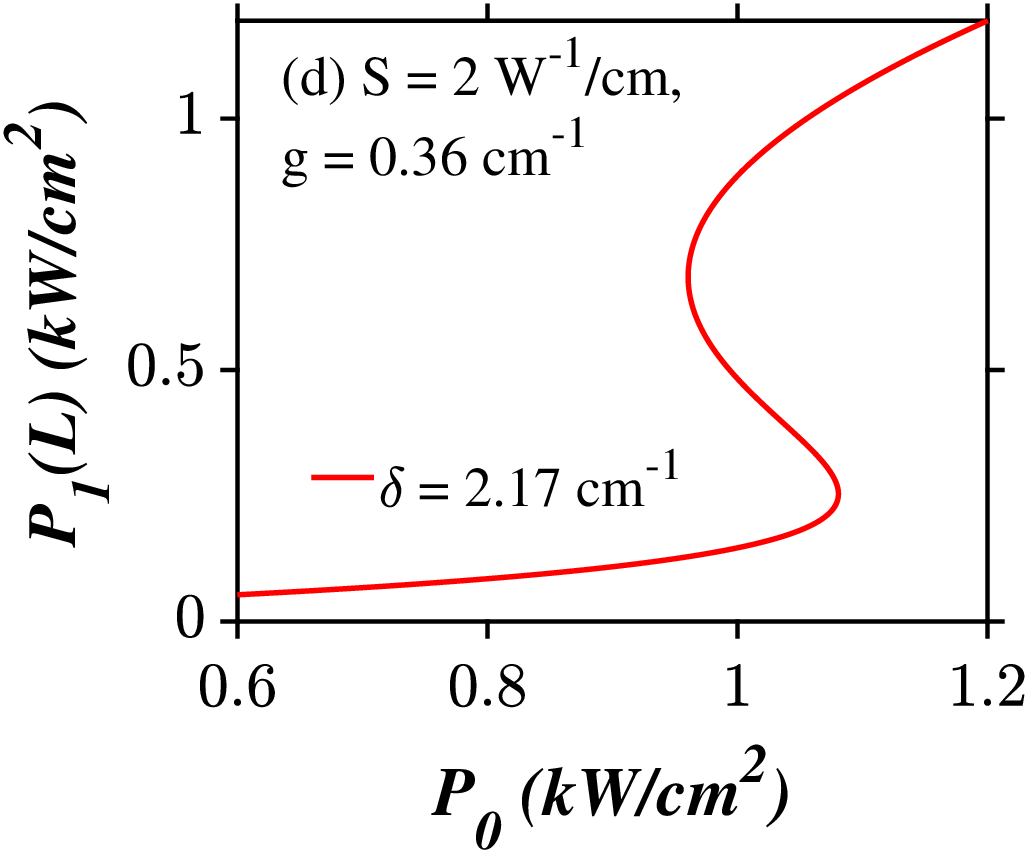}\\\includegraphics[width=0.25\linewidth]{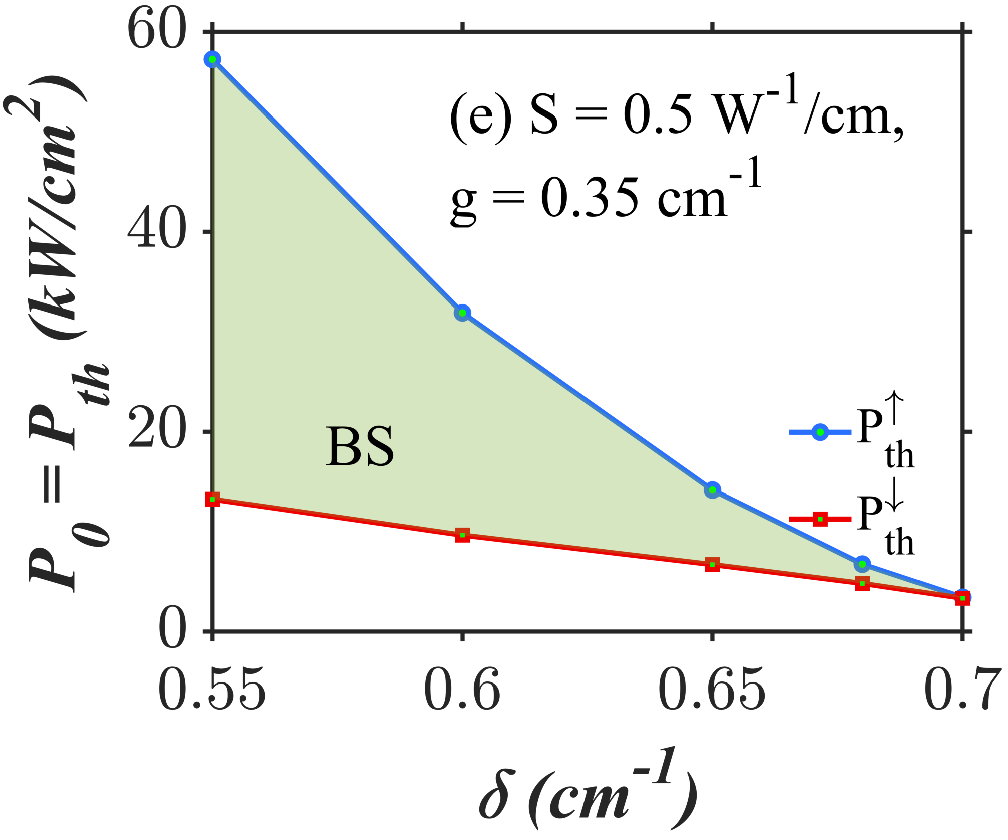}\includegraphics[width=0.25\linewidth]{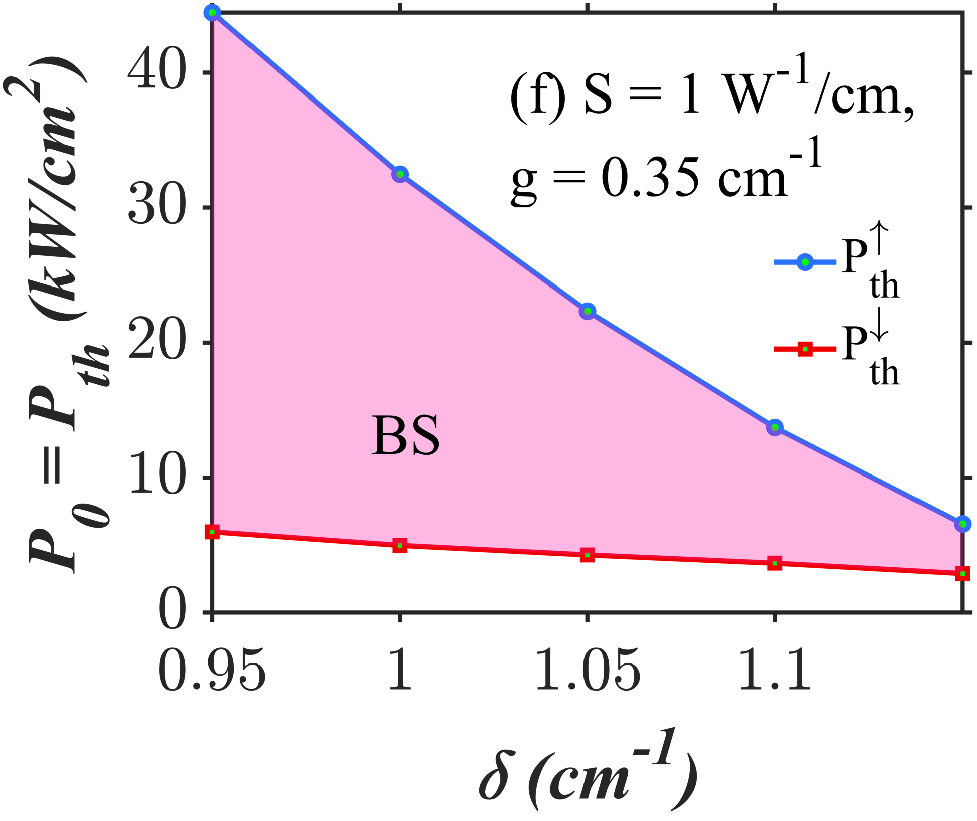}\includegraphics[width=0.25\linewidth]{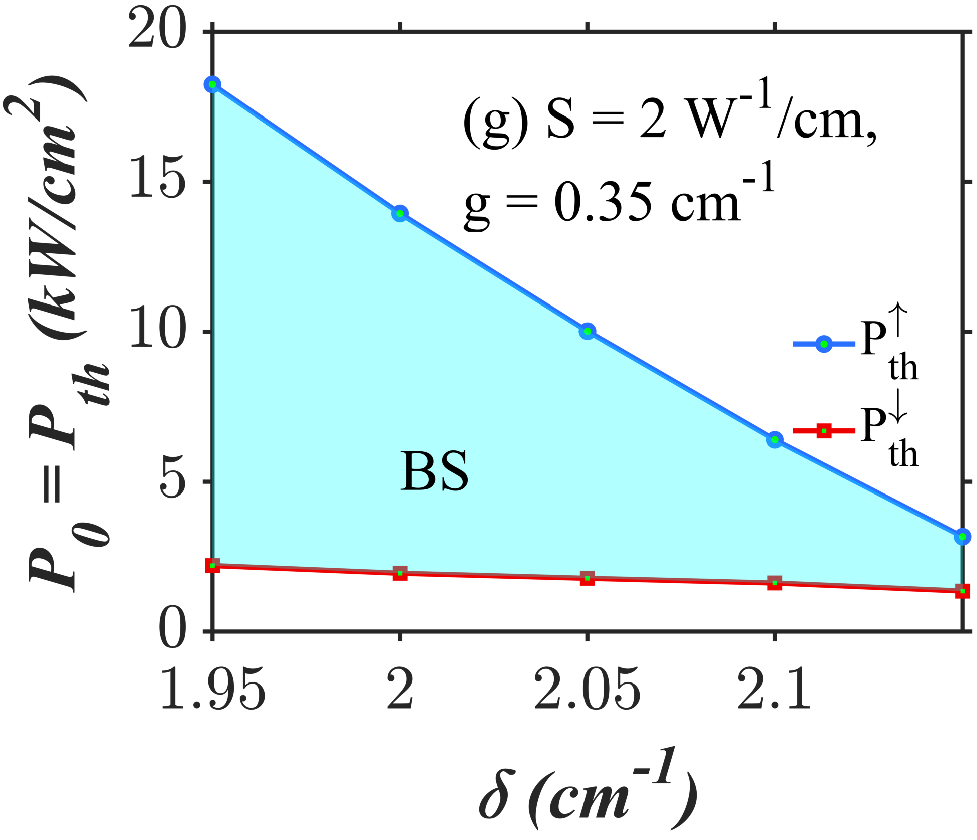}\includegraphics[width=0.25\linewidth]{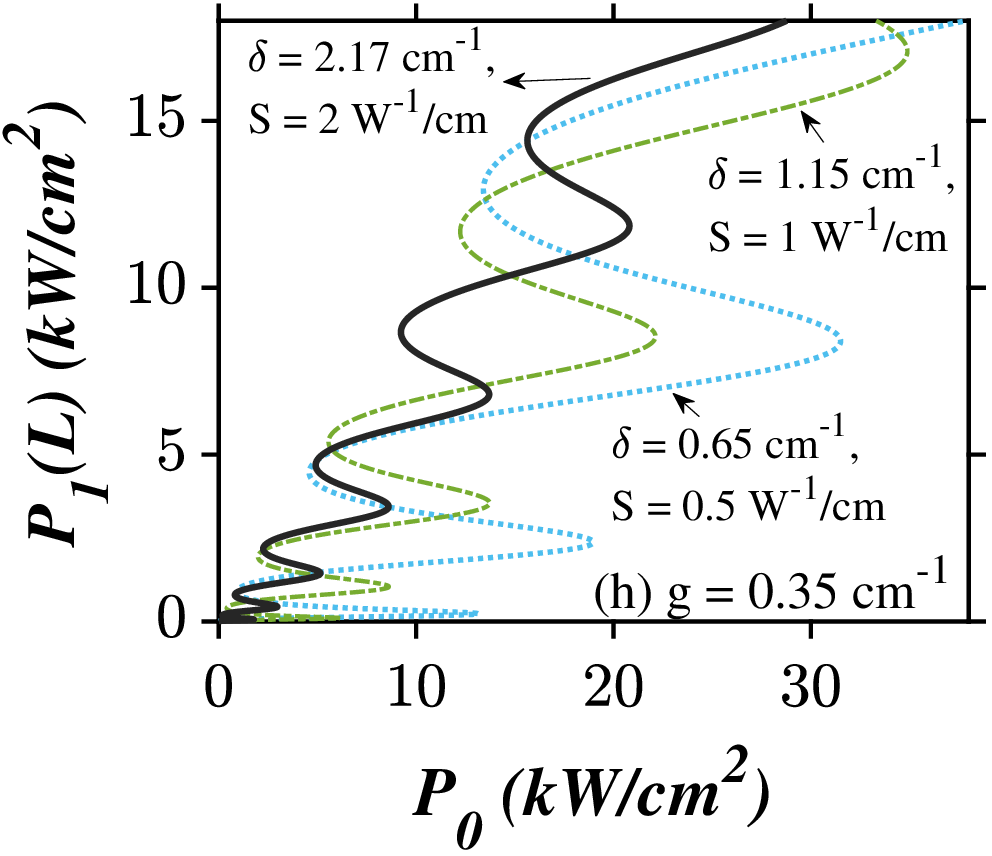}
		\caption{Frequency detuning induced ultralow power S-shaped OB curves in an unbroken PTFBG with SNL. The direction of light incidence is right. (a) -- (d) The system parameters are the same as in Fig. \ref{fig3}(a) -- (d), respectively. (e) -- (g) Variations in switching intensities and hysteresis width against the detuning and NL.  (h) low-power S-shaped OM curves at $L = 70$ $cm$ and $\kappa = 0.4$ $cm^{-1}$.}  
	\label{fig8}
	\end{figure*}
	\begin{table*}[t]
		\caption{Summary of variations in the hysteresis curves shown by PTFBG with SNL vs detuning parameter in various operating regimes. Here, LI and RI represent the left and right light incidence conditions, respectively.}
		
	\begin{tabular}{ c c c c c c c}
			\hline
			\hline
			{Nature of }	& {Classification of }& {Range of } & {Nature of variations} &{Operating}&{Light incidence} &{Figure} \\
			
			{OB/OM }	& {the curves }& {$\delta$} & { in the output intensities } & {regime }&{condition}&{no.}\\
			{curves}	& {}& {} & {} &{}&{} &{} \\
			{}	& {}& {} & {} &{}&{}&{}  \\
			\hline
			
			{Ramp-like OB }	&{Type -- I}&{$\delta$ = 0} & {sharp} & {Conventional }&{}&{Fig. 1}    \\
			{and OM curves}	& {}& {} & {} &{case}&{}&{}  \\
			{}	& {}& {} & {} &{}&{}&{}  \\
			&	{}& {}& {}& {unbroken} &{LI }&{Fig. 2}\\
			&	{}& {}& {}& {unbroken  }	&{RI  }&{Figs. 5(a), (c), (e),} \\
			&	{}& {}&{}&{} &{}&{6(a), (b), 7(a) and (b)}
			\\
			{}	& {}& {} & {} &{}&{}&{}  \\
			
			{Mixed OB and } & 	{Type -- II }&{$0<\delta<\delta_{min}$} & {sharp at}& {unbroken}&{LI} &{Fig. 4} \\
			{OM curves}	& {}& {} & {low input intensities} &{}&{} &{} \\
			{}	& {}& {} & {} &{}&{}&{}  \\
			{}&{}&{ }&{gradual at}&{unbroken}&{RI}&{Figs. 5(b), (d), (f),}\\
			
			{}	&{}&{}&{higher input intensities}&{}&{}&{6(c), (d), 7(c) and (d)}\\
			{}	& {}& {} & {} &{}&{}&{}  \\

			{S-shaped OB} & 	{Type -- III }&{$\delta_{min}<\delta<\delta_{max}$ }& {gradual}&{unbroken} &{ LI}&{Fig. 3}
			\\{and OM curves}	&{}&{}&{}&{unbroken}&{ RI}&{Fig. 8}\\
			
			\hline
			\hline
		\end{tabular}
		\label{tab3}
		
	\end{table*}
	\begin{table*}[t]
		\caption{Role of different control parameters on the critical switch up ($P_{th}^{\uparrow}$), down ($P_{th}^{\downarrow}$) intensities and hysteresis width pertaining to the hysteresis curves shown by PTFBG with SNL in different operating regimes. Here, LI and RI represent the left and right light incidence conditions, respectively.}
		\begin{center}
			\begin{tabular}{c c c c c c c}
				\hline
				\hline
				{Type of}&{Increase in}& {Impact on} & {Impact on} & {Operating}&{Light incidence}&{Figure} \\
				{OB/OM }&{the control }&{the switching }&{the hysteresis}&{regime }&{condition}&{no.}\\
				{curve}&{parameters}&{intensities}&{ width}&{ }&{}&{}\\
				\hline
				
				{Type -- I}&{S}&{increases}& {increases} &{conventional}&{}&{Fig. 1}\\
				{}&{}&{}&{}&{case}&{}&{}\\
				{}&{}&{}&{}&{}&{}&{}\\

				{Type -- I}&{g}&{increases}& {increases} &{unbroken}&{LI}&{Fig. 2 }\\
				{}&{}&{}&{}&{unbroken}&{RI}&{Figs. 5(a), (c) and (e)}\\
				{}&{}&{}&{}&{}&{}&{}\\

				{Type -- I}&{S}&{increases}& {increases} &{unbroken}&{LI}&{Fig. 2}\\
				{}&{}&{}&{}&{unbroken}&{RI}&{Figs. 5(a), (c) and (e)}\\
				{}&{}&{}&{}&{}&{}&{Figs. 6(a), (b), 7(a) and (b)}\\
				{}&{}&{}&{}&{}&{}&{}\\
				
				{Type -- II}&{g and S}&{increases}&{increases}&{unbroken}&{LI}&{Fig. 4}\\
				{}&{}&{}&{}&{}&{}&{}\\
				{Type -- II}&{$\delta$}&{decreases}&{decreases}&{unbroken}&{LI}&{Figs. 4(a), (b) and (c)}\\
				{}&{}&{}&{}&{unbroken}&{RI}&{Figs. 5(b),(d) and (f)}\\
				{}&{}&{}&{}&{}&{}&{Figs. 6(c),(d), 7(c) and (d)}\\
				{}&{}&{}&{}&{}&{}&{}\\
				{Type - III}&{$\delta$, $S$,}&{decreases}&{decreases}& {unbroken} &{LI }&{Fig. 3} \\
				{}&{and $g$}&{}&{}& {unbroken}&{RI}&{Fig.  8}  \\		
				\hline\hline
			\end{tabular}
			\label{tab4}
		\end{center}
	\end{table*}
	
	The search for realizing S-shaped OB (OM) curves at ultra-low intensities motivated us to study the switching dynamics of an unbroken PTFBG with SNL for positive values of the detuning parameter under right light incidence conditions. The system offers control over the switch-up and down intensities ($P_{th}^{\uparrow}$ and $P_{th}^{\downarrow}$ ) via an independent tuning of one or more system parameters, as shown in Figs. \ref{fig8}(a) -- \ref{fig8}(d). In the first approach, they get reduced via the frequency detuning, as shown in Figs. \ref{fig8}(a) -- (c). The higher the difference between the operating and Bragg wavelengths, the lesser is their values, and the narrower is the hysteresis width.
	
	 Tuning the NL parameter to a high value ($S \ge 1$ $W^{-1}/cm$) can perform the equivalent job of reducing the switch-up and down intensities at fixed values of the gain and loss parameter, as shown in Figs. \ref{fig8}(b) and  (c). For example, their values are less than 7 and 2.5 $kW/cm^2$ at $S = 1$ and 2 $W^{-1}/cm$, respectively.  Similar to the descriptions made in Sec. \ref{Sec:4B}, the S-shaped OB curves occurring for a set of specific values of the detuning parameter ($\delta_{min} < \delta < \delta_{max}$), vary with the value of NL. Numerical investigations reveals that light incidence direction (left and right) does not alter the values of detuning parameters at which S-shaped OB occurs. In other words, the spectral span depends only on the NL parameter and is independent of the light launching directions. Note that the hysteresis width and switching intensities are the lowest and highest at $\delta_{max}$ and $\delta_{min}$, respectively.

	 An alternate solution to reduce the switch-up and down intensities is to increase the gain and loss levels (from $g = 0.35$ to 0.36 $cm^{-1}$) at a fixed value of the detuning parameter ($\delta$). In this fashion, they reduce to a level of $<$ 1.1 $kW/cm^2$, provided that the value of NL is high ($S = 2$ $W^{-1}/cm$), as portrayed by the dash-dotted curve in Fig. \ref{fig8}(d).

 In equivalent normalized units, the curve indicated by the solid line in Fig. \ref{fig8}(d) features switching intensities less than 0.0011. In the literature, the lowest ever switching intensities recorded so far in the context of PTFBGs lies in the range $0.04<P_0<0.05$ (theoretically) \cite{sudhakar2022inhomogeneous, sudhakar2022low}. But the curve indicated by the solid line in Fig. \ref{fig8}(d) features switching intensities less than 0.0011. To the best of our knowledge, this should be the lowest switching intensities ever realized in the context of PTFBGs. 
	
	How does the reversal in the light incidence direction lower the switching intensities dramatically? There seem to exist a few scientific articles that address this natural query. When the light gets launched from the other input surface of the PTFBG, the two counter-propagating modes interact constructively in the gain regions \cite{kulishov2005nonreciprocal}. The asymmetric nature of the complex RI distribution prohibits a constructive interaction between the modes for the left light incidence condition. Even though this interactive picture traces its origin to the linear domain, it is still applicable to nonlinear PTFBGs. Studies on the spatial distribution of the total electric field (superposition of the forward and backward traveling waves) suggest that the maxima of the optical field lie in the gain for the right light incident conditions and vice-versa \cite{komissarova2019pt}. The nonreciprocal switching at ultra-low power intensities arises from such a mutual arrangement of the optical field maxima in the gain-loss structure. 
	
	\subsubsection{Low power S-shaped OM}
	
	 Ultralow-power S-shaped OB shown in the top panel of Fig. \ref{fig8} transforms into low-power S-shaped OM curves upon  increasing the device length to $L = 70$ $cm$, as shown in Fig. \ref{fig8}(h). The first hysteresis curve features the narrowest hysteresis width. The hysteresis width of the successive stable branches increases with an increase in the intensities. The interplay among the detuning, NL, gain, and loss parameters leads to a dramatic reduction in switch-up and down intensity values of various branches. In short, the individual roles of different system parameters on the switching intensities remain the same irrespective of the light incidence condition. Comparing the plots in Fig. \ref{fig3}(h) and Fig. \ref{fig8}(h), we conclude that the switch-up and down intensities required to generate the OM curves reduce  more than ten times (approximately) under a reversal in the light incidence direction. It is worthwhile to mention that the model presented here is not confined to the FBGs, but applies to periodic devices like photonic crystals that share a closer resemblance in their bandgap structure and operation to FBGs.

 		\section{Conclusions}
 	\label{Sec:6}
	The outcomes of the present work are summarized in Table \ref{tab3}. Instead of admitting an S-shaped hysteresis curve, conventional and unbroken FBGs exhibited ramp-like OB and OM curves. We discovered that the switching intensities of different stable branches in a ramp-like OB (OM) curve increases with an increase in the gain-loss parameters, as seen in Table \ref{tab4}.  We found that operating the unbroken PTFBG at wavelengths far away from the synchronous wavelength restored the S-shaped OB curves, which experienced a decrease in the switching intensities with an increase in the NL, detuning, and gain-loss parameters. The system admitted mixed OM curves in which ramp-like first stable states preceded the S-shaped stable states for the values of detuning parameter closer to the Bragg wavelength.
	
	 Dramatic reduction in the switch-up and down intensities of the various OB (OM) curves occurred under a reversal in light incidence direction. In particular, they fell below $1.1$ $kW/cm^2$ in the case of an S-shape OB curve with higher values of detuning, NL, and gain-loss parameters, which must be the lowest-switching intensities in the context of PTFBGs. The numerical results presented here confirm that the presence of SNL in a PTFBG opens a road map to control light with light in diverse fashions. Also, they indicate that the PTFBGs are not only interesting from a theoretical perspective, as they appear to be attractive platforms for the practical realization of ultra-low power AOS, thanks to the number of independent approaches they offer to control the switching intensities. A potential avenue for subsequent research could involve constructing a system comprising two tunnel-coupled gratings with opposite directions. This setup, as we hope, could facilitate the exploration of the interplay between mutually non-reciprocal transmission directions in the coupled gratings \cite{lepri2013symmetry}.

	\section*{Acknowledgements}
	SVR is supported by the  Department of Science and Technology (DST)-Science and Engineering Research Board (SERB), Government of India, through a National Postdoctoral Fellowship (Grant
	No. PDF/2021/004579). AG acknowledges the  support from the University Grants Commission (UGC), Government of India, for providing a Dr. D. S. Kothari Postdoctoral Fellowship (Grant
	No. F.4-2/2006 (BSR)/PH/19-20/0025). ML wishes to thank the DST-SERB for the award of a DST- SERB National Science Chair in which AG is now a Visiting Scientist (Grant No. NSC/2020/000029).

\end{document}